\documentclass[twocolumn,twocolappendix]{aastex63}
\newcommand{\Bkavg}{0.538}
\newcommand{\BAavg}{0.149}

\usepackage{comment}

\received{}
\revised{}
\accepted{}

\shorttitle{zCOSMOS-deep: the $2175~\textrm{\AA}$ feature}
\shortauthors{Kashino et al.}

\graphicspath{{./}{figs/}{figures/}}

\begin{document}

\title{The 2175~{\AA} Dust Feature in Star-forming Galaxies at $1.3 \le z \le 1.8$: The Dependence on Stellar Mass and Specific Star Formation Rate}

\correspondingauthor{Daichi Kashino}
\email{kashino.daichi@b.mbox.nagoya-u.ac.jp}\tabletypesize{\footnotesize}

\author[0000-0001-9044-1747]{Daichi Kashino}
\affiliation{Department of Physics, ETH Z{\"u}rich, Wolfgang-Pauli-Strasse 27, CH-8093 Z{\"u}rich, Switzerland}
\affiliation{Institute for Advanced Research, Nagoya University, Furocho, Chikusa-ku, Nagoya, 464-8601, Japan}

\author[0000-0002-6423-3597]{Simon J.~Lilly}
\affiliation{Department of Physics, ETH Z{\"u}rich, Wolfgang-Pauli-Strasse 27, CH-8093 Z{\"u}rich, Switzerland}

\author[0000-0002-0000-6977]{John D.~Silverman}
\affiliation{Kavli Institute for the Physics and Mathematics of the Universe, The University of Tokyo, Kashiwa, Japan 277-8583 (Kavli IPMU, WPI), Japan}
\affiliation{Department of Astronomy, School of Science, The University of Tokyo, 7-3-1 Hongo, Bunkyo, Tokyo 113-0033, Japan}

\author[0000-0002-7093-7355]{Alvio Renzini}
\affiliation{INAF--Osservatorio Astronomico di Padova, Vicolo dell'Osservatorio 5, I-35122, Padova, Italy}

\author[0000-0002-3331-9590]{Emanuele Daddi}
\affiliation{
CEA, Irfu, DAp, AIM, Universit\'e Paris-Saclay, Universit\'e de Paris, CNRS, F-91191 Gif-sur-Yvette, France}

\author[0000-0002-8900-0298]{Sandro Bardelli}
\affiliation{
INAF--Osservatorio di Astrofisica e Scienza dello Spazio di Bologna, via Gobetti 93/3, I-40129, Bologna, Italy
}

\author[0000-0002-9336-7551]{Olga Cucciati}
\affiliation{
INAF--Osservatorio di Astrofisica e Scienza dello Spazio di Bologna, via Gobetti 93/3, I-40129, Bologna, Italy
}

\author[0000-0001-9187-3605]{Jeyhan S.~Kartaltepe}
\affiliation{
School of Physics and Astronomy, Rochester Institute of Technology, 84 Lomb Memorial Drive, Rochester, NY 14623, USA
}

\author[0000-0002-1047-9583]{Vincenzo Mainieri}
\affiliation{
European Southern Observatory, Karl-Schwarzschild-Strasse 2, Garching bei M{\"u}nchen, Germany}

\author[0000-0003-0858-6109]{Roser Pell{\'o}}
\affiliation{
Aix Marseille Universit{\'e}, CNRS, CNES, LAM (Laboratoire d'Astrophysique de Marseille), UMR 7326, F-13388 Marseille, France
}

\author{Ying-jie Peng}
\affiliation{
Kavli Institute for Astronomy and Astrophysics, Peking University, 5 Yiheyuan Road, Beijing 100871, China}
\affiliation{
Department of Astronomy, School of Physics, Peking University, 5 Yiheyuan Road, Beijing 100871, China}

\author[0000-0002-1233-9998]{David B.~Sanders}
\affiliation{
Institute for Astronomy, University of Hawaii, 2680 Woodlawn Drive, Honolulu, HI 96822, USA
}

\author[0000-0002-5845-8132]{Elena Zucca}
\affiliation{
INAF--Osservatorio di Astrofisica e Scienza dello Spazio di Bologna, via Gobetti 93/3, I-40129, Bologna, Italy
}


\begin{abstract}

We present direct spectroscopic measurements of the broad 2175~{\AA} absorption feature in 505 star-forming main-sequence galaxies at $1.3\le z\le 1.8$ using individual and stacked spectra from the zCOSMOS-deep survey.  Significant 2175~{\AA} excess absorption features of moderate strength are measured, especially in the composite spectra.
The excess absorption is well described by a Drude profile. The bump amplitude expressed in units of $k(\lambda)=A(\lambda)/E(B-V)$, relative to the featureless Calzetti et al. law, has a range $B_k\approx0.2\textrm{--}0.8$.
The bump amplitude decreases with the specific star formation rate (sSFR), while it increases moderately with the stellar mass.
However, a comparison with local ``starburst'' galaxies shows that the high-redshift main-sequence galaxies have stronger bump features, despite having a higher sSFR than the local sample. Plotting the bump strength against the $\Delta\log\mathrm{sSFR}\equiv \log \left( \mathrm{SFR}/\mathrm{SFR_{MS}}\right)$ relative to the main sequence, however, brings the two samples into much better concordance. This may indicate that it is the recent star formation history of the galaxies that determines the bump strength through the destruction of small carbonaceous grains by supernovae and intense radiation fields coupled with the time delay of $\sim1~\mathrm{Gyr}$ in the appearance of carbon-rich asymptotic giant branch stars.

\end{abstract}



\section{Introduction}

A broad excess in the extinction curve at a rest wavelength $\lambda_\mathrm{rest}\approx2175$~{\AA} (often called the ``ultraviolet (UV) bump'') is the strongest signature of dust in the interstellar medium (ISM).  It has attracted attention since its discovery by \citet{1965ApJ...142.1683S} as a unique probe of the nature of dust in galaxies (see \citealt{2020ARA&A..58..529S} for the latest review).
Direct measurements of the extinction curves toward individual stars are limited in the Milky Way (MW) and a few very nearby extragalactic objects.
The strong UV bump is ubiquitously seen in the extinction curves of stars in the MW \citep{1986ApJ...307..286F} and the Large Magellanic Cloud (LMC).  The extinction curves in the LMC2 supershell region, which is close to 30 Doradus (a starbursting giant H\,{\sc ii} region in the LMC), shows a noticeably weaker bump than that in the average LMC extinction curve \citep{1986AJ.....92.1068F,2003ApJ...594..279G}.  The 2175~{\AA} feature was also found to be weaker in M31 \citep{1996ApJ...471..203B} and almost absent in the Small Magellanic Cloud (SMC; \citealt{1984A&A...132..389P}).
These observations indicate that the nature of interstellar dust grains depends significantly on the local environment.

On galaxy scales, the impact of dust on the integrated spectrum is described by an \textit{attenuation} curve, which includes the effects due to a more complicated spatial arrangement of dust and sources of light than the simple case of a single star behind a dust screen. Nearby starburst galaxies have been found to have no significant UV bump in their attenuation curves, like the SMC \citep{1994ApJ...429..582C,1997ApJ...487..625G,2000ApJ...533..682C}.
Using radiative transfer calculations, \citet{2000ApJ...528..799W} showed that the so-called Calzetti attenuation law could be reproduced by an SMC-like extinction curve combined with a clumpy shell configuration.  \citet{2003ApJ...587..533V} showed that dust attenuation in Lyman break galaxies at high redshifts ($z>2$) is also consistent with that predicted with dust that has SMC-like, rather than MW-like, characteristics.
The \citet{2000ApJ...533..682C} attenuation curve with no bump feature has thus been commonly assumed for both local and distant star-forming galaxies.

More recent observations, however, have detected and even accurately measured the bump feature for low- and high-redshift star-forming galaxies based on both photometry and spectroscopy \citep{2007A&A...472..455N,2009A&A...499...69N,2011A&A...533A..93B,2012A&A...545A.141B,2015ApJ...800..108S,2017ApJ...851...90B,2018ApJ...859...11S,2020ApJ...888..108B,2020ApJ...899..117S}.  
Notable work by \citet{2009A&A...499...69N} used a statistical sample of $\sim200$ star-forming galaxies at $1<z<2.5$ to show that at least 30\% of the sources exhibit a significant attenuation bump feature in their spectra and to directly determine the UV bump profiles using stacks.
Another approach has been to fit a galaxy's spectral energy distribution (SED) measured in multiple photometric bands with stellar population synthesis models while applying different attenuation laws in which UV bump features of different strengths are implemented \citep[e.g.,][]{2011A&A...533A..93B,2018ApJ...859...11S}.  All of these studies, however, have indicated that the strengths of the UV bump feature in $z\gtrsim1$ galaxies are typically much weaker than observed in the MW extinction curve, usually of intermediate strength between the LMC2 supershell and the SMC extinction curves, with the latter having no bump at all.  

Measurement of the UV bump across a range of redshifts is important for understanding the formation of dust grains through cosmic time. 
The bump feature is known to be reproduced by resonant absorption by carbon $sp^2$ bonds that efficiently occurs on the surfaces of carbonaceous small grains (grain size $a \lesssim 10 ~\mathrm{nm} \ll \lambda$; \citealt{1971Natur.229..237G,1984ApJ...285...89D,2009MNRAS.394.2175P}). However, the carrier(s) of the bump feature is still under debate, and other potential species have also been proposed \citep[e.g.,][]{2005Sci...307..244B}.
Regardless of the precise identification of the bump carrier, analysis of the likely evolution of the distribution of grain sizes in a galaxy has succeeded in reproducing the observed extinction curves \citep[e.g.,][]{2013MNRAS.432..637A,2014MNRAS.440..134A,2015MNRAS.447.2937H}.
This suggests that the bump strength is linked to the relative abundance of small dust grains.  The main processes determining the grain size distribution include dust production by stellar sources (see \citealt{2011A&ARv..19...43G} for a review), grain growth through metal accretion \citep[e.g.,][]{1998ApJ...501..643D,2011EP&S...63.1027I}, grain destruction in the hot ISM \citep{1980ApJ...239..193D,1989IAUS..135..431M}, coagulation \citep[e.g.,][]{2009A&A...502..845O}, shattering \citep[e.g.,][]{2004ApJ...616..895Y},\footnote{See, e.g., \citet{2015MNRAS.447.2937H} for a brief summary of all of these processes on dust in different size regimes.} and possible selective removal by galactic winds \citep[e.g.,][]{2015ApJ...810...39B}.
The efficiencies of many of these different processes will depend on the properties of the host galaxies. 
However, our knowledge of the link between the galaxy properties and the UV bump characteristics is currently limited, particularly at high redshifts.  

In this work, we directly measure the UV bump profiles for a large sample of star-forming galaxies at $1.3\le z \le 1.8$ and correlate the derived bump strength with the global properties of the galaxies, specifically the stellar mass and star-formation rate (SFR).  We utilize the rest-frame UV spectra that have been obtained by the VIsible Multi-Object Spectrograph (VIMOS) mounted on the Very Large Telescope (VLT) UT3 in the zCOSMOS-deep survey (\citealt{2007ApJS..172...70L}; S.~J.~Lilly et al. 2021, in preparation).

The paper is organized as follows.
Section \ref{sec:data} presents an overview of the observations and describes the sample selection.
Section \ref{sec:methodology} describes our spectral analysis, including the SED fitting for the sample.
The results are presented in Section \ref{sec:results}.
These are discussed in Section \ref{sec:discussion}, and Section \ref{sec:summary} provides a summary of the paper.
This paper uses a standard flat cosmology $(h=0.7, \Omega_\mathrm{M}=0.3, \Omega_\Lambda=0.7)$. Magnitudes are quoted on the AB system.
A \citet{2003PASP..115..763C} initial mass function (IMF) is used throughout.

\section{Data}
\label{sec:data}
We investigate the possible presence of the UV bump feature in dust attenuation curves using rest-frame UV spectra obtained through the zCOSMOS-deep project.  This section gives a brief summary of the survey and describes the sample selection.

\subsection{Observations}
\label{sec:observations}
The zCOSMOS-deep redshift survey has observed $\sim10^4$ galaxies in the central $\sim 1~\mathrm{deg^2}$ of the COSMOS field \citep{2007ApJS..172....1S}.  Observations were carried out with the VIMOS \citep{2003SPIE.4841.1670L} mounted on the 8m VLT/UT3 telescope from 2005 to 2010.  Observations used the low-resolution (LR) blue grism with $1.\!\!\arcsec0$ slits, yielding a spectral resolution of $R \sim 180$ and a spectral range of $\approx3600\textrm{--}6700~\textrm{\AA}$. All masks are observed with the slits oriented north-south. 

The selection of the targets was performed based on a then-current version of the COSMOS photometric catalog.  All of the objects were color-selected to preferentially lie at high redshifts, mostly through a $BzK$ method \citep{2004ApJ...617..746D} with a nominal $K_\mathrm{AB}$ magnitude cut at 23.5, supplemented by the $ugr$ selection \citep{2004ApJ...604..534S}.  An additional blue magnitude selection $B_\mathrm{AB}<25.25$ was adopted for most objects.  These selection criteria yield a set of star-forming galaxies that lie mostly in the redshift range $1.3<z<3.5$ \citep{2007ApJS..172...70L}.  In total, the survey field covers a $0.92\times0.91~\mathrm{deg^2}$ region centered on $\textrm{R.A.}=10^\mathrm{h}00^\mathrm{m}43^\mathrm{s}$, $ \textrm{decl.}=+02\arcdeg 10\arcmin 23\arcsec$.  The central $0.60\times0.62~\mathrm{deg^2}$ region is referred to as the full sampling area, where the sampling rate achieved is $\approx57\%$, while it is $30\%$ in the outer region.

Data reduction was carried out using the VIPGI software \citep{2005PASP..117.1284S}. Redshift measurements were visually inspected using 2D and 1D reduced spectra by identifying multiple prominent spectral features.
In general, we often rely on strong absorption lines, such as Si\,{\sc iv}~$\lambda$1393,~1402, C\,{\sc iv}~$\lambda$1548,~1550, Fe\,{\sc ii}~$\lambda$1608, and Al\,{\sc ii}~$\lambda$1670 at $z\sim 2\textrm{--}3$.  The Ly$\alpha$ emission line and break are key features at $z>2.1$.  At lower redshifts ($z\sim1.3\textrm{--}2.0$), some redder absorption lines are important for the redshift identification, such as Fe\,{\sc ii}~$\lambda$2344,~2374,~2382,~2586,~2600 and Mg\,{\sc ii}~$\lambda$2796,~2803.
The spectroscopic redshifts have been determined purely from the spectra, independently of a photometric redshift or other information about the objects. A typical velocity error is $\approx420~\mathrm{km~s^{-1}}$ ($\sigma(z)/(1+z)\approx0.0014$) in the redshifts including the uncertainties in the wavelength calibration.  
We excluded the flux data in the wavelength ranges that are highly impacted by sky lines (5536--5617, 5836--5976, 6189--6399~{\AA}) and those impacted by zeroth-order contamination of individual spectra.  We will describe further spectrophotometric correction for the slit loss in Section \ref{sec:flux_calibration}.

\subsection{Sample Selection}
\label{sec:selection}

We constructed the sample from the full catalog of the zCOSMOS-deep survey (S.~J.~Lilly et al. 2021, in preparation).
The redshift range for the current analysis is limited to $1.3\le z \le 1.8$ to ensure that the spectral window of the VIMOS LR blue grism covers the rest-frame wavelength interval around the 2175~{\AA} feature.  To spectrophotometrically calibrate the spectra and perform SED fitting, we limited the sample to having a clear photometric counterpart in the COSMOS2015 photometric catalog \citep{2016ApJS..224...24L} and excluded a handful of targets that do not have one.  
Galaxies detected in X-rays were also excluded using the column \texttt{type} in the COSMOS2015 catalog to remove possible active galactic nuclei from the sample.

For the majority of the sample, we use the spectroscopic redshift ($z_\mathrm{spec}$) measurements from zCOSMOS-deep.  We use all objects with a very secure zCOSMOS-deep redshift ($\mathrm{class}=3$ or 4) within the selection range.  For those with confidence $\mathrm{class}=2$, we use only those that are consistent to within $|z_\mathrm{phot}-z_\mathrm{spec}|/(1+z_\mathrm{spec})\le0.1$ of the photometric redshift in the COSMOS2015 catalog.  The redshifts in these two categories are both estimated to be $\ge99\%$ reliable (S.~J.~Lilly et al. 2021, in preparation).

\begin{figure}[tbp]
\begin{center}
\includegraphics[width=3.5in]{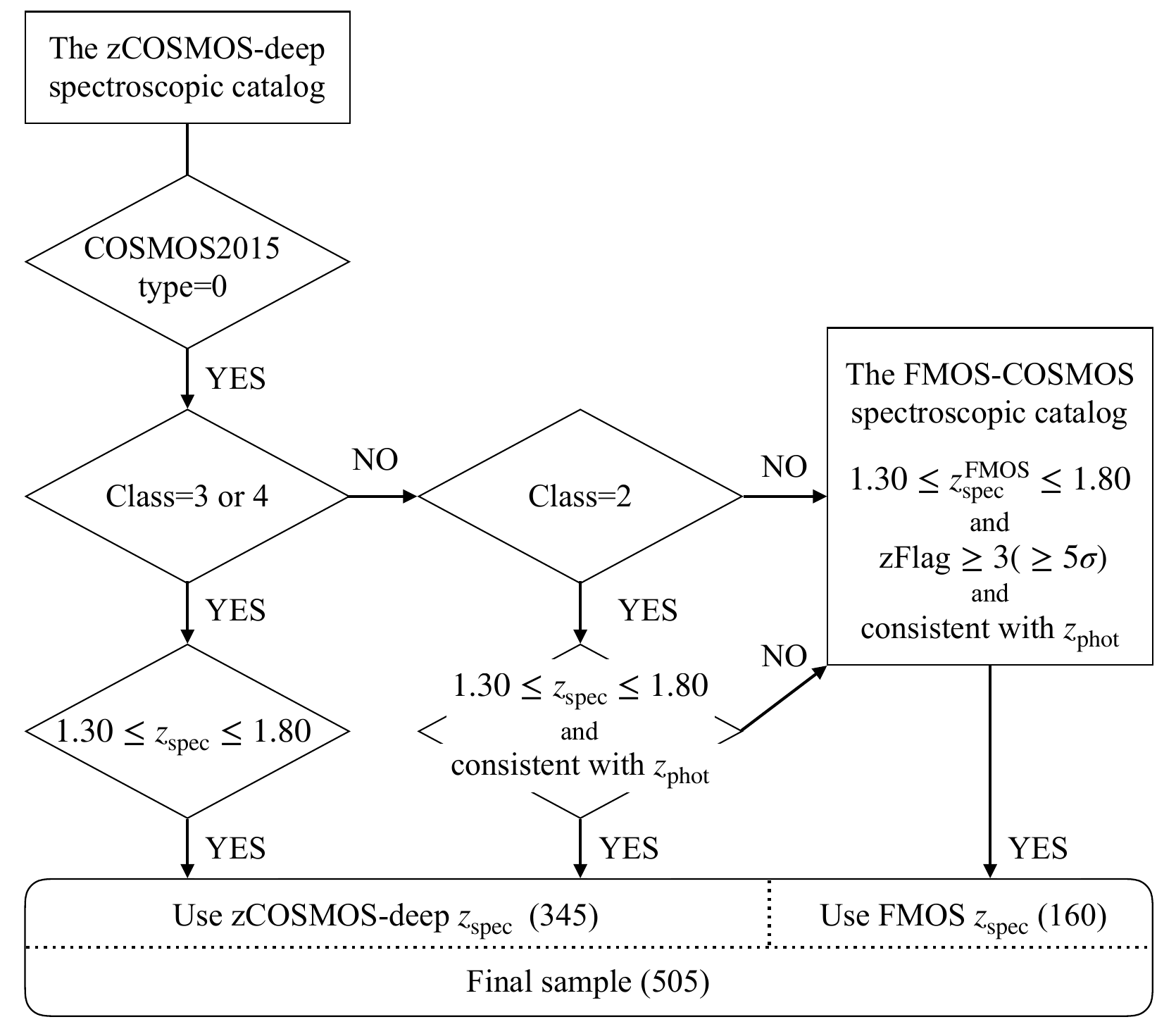}
\caption{Flowchart of the sample selection.  The consistency with $z_\mathrm{phot}$ is examined based on a threshold of $|z_\mathrm{phot}-z_\mathrm{spec}|/(1+z_\mathrm{spec})\le 0.1$ against the photometric redshift in the COSMOS2015 catalog.
\label{fig:sample_selection}}
\end{center}
\end{figure}

For some sources without reliable zCOSMOS-deep redshifts, we use the $z_\mathrm{spec}$ measurements from the FMOS-COSMOS survey \citep{2015ApJS..220...12S,2019ApJS..241...10K}, which is a near-infrared (IR) spectroscopic survey of star-forming galaxies and covers most of the zCOSMOS-deep survey field. These spectroscopic redshifts are based on the detection of the H$\alpha$ emission line and some additional rest-frame optical lines (e.g., [N\,{\sc ii}]$\lambda6584$, H$\beta$, and [O\,{\sc iii}]$\lambda$5007) in the $H$- and $J$-band medium-resolution ($R\sim3000$) spectra.
We accepted the FMOS $z_\mathrm{spec}$ measurements if the quality flag ($z$Flag) is $\ge3$ (which corresponds to detecting $\ge1$ line at $\ge5\sigma$) and $z_\mathrm{spec}$ is consistent with $z_\mathrm{phot}$. Figure \ref{fig:sample_selection} shows the flowchart of selecting the sample and determining which zCOSMOS-deep or FMOS-COSMOS measurement of $z_\mathrm{spec}$ is adopted.

Finally, we excluded 17 sources for which the UV continuum is barely detected or that suffered from severe spectral contamination or presented strong broad emission lines.

\begin{figure}[tbp]
\begin{center}
\includegraphics[width=3.5in]{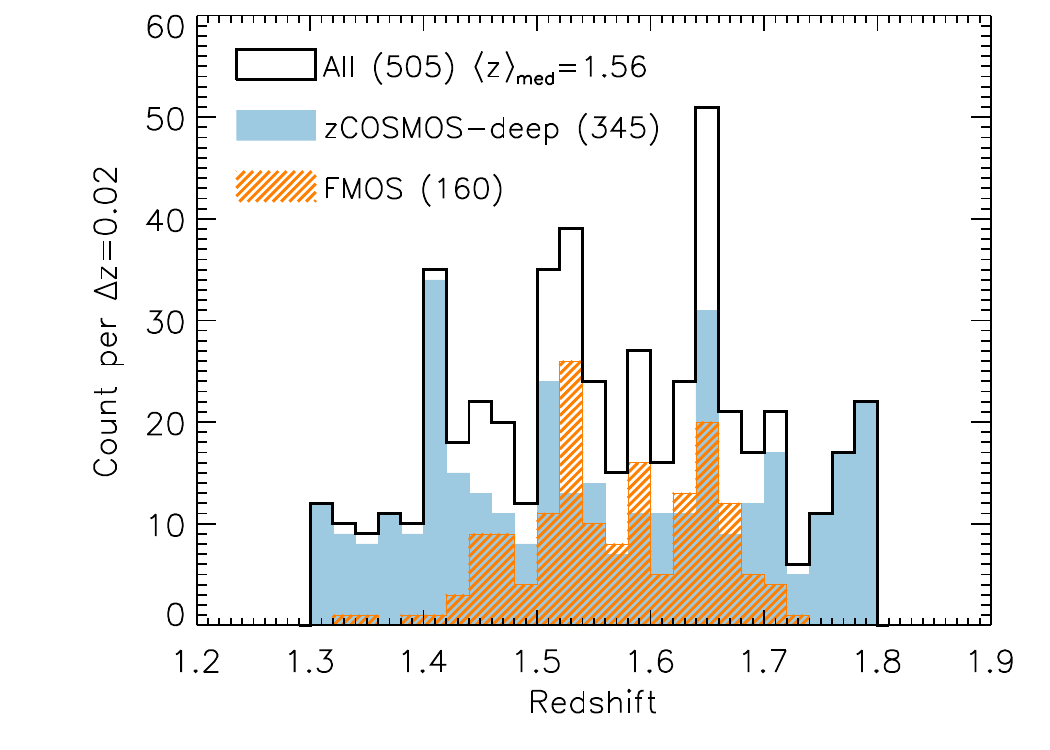}
\caption{Spectroscopic redshift distribution of our zCOSMOS-deep sample of 505 star-forming galaxies at $1.3\le z_\mathrm{spec}\le 1.8$.  The filled blue histogram indicates the subset ($N=345$) for which the zCOSMOS-deep $z_\mathrm{spec}$ measurements were adopted, and the orange hatched histogram indicates the additional sources ($N=160$) whose $z_\mathrm{spec}$ comes from the FMOS-COSMOS catalog.
\label{fig:histogam_zspec}}
\end{center}
\end{figure}

The final sample consists of 505 galaxies, of which $z_\mathrm{spec}$ comes from the FMOS-COSMOS catalog for 160 galaxies.  Figure \ref{fig:histogam_zspec} shows the redshift distribution of our sample, where the subset of 345 galaxies whose $z_\mathrm{spec}$ is based on the zCOSMOS-deep catalog shows a nearly homogeneous coverage of the redshift range of interest.  The additional component of 160 sources based on the FMOS-COSMOS catalog is mostly limited to the range $1.4\lesssim z \lesssim1.7$.  The median redshift of the entire sample is $\left<z\right>_\mathrm{med}=1.556$.  As shown below in Section \ref{sec:M_and_SFR}, the galaxies in our sample are typical main-sequence (MS) galaxies in the stellar mass range of $9.5 \lesssim \log(M_\ast/M_\odot) \lesssim 11$.

\subsection{Local Sample}
\label{sec:local}

Our goal is to measure the excess absorption due to the UV bump with respect to the so-called Calzetti law that is often adopted as a baseline attenuation curve.  We thus compiled nearby ($\left<z\right>=0.012$) starburst galaxies from \citet{1994ApJ...429..582C}, \citet{1997AJ....113..162C}, and \citet{2000ApJ...533..682C}.  The Calzetti et al. series of papers utilized a total of 47 local ``starburst'' galaxies to infer the featureless attenuation curve.\footnote{The sample sizes in three Calzetti et al. (1994, 1997, and 2000) papers are, respectively, 39, 24, and 8.  Excluding duplicates, the number of unique starburst galaxies is 47.}  
For 23 of these, \citet{1994ApJ...429..582C} measured the excess absorption in their spectra and showed that the bump feature is, on average, negligible compared to that seen in the MW and LMC extinction laws (see their Figure 13).  We will utilize these measurements together with the available reddening $E(B-V)$ measurements (see Table 1 of \citealt{1994ApJ...429..582C}) for comparison with our own results.

For these local sources, we adopted the estimates of the stellar mass ($M_\ast$) and SFR inferred from Galaxy Evolution Explore (GALEX) and Wide-field Infrared Survey Explore (WISE) photometry by \citet{2019ApJS..244...24L}.  The $M_\ast$ and SFR estimates are available for 28 sources.  For the majority of these (25/28), for which both GALEX and WISE data are available, the SFR is estimated from a linear combination of the observed luminosities in the GALEX far-UV (FUV) and WISE4 (22~$\mu$m) bands, and the stellar mass is converted from WISE1 (3.4~$\mu$m) with the SFR/WISE1-dependent mass-to-luminosity ratio.  For the remaining three with no GALEX data, the SFR is estimated only from WISE4, and the mass is converted from WISE1 with the WISE4$-$WISE1 color-dependent mass-to-luminosity ratio.  The average MS and the scatter ($\approx0.3$~dex) of all local galaxies in \citet[][see Equation 19]{2019ApJS..244...24L} are consistent with other literature results.

Throughout the paper, we utilize the 34 local sources that have either $M_\ast$/SFR or bump strength measurements.  Of these, both the $M_\ast$/SFR and bump measurement are available for 17, only the $M_\ast$/SFR estimates are available for 11, and only the bump measurement is available for six.

\section{Methodology}
\label{sec:methodology}

In this section, we describe the method used to measure the 2175~{\AA} bump feature in the spectra. We also describe how other galaxy properties are derived. In the following subsections, we will first introduce the formulation of the attenuation curves and the approach to determining the attenuation curve from precisely spectrophotometrically calibrated spectra.

We will then move to analyzing the VIMOS spectra themselves.  Every single spectrum is carefully corrected for slit losses in a way that will not introduce artificial effects that may mimic the UV bump.  After the correction, we then characterize the shapes of the UV continua of the individual galaxies and also employ stacking analysis to increase the signal-to-noise ratios (S/Ns) and precisely measure the UV bump profiles of aggregates of galaxies.  Each step of the analysis is described in detail below.

\subsection{Dust Attenuation Prescription}
\label{sec:dust_prescription}

As usual, we define the attenuation at a given wavelength in magnitudes, $A(\lambda)$, to be 
\begin{equation}
    A(\lambda) = 2.5\log \left(\frac{ f_\lambda^\mathrm{int}(\lambda)}{f_\lambda^\mathrm{obs}(\lambda) }\right),
\end{equation}
where $f_\lambda^\mathrm{int}$ and $f_\lambda^\mathrm{obs}$ are the intrinsic flux density (before dust obscuration) and the observed flux density (after dust obscuration), respectively.  

The wavelength dependence of dust attenuation is formulated in different ways in the literature (see \citealt{2018ApJ...859...11S} for a brief summary).  Since the effects of dust are multiplicative, the observed $A(\lambda)$ is normalized by some measure of the overall amount of extinction present (such as $E(B-V)$) to obtain the wavelength dependence of the extinction, independent of the overall amount of extinction. Throughout the paper, we will refer to $k(\lambda)\equiv A(\lambda)/E(B-V)$ as an extinction curve if it is measured toward individual stars and an attenuation curve if it is for the integrated spectra of galaxies.

In this work, we will assume that the dust attenuation curves $k(\lambda)$ of high-redshift star-forming galaxies have two components: one with a smooth dependence on wavelength, which is assumed to be common to all sources, and another ``excess'' component from the UV bump feature at $\lambda_\mathrm{rest}\approx2175$~{\AA}, whose strength we wish to determine.  For the smooth component, we adopt the \citet{2000ApJ...533..682C} attenuation curve, $k_\mathrm{Cal}(\lambda)$, and thus denote the observed attenuation curve as a sum of $k_\mathrm{Cal}(\lambda)$ and an excess ($k_\mathrm{bump}(\lambda)$) due to the UV bump:
\begin{eqnarray}
    k(\lambda) = k_\mathrm{Cal}(\lambda)+k_\mathrm{bump}(\lambda).
\end{eqnarray}
For the bump component, we employ a Drude profile with a parameterization of \citet{1990ApJS...72..163F},
\begin{eqnarray}
    k_\mathrm{bump}(\lambda)=c_3 D(x;x_0,\gamma), \label{eq:kbump}\\
    D(x;x_0,\gamma)=\frac{x^2}{(x^2-x_0^2)^2+x^2\gamma^2}, \label{eq:drude}
\end{eqnarray}
where $x$, $x_0$, and $\gamma$ are expressed in units of inverse wavelength; $x_0$ is the inverse of the central wavelength ($x_0=\lambda_0^{-1}$), and $\gamma$ is the FWHM of the UV bump in $\lambda^{-1}$ space.  The FWHM in wavelength is therefore given as $w_\lambda\approx \gamma/\lambda_0^2$.  
The parameter $c_3$ is related to the peak amplitude of the excess, $B_k$, as $B_k\equiv c_3/\gamma^2$, 
measured from the $k(\lambda)$ attenuation curve.  
The fraction of the total absorption at 2175~{\AA} that is due to the bump feature, which is denoted as $f_\mathrm{bump}$, is therefore given by $f_\mathrm{bump}=B_k/(B_k + k_\mathrm{Cal}(2175~\textrm{\AA}))$, where $k_\mathrm{Cal}(2175~\textrm{\AA})$ is fixed to have a value of 8.48.

Observationally, we require the knowledge of $E(B-V)$ to derive $k(\lambda)$, and thus $k_\mathrm{bump}(\lambda)$ and $B_k$, from the ``observed'' $A(\lambda)$.  We could instead directly measure the absolute level of the excess absorption, $A_\mathrm{bump}(\lambda)$ in $A(\lambda)$. 
We therefore define $B_A$ as the amplitude of the bump in the observed $A(\lambda)$.  By definition, this means that $B_A = B_k \times E(B-V)$.  

It is important to note that the measurement of $B_k$ depends on the assumed shape of the baseline smooth attenuation curve over a wide range of wavelengths (i.e., $k_\mathrm{Cal}(\lambda)$ in our case). For a given observed UV continuum, with a certain slope and bump signature, an application of a baseline curve that has a steeper rise toward the FUV, such as the SMC extinction curve, will result in a lower estimated $E(B-V)$ and thus require a higher $B_k$ to account for the same level of the absolute excess absorption seen in $A(\lambda)$.  On the other hand, the estimate of $B_A$ does not depend much on the shape of the baseline curve.

Our primary goal is to measure the strength of the UV bump that manifests itself in the attenuation curves, i.e., $B_k$, as a function of galaxy properties.  However, we will also show the corresponding $B_A$ measurements for reference, as this is the more fundamental observable quantity and largely independent of the assumptions about the shape of the smooth attenuation curve.  We will later discuss the effects of possible changes in the overall UV slope of the attenuation curve in Section \ref{sec:discussion:slope}.

\subsection{SED Fitting} 
\label{sec:SEDfitting}

In order to measure the UV bump in an individual spectrum or stack of spectra, we require a model for the underlying ``intrinsic'' spectrum.  Our approach to determine this model is as follows.  We first fit the observed photometric SED of the galaxies with stellar population models, including the effects of a featureless attenuation curve, i.e., $k_\mathrm{Cal}(\lambda)$, without any bump feature. This SED fit excludes all of the photometric bands in the rest-frame region of the bump feature. If we then compare the observed spectrum with this fitted model SED through the region of the bump, we will be able to detect whether there is any excess absorption relative to the model spectrum, i.e., whether there is any excess absorption that can be attributed to a bump feature in the attenuation law.  

One could, of course, have used the excluded photometric bands in the region of the bump for this purpose \citep{2009A&A...507.1793N,2011A&A...533A..93B,2012A&A...545A.141B}, but by using the spectra, a high-resolution spectrum of the absorption bump can be obtained, in which, for instance, the effect of discrete absorption or emission lines in the galaxy, both of variable strength, can be easily isolated and the shape of the underlying bump feature precisely determined. This method of course requires a high degree of spectrophotometric consistency between the photometry and the spectra, and this will be constructed in Section \ref{sec:flux_calibration} below.

In this section, we first describe our photometric SED fitting to obtain the model spectra that are used to measure the bump features.  We employ the software LePhare \citep{1999MNRAS.310..540A,2006A&A...457..841I} with a template library containing synthetic spectra generated using the population synthesis model of \citet{2003MNRAS.344.1000B} assuming a \citet{2003PASP..115..763C} IMF. 
We considered 12 models, combining a constant star-formation history (SFH) and delayed SFHs ($t/\tau^2 e^{-t/\tau}$ with $\tau=0.1$--3~Gyr) with two metallicities ($Z=0.008$ and 0.004) applied.  The contribution of the emission lines (Ly$\alpha$, [O\,{\sc ii}]$\lambda$3727, H$\beta$, [O\,{\sc iii}]$\lambda\lambda$4959,~5007, and H$\alpha$) in the different filters is included following the recipe described in \citet{2009ApJ...690.1236I}.  As explained in Section \ref{sec:dust_prescription}, we considered a single featureless attenuation curve from \citet{2000ApJ...533..682C} so that none of the resulting model spectra can possibly contain a bump feature.  
We allowed the reddening value to vary among $E(B-V)=0, 0.01, 0.02, 0.06, 0.08$, and 0.10--0.70 in fine steps of 0.025.  The effects of possible variations in the overall shape of the baseline attenuation curve will be discussed later in Section \ref{sec:discussion:slope}.

We used photometry from the COSMOS2015 catalog \citep{2016ApJS..224...24L} measured within 30 broad-, intermediate-, and narrowband filters from GALEX near-UV to Spitzer/IRAC ch2 ($4.5~\mathrm{\mu m}$), as listed in Table 3 of \citet{2016ApJS..224...24L}.  For CFHT, Subaru, and UltraVISTA photometry, we used the fluxes measured in a $3\arcsec$ diameter aperture and applied the offsets provided in the catalog to convert them to the total fluxes. 

As noted above, a key feature of the analysis is that we exclude all of the photometric bands whose rest-frame central wavelengths are within $1960\textrm{\AA}\le\lambda_\mathrm{cen}/(1+z)\le2440\textrm{\AA}$ to ensure that the SED fitting is not affected by the SED of the galaxy in the region of the 2175~{\AA} bump feature.  The number of filters excluded is four or five, depending on the redshift.  The resulting SED fit is therefore quite independent of whether the actual attenuation curve of the galaxy in question does or does not have a UV bump feature.

One complicating fact is the presence of several sharp spectral features in the galaxy spectra, arising from absorption and emission in the ISM, which are not included in the stellar templates.  This was dealt with in the SED fitting by applying small corrections to the photometry of the filters whose rest-frame central wavelengths are in the range 1500--2900~{\AA}.  Applying this correction is important in order to obtain the best possible continuum model for comparison with the observed spectrum.
These corrections were determined using a stacked spectrum of the entire sample (see Section \ref{sec:stacking}) to estimate the effects of these spectral lines for each of the photometric bands at the redshift of a given galaxy.  Most individual spectra indeed do not have high enough S/Ns to allow accurate determination of the effects of the ISM emission and absorption features.  Therefore, we here ignore the possible galaxy-to-galaxy variations of the strengths of these features.  Although these effects of the ISM features are small ($<5\%$) in most photometric bands, some strong absorption lines in the red side (i.e., Fe\,{\sc ii}~$\lambda$2344,~2374,~2382,~2586,~2600 and Mg\,{\sc ii}~$\lambda$2796,~2803) could cause $\gtrsim10\%$ reduction of the fluxes in the intermediate and/or narrow bands that sample these particular wavelengths.

\begin{figure*}[t]
\begin{center}
\includegraphics[width=\textwidth]{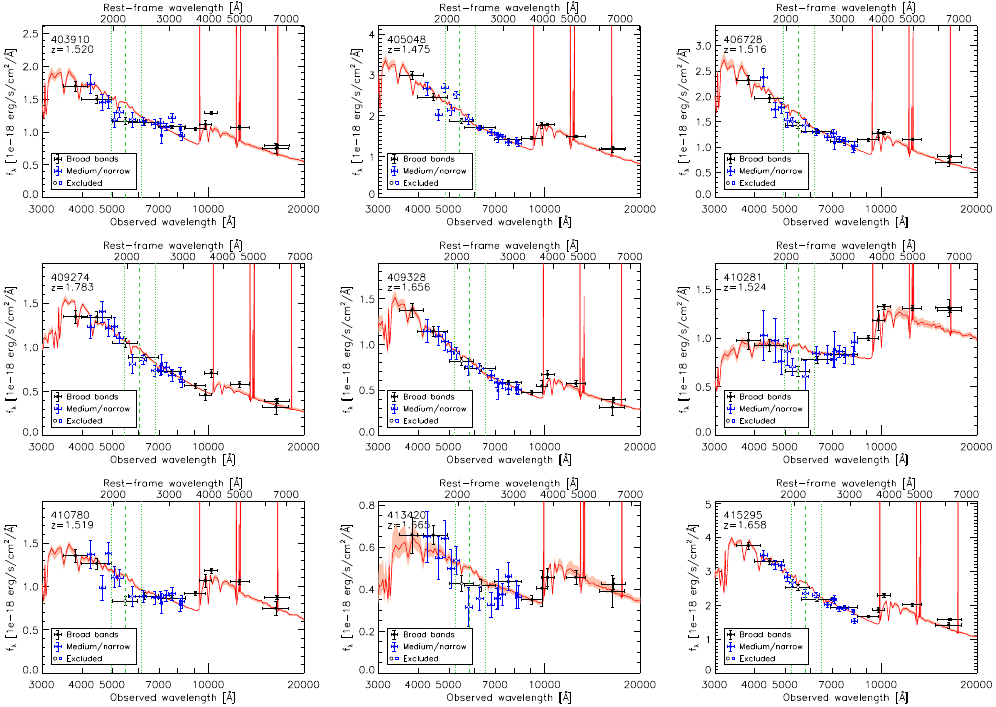}
\caption{Examples of SED fits with LePhare.  The red lines indicate the PDF-weighted median model spectra with the 16th-84th percentiles (light red).  Symbols indicate the photometric fluxes in the broad bands (black circles) and the medium and narrow bands (blue squares).  The photometric bands between the vertical dotted green lines are excluded from SED fitting (open symbols) so that the possible excess absorption by the bump feature does not affect the SED fit.  The vertical dashed lines mark the peak wavelength of the UV bump ($\lambda_\mathrm{rest}=2175$~{\AA}).  The object ID and redshift are noted in the top left corner of each panel.
\label{fig:sedfits}}
\end{center}
\end{figure*}

For each galaxy, we then analyze the likelihood, $\propto \exp(-\chi^2/2)$, that is associated with every possible model in the grid of models.  Specifically, we computed the marginalized probability distribution function (PDF) of the flux value, $f^\mathrm{mod}_\mathrm{\lambda}(\lambda)$, at each wavelength grid using the likelihoods and then derived the model spectrum as the median values of the marginalized PDFs.  Figure \ref{fig:sedfits} presents some examples of SED fits, demonstrating that the overall SEDs are well fitted by the models.  In some cases, the UV bump feature is seen as a gap between the photometric fluxes and the model spectrum.
To measure the bump in units of $k(\lambda)$, we also estimated the reddening $E(B-V)$ as the median value of the marginalized PDF.  Throughout the paper, the bump measurements are made by using the PDF-weighted median spectra and $E(B-V)$.  The results of the bump measurements are consistent within 1\% with those obtained using the best-fit model spectra and the corresponding best-fit $E(B-V)$ values that provide the maximum likelihood.  However, the use of the likelihoods enables us to account for the uncertainties in the models, not only the errors in the observed spectra.  We will describe the estimation of the errors in the $B_k$/$B_A$ measurements using the marginalized PDFs in Section \ref{sec:uncertainties}.

\subsection{Stellar Masses and SFRs} 
\label{sec:M_and_SFR}

For estimating the stellar masses of the galaxies, we again rely on SED fitting with LePhare.  Throughout the paper, the stellar mass represents the mass of living stars at the time of observation and is estimated as the median value from the marginalized PDF based on the likelihoods computed for all possible models.

In the last subsection, our purpose was to obtain the model spectra that fit to the UV continuum of the individual galaxies. However, it is well known that unconstrained SED fitting often leads to unrealistically young ages, and thus low stellar masses, for $z\gtrsim 1$ star-forming galaxies when the age is left as a free parameter.\footnote{Note that here the age represents the time elapsed since the onset of the SFH, which is thus the age of the oldest stellar population present in the galaxy.}  This is because the SED of such galaxies is dominated by the youngest stellar populations, which outshine the older ones \citep[e.g.,][]{2010MNRAS.407..830M}.
It has been demonstrated that limiting the starting times of star formation in such SED fitting to $z\sim3\textrm{--}5$ better recovers the stellar masses of high-$z$ star-forming galaxies \citep[e.g.,][]{2012MNRAS.422.3285P}.

We therefore impose a lower limit on the age of a given galaxy to estimate the stellar mass.  The fitting procedure and photometric data are the same as those used to derive the model spectra in the last subsection, but now we restrict the SFH to begin before a redshift of 3 (i.e., \texttt{ZFORM\_MIN}~$=3$).  The resulting stellar masses of the galaxies are, on average, $\approx0.3~\mathrm{dex}$ larger than those derived with age as a free parameter.  This level of the systematic offset is consistent with what has been found in previous work \citep{2012MNRAS.422.3285P}.

The total SFR is estimated from the UV luminosity as follows.  Dust attenuation is accounted for based on the UV slope $\beta_\mathrm{UV}$ of the rest-frame UV continua, which is defined as $f_\lambda \propto \lambda^{\beta_\mathrm{UV}}$.  We measured the rest-frame FUV (1600~{\AA}) flux density and $\beta_\mathrm{UV}$ by fitting a power-law function to the broad- and intermediate-band fluxes within $1200\textrm{\AA}\le \lambda_\mathrm{cen}/(1+z)\le2700\textrm{\AA}$ but excluding the bands around the 2175~{\AA} feature, i.e., $1960\textrm{\AA}\le\lambda_\mathrm{cen}/(1+z)\le2440\textrm{\AA}$.  The slope $\beta_\mathrm{UV}$ is converted to the attenuation at 1600~{\AA} following \citet{2000ApJ...533..682C}:
\begin{eqnarray}
A_{1600}=4.85+2.31 \beta_\mathrm{UV}.
\label{eq:AFUV}
\end{eqnarray}
The dust-corrected UV luminosity density, $L_{1600}$, at rest frame 1600~{\AA} is then converted to SFR using a relation from \citet{2004ApJ...617..746D},
\begin{equation}
    \mathrm{SFR_{UV,corr}}(M_\odot~\mathrm{yr^{-1}})=\frac{L_{1600} \left( \mathrm{erg~s^{-1}~Hz^{-1}}\right)}{1.7\times8.85\times10^{27}},
    \label{eq:UVSFR}
\end{equation}
where a factor of 1/1.7 is applied to convert from a \citet{1955ApJ...121..161S} IMF to a \citet{2003PASP..115..763C} IMF.

\begin{figure}[tbp]
\begin{center}
\includegraphics[width=3.5in]{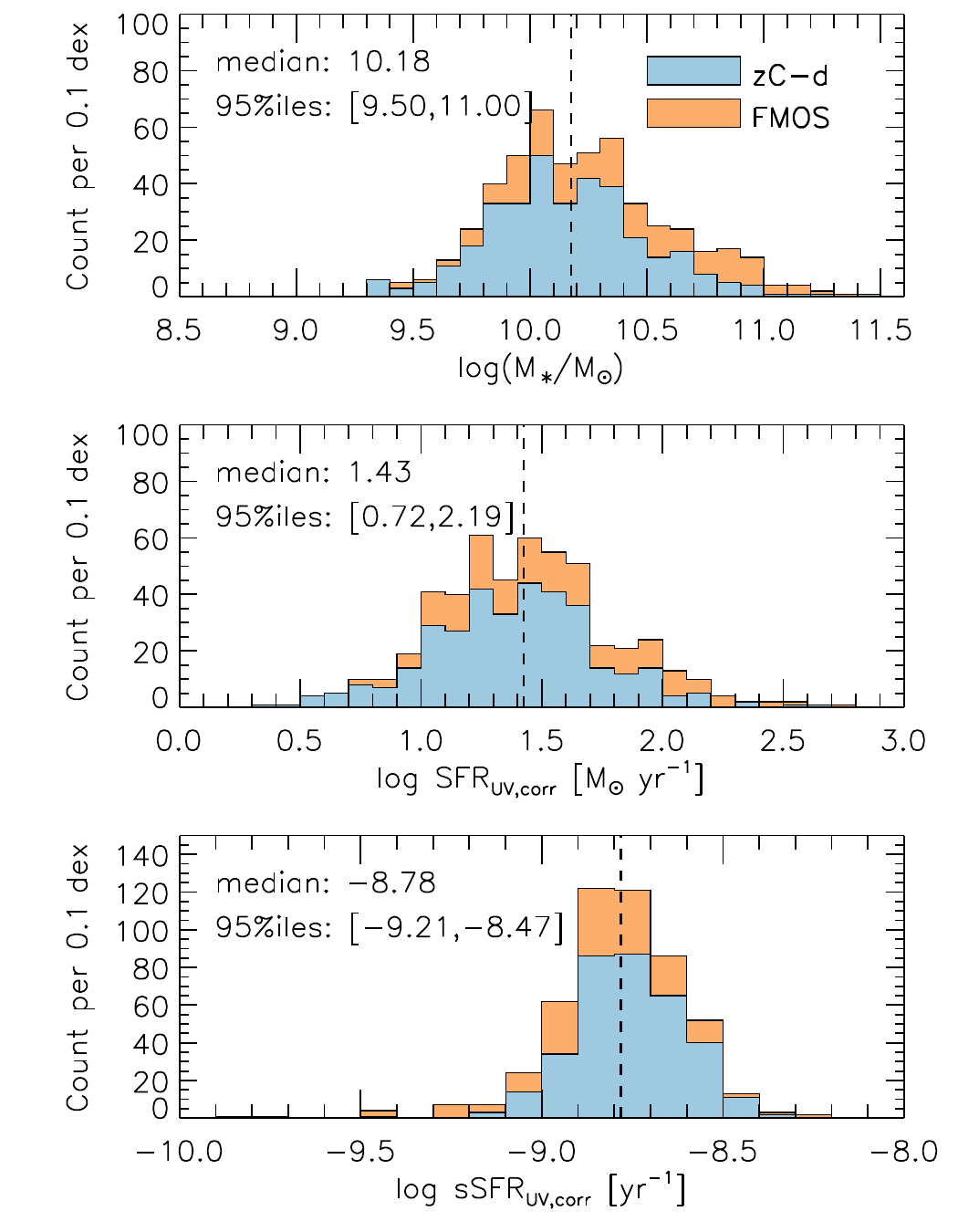}
\caption{Histograms of stellar mass, SFR, and sSFR for our sample, all showing the number count per 0.1~dex.  The SFRs are based on dust-corrected UV luminosity.  The blue histograms show the subset of 345 galaxies whose $z_\mathrm{spec}$ comes from the zCOSMOS-deep catalog, while the orange areas correspond to the additional 160 sources for which the $z_\mathrm{spec}$ measurement from the FMOS-COSMOS catalog was adopted.  The vertical dashed lines indicate the median value of the entire sample, which is also noted in each panel with the 2.5th--97.5th percentiles.
\label{fig:histogams_sample}}
\end{center}
\end{figure}

Figure \ref{fig:histogams_sample} summarizes the distribution of stellar mass, SFR, and specific SFR (sSFR; $=\mathrm{SFR}/M_\ast$) for our sample.  Our sample includes galaxies with $9.5 \lesssim \log(M_\ast/M_\odot) \lesssim 11.0$ and $0.72\lesssim \log \mathrm{SFR}_\mathrm{UV,corr}~[M_\odot~\mathrm{yr^{-1}}]\lesssim2.19$ (the 2.5th--97.5th percentiles) with median values of $\left< \log(M_\ast/M_\odot)\right>_\mathrm{med}=10.18$ and $\left<\mathrm{SFR}_\mathrm{UV,corr}\right>_\mathrm{med}=10^{1.43}~M_\odot~\mathrm{yr^{-1}}$.  The sSFR ranges across $-9.2\lesssim \log~\mathrm{sSFR}_\mathrm{UV,corr}~[\mathrm{yr^{-1}}]\lesssim-8.4$ with a median $\left< \mathrm{sSFR}_\mathrm{UV,corr}\right>_\mathrm{med}=10^{-8.78}~\mathrm{yr^{-1}}$.
Note that the subset that is based on the FMOS $z_\mathrm{spec}$ measurements has compensated at some level for the lack of the massive dusty population. The FMOS subset is biased toward higher stellar masses ($+0.2~\mathrm{dex}$) and SFR ($+0.07$~dex) and lower sSFR ($-0.05~\mathrm{dex}$) relative to the remaining subset.

\begin{figure}[tbp]
\begin{center}
\includegraphics[width=3.5in]{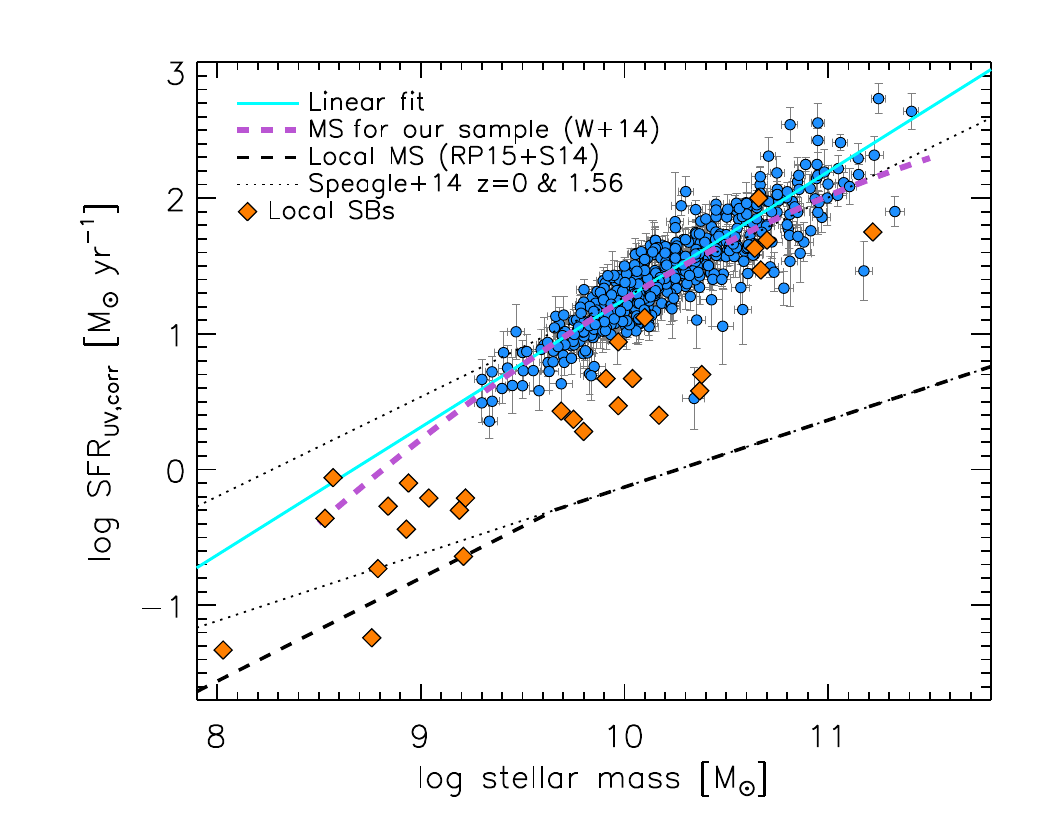}
\caption{Stellar mass vs. SFR based on the dust-corrected UV luminosity.  The cyan solid line indicates the linear fit to the data points.  Orange diamonds indicate local starburst galaxies (see Section \ref{sec:local}).  For reference, the MS relations from the literature are shown.  The thin dotted lines indicate the relations at the median redshift of the sample ($z=1.56$) and at $z=0.01$ based on the time-dependent parameterization derived by \citet{2014ApJS..214...15S}.  The purple dashed line indicates a relation derived within $z=1.5\textrm{--}2.0$ by \citet{2014ApJ...795..104W}, which will be used to calculate $\Delta \log \mathrm{sSFR}$ for galaxies in our sample (see Section \ref{sec:discussion:comparison}).  The black dashed line indicates the MS relation for local sources, combining the formulations from \citet[][$M_\ast\ge10^{9.66}~M_\odot$]{2014ApJS..214...15S} and \citet[][$<10^{9.66}~M_\odot$]{2015ApJ...801L..29R}.  Note that the local starburst galaxies have lower sSFRs than the high-redshift sample but greater $\Delta \log \mathrm{sSFR}$, on average, relative to the evolving MS. 
\label{fig:Mstar_vs_SFRUV}}
\end{center}
\end{figure}

Figure \ref{fig:Mstar_vs_SFRUV} shows the distribution of the $M_\ast$ and UV-based dust-corrected SFR for the sample, together with the local starburst sample (Section \ref{sec:local}).  Our galaxies lie closely along a linear relation (cyan line in the figure)
\begin{equation}
    \log~\mathrm{SFR}_\mathrm{UV,corr} = 1.25 + 0.94(\log M_\ast/M_\odot-10)
\end{equation}
with a standard deviation of $0.18~\mathrm{dex}$.  
For reference, the MS $M_\ast$--SFR relations from the literature are shown and are all converted to a \citet{2003PASP..115..763C} IMF if necessary.  Our sample is in good agreement with the relation derived at $z=1.5\textrm{--}2.0$ by \citet{2014ApJ...795..104W}.  The \citet{2014ApJ...795..104W} relation precisely matches in normalization at $M_\ast\sim10^{10}~M_\odot$ the relation at the median redshift ($z=1.56$) based on the time-dependent parameterization derived by \citet{2014ApJS..214...15S} while better recovering the low-mass end slope.  The agreement of our sample with the literature MS relations demonstrates that our sample is representative of MS galaxies at these epochs.
We will use the \citet{2014ApJ...795..104W} relation to compute the offsets ($\Delta \log \mathrm{sSFR}\equiv \log\mathrm{sSFR}/\left<\mathrm{sSFR}\right>_\mathrm{MS}$) from the MS for our sample in Section \ref{sec:discussion:comparison}.
For local sources, we constructed the MS relation combining the \citet{2014ApJS..214...15S} relation at $z=0.01$ ($M_\ast\ge10^{9.66}~M_\odot$) and the one from \citet[][$<10^{9.66}~M_\odot$]{2015ApJ...801L..29R} for better reproducing the low-mass slope.

It should be noted that our sample of galaxies at high redshift has higher sSFRs, by roughly $0.5~\mathrm{dex}$, than the local ``starburst'' sample.  However, relative to the evolving MS, it is the local starburst sample that has the higher $\Delta \log \mathrm{sSFR}$, on average.  One might surmise that the high-redshift galaxies have had a more steady recent SFH than the local starbursts, which have indeed likely experienced a recent short-lived ``burst'' in SFR.  This distinction will become significant in our interpretation (Section \ref{sec:discussion:interpretation}).

A final point to note is that the dust correction to the UV luminosity can be quite large ($A_{1600} \gtrsim 2.5$~mag) in some galaxies in our sample.  We have thus checked that the conclusions do not change even if we replace the dust-corrected UV-based SFRs (SFR$_\mathrm{UV,corr}$) with the UV$+$IR-based SFRs (SFR$_\mathrm{UV+IR}$; if available), which are the sum of the UV-based SFRs but not corrected for dust attenuation and the SFR based on the total IR luminosity.  For the latter, we utilized the public ``super-deblended'' catalog presented in \citet{2018ApJ...864...56J}.  These two SFRs are in broad agreement, and no substantial population of heavily obscured starbursting galaxies was found in our sample.  We present SFRs using the IR photometric data and the corresponding results from the reanalysis in Appendix \ref{sec:Appendix:reanalysis}.

\subsection{Flux Calibration of the VIMOS Spectra}
\label{sec:flux_calibration}

In this study, we will measure the excess absorption based on the comparison between the observed zCOSMOS spectra (and stacks of spectra) and the model spectra that were obtained from the broad-, medium-, and narrowband SED fitting.  
It should be recalled that these SED fits neither included a UV bump feature in the assumed attenuation curve nor considered photometric filters that were in the vicinity of the wavelength of the UV bump feature.  
The comparison requires an accurate spectrophotometric calibration of the observed spectrum and, in particular, correction of the slit losses that are likely to depend on the wavelength.  This correction is also based on the broad- and medium-band photometry and separately calculated for each of the individual spectra that are produced from the standard zCOSMOS-deep reduction process. These spectra do have a nominal flux calibration applied to all spectra based on standard star observations, but they do not include effects such as slit width, inaccuracies in slit centering, and the effects of atmospheric dispersion.  It is clear that a key requirement is that this correction must not induce any spurious feature in the spectra that may mimic any excess absorption due to the UV bump.  

\begin{figure}[tbp]
\begin{center}
\includegraphics[width=3.3in]{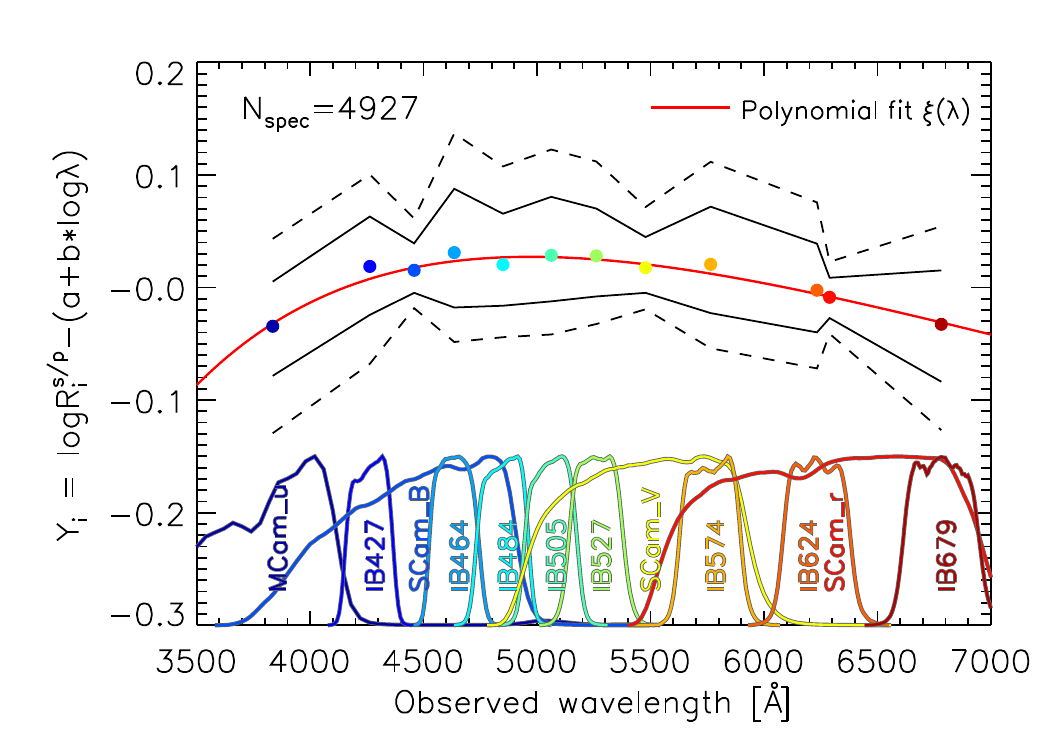}
\caption{Altered flux ratios $Y_i$ defined as Equation (\ref{eq:Y_i}).
The circles indicate the median values of $Y_i$ from 4972 spectra at the central wavelength of each filter, whose transmission curve is shown in the bottom part of the panel with the corresponding color.  The black solid and dashed lines indicate, respectively, the 16th--84th and 5th--95th percentiles.  The third-order polynomial fit to the median values, shown by the red curve, is adopted as $\xi(\lambda)$ in Equation (\ref{eq:flux_ratio}).
\label{fig:flux_ratios}
}
\end{center}
\end{figure}

In constructing this correction, we make use of $\sim5000$ spectra available in zCOSMOS-deep, regardless of whether there was a successful measurement of the spectroscopic redshift and whether the object lies in the redshift range used in the present study.  For each object, we use the available photometry in four broad bands (CFHT/MegaCam $u^\ast$ and Subaru/Suprime-Cam $B$, $V$, and $r$) and eight intermediate bands (Subaru/Suprime-Cam IA427, IA464, IA484, IA505, IA527, IA574, IA624, and IA679), which lie within the spectral coverage of the VIMOS LR blue grism used in zCOSMOS-deep.  As was done for SED fitting (Section \ref{sec:SEDfitting}), we adopted the photometric flux measurements from a $3\arcsec$ diameter aperture from the COSMOS2015 catalog \citep{2016ApJS..224...24L} and converted these to total fluxes with the appropriate offsets.  
We refer to the photometric flux in the $i$th filter as $F^\mathrm{phot}_i$ and the corresponding central wavelength as $\lambda_i$.  We then also compute the ``pseudo'' broad- and intermediate-band fluxes based on the VIMOS spectra in the same filter passbands using the filter transmission curves (see Figure \ref{fig:flux_ratios}).  We refer to the fluxes measured on the input spectra using the $i$th filter curve as $F^\mathrm{spec}_i$.  
The flux ratio $R^\mathrm{s/p}_i \equiv F^\mathrm{spec}_i/F^\mathrm{phot}_i$ is therefore a measure of the slit loss at the different wavelengths, modulo the effects of observational noise in both the imaging and spectral measurements.  Here $R^\mathrm{s/p}_i$ is calculated for all filters, regardless of their rest-frame wavelength.

Our goal here is to derive a smooth functional form of the spectrophotometric correction that adequately describes the flux ratios $R^\mathrm{s/p}_i$ for each spectrum.
We therefore assume that the correction for a single spectrum consists of (i) an overall constant normalization, (ii) a $\lambda$-dependent term whose shape is assumed to be common to all spectra but whose amplitude may vary from spectrum to spectrum, and (iii) an additional smooth $\lambda$-dependent term that is to be determined individually for each spectrum.  

The additional $\lambda$-dependence should have the simplest form that describes any residual monotonic trend of the correction in each spectrum.  After some tests, we chose the following functional form for the total correction for each spectrum, in which the third term is a simple power law in wavelength:
\begin{equation}
    \log R^\mathrm{s/p}(\lambda) = p_1 + p_2 \xi(\lambda) + p_3 \log \left(\lambda/5000~\textrm{\AA}\right),
    \label{eq:flux_ratio}
\end{equation}
where $\xi(\lambda)$ is a function of $\lambda$ that is determined from (and then applied to) all of the spectra collectively, and the three parameters, 
$p_1$, $p_2$, and $p_3$, are determined for each spectrum individually by fitting the flux ratio $R^\mathrm{s/p}_i = F^\mathrm{spec}_i/F^\mathrm{phot}_i$.  The parameters $p_1$ and $p_2$ determine, respectively, the overall scaling and the amplitude of $\xi(\lambda)$.   We found that the inclusion of the third term, which is arguably rather arbitrary, significantly improved the representation of $R^\mathrm{s/p}(\lambda)$. 

We first need to derive the shape of the common $\lambda$-dependence, $\xi(\lambda)$.  We define $Y_i$ by dividing each $R^\mathrm{s/p}_i$ to remove the overall scaling and tilt as follows:
\begin{equation}
    {Y}_i = \log R^\mathrm{s/p}_i - (a+b\log\lambda_i)
    \label{eq:Y_i}
\end{equation}
where the term $(a+b\log\lambda_i)$ is a linear fit to $\log \lambda_i$--$\log R^\mathrm{s/p}_i$.  We then compile the $Y_i$ of all of the individual spectra to derive the $\xi(\lambda)$ function.

Figure \ref{fig:flux_ratios} shows the average $\left< Y_i\right>$ as a function of wavelength.  
The symbols indicate the median values in the filters at their corresponding central wavelengths, and the black solid and dashed lines indicate the 16th--84th and 5th--95th percentiles of the range of $Y_i$.
To compute these medians and percentiles, we limited the spectra to those with $\mathrm{S/N}\ge5$ in $F^\mathrm{phot}_i$ while imposing a quite relaxed limit ($\mathrm{S/N}\ge1$) on $F^\mathrm{spec}_i$ to avoid biasing the spectra toward those with smaller flux losses.  The average $\left< Y_i\right>$ shows a moderately curved shape decreasing (i.e., representing increasing spectral losses) at both extremities of the spectral window.  This median trend in $\left< Y_i\right>$ is well described by a third-order polynomial function of $\log \lambda$ (red curve in the figure).  This overall trend is then adopted as our $\xi(\lambda)$ function.  

It is worth noting that the derived $\xi(\lambda)$ function may have relatively large uncertainties at the shortest wavelengths ($\lambda \lesssim 4000$~{\AA}) because, as shown in Figure \ref{fig:flux_ratios}, the flux loss becomes rapidly worse at $<4000$~{\AA}, and we have only the CFHT/MegaCam $u^\ast$ band there.  Even worse, the $u^\ast$ filter has a blue wing that extends down to 3300~{\AA}, largely beyond the wavelength coverage.  This likely induces additional uncertainties in comparisons between $F^\mathrm{phot}(u^\ast)$ and $F^\mathrm{spec}(u^\ast)$.  However, these uncertainties in the flux calibration have negligible effects on the bump measurements because the central wavelength of the bump (2175~{\AA}) is observed at $>5000$~{\AA}.  We will refer to this concern later when we present the stacked spectrum in Section \ref{sec:stacking}.

\begin{figure}[tbp]
\begin{center}
\includegraphics[width=3.3in]{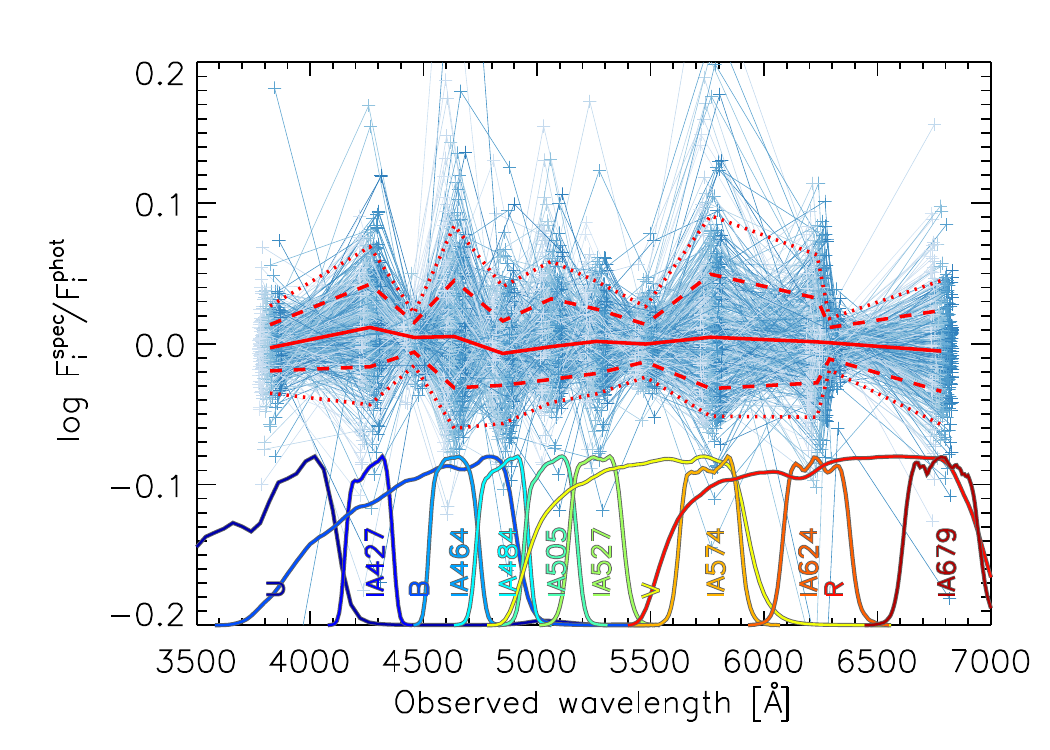}
\includegraphics[width=3.3in]{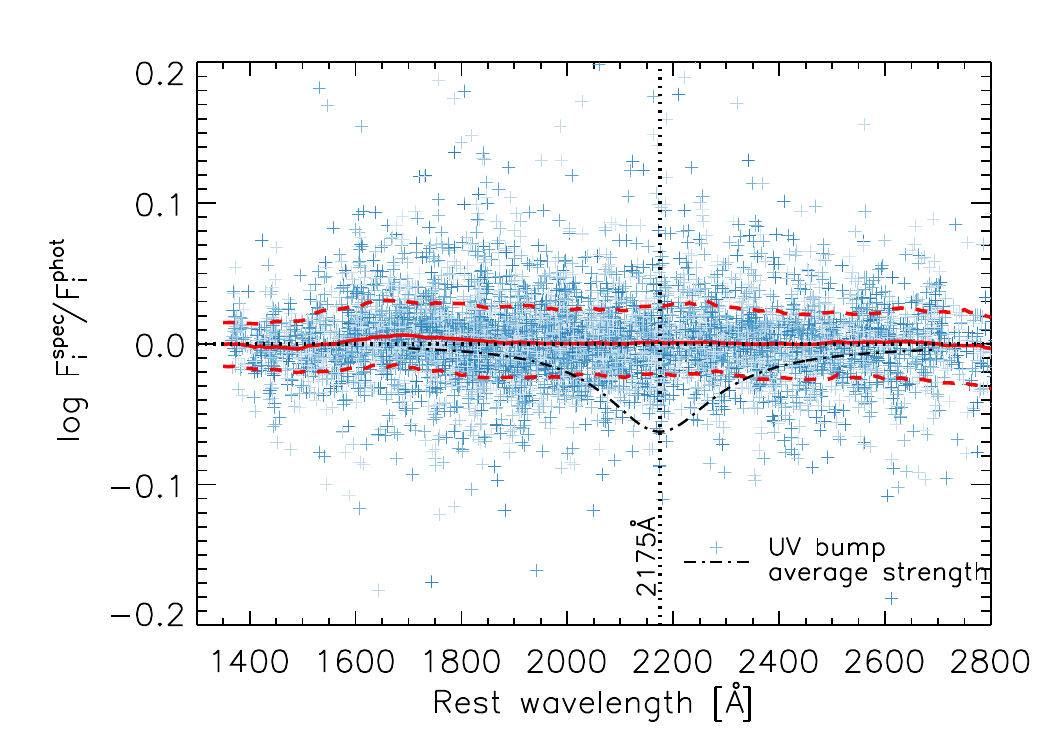}
\caption{Upper panel: individual flux ratios, $F^\mathrm{spec}_i/F^\mathrm{phot}_i$, after correction for the sample of 505 galaxies.  Symbols for each single spectrum are connected by a line.  Different spectra are colored differently for display purposes.  The red solid, dashed, and dotted lines indicate, respectively, the median values, 16th--84th percentiles, and 5th--95th percentiles.  Lower panel: same flux ratios but as a function of rest-frame wavelength.  The red solid and dashed lines indicate running medians and 16th--84th percentiles with a window size of 200~{\AA}.  The horizontal and vertical lines indicate $F^\mathrm{spec}_i/F^\mathrm{phot}_i=1$ and the center wavelength of the UV bump (2175~{\AA}), respectively.  The dotted-dashed curve indicates the typical level of flux absorption caused by the UV bump feature measured in the stack of the whole sample (see Section \ref{sec:stacking}).  
\label{fig:flux_ratios_sample}}
\end{center}
\end{figure}

Given $\xi(\lambda)$, we can then determine $p_1$, $p_2$, and $p_3$ in Equation (\ref{eq:flux_ratio}) for each spectrum and correct the individual spectra by dividing by the resulting $\log R^\mathrm{s/p}(\lambda)$.   Figure \ref{fig:flux_ratios_sample} shows that the pseudo broad- and intermediate-band fluxes recomputed in the flux-corrected spectra are now in excellent agreement with the photometric fluxes.  The median values of the corrected $F^\mathrm{spec}_i$-to-$F^\mathrm{phot}_i$ ratios in each band are all within $\pm0.012~\mathrm{dex}$, presumably reflecting small residual errors in the photometric calibration of the photometry.  The scatter at a given wavelength is 0.01--0.05~dex and larger, relatively, in the intermediate bands because of their generally lower S/N.  

The lower panel of Figure \ref{fig:flux_ratios_sample} shows the same data but now shifted to the rest-frame wavelength.  There is no systematic trend in the corrected flux ratios as a function of the rest-frame wavelength.  The running medians with a window size of 200~{\AA} are all within $0.01~\mathrm{dex}$ over the entire wavelength range of interest, and, in particular, they are quite close to zero (i.e., corrected $F^\mathrm{spec}/F^\mathrm{phot}=1$) around the rest-frame wavelength of the UV bump (2175~{\AA}).  Simply as a reference, we show in this lower panel the excess absorption due to the UV bump that will be observed in the stacked spectrum of the entire sample (Section \ref{sec:stacking}).   This comparison simply shows that the accuracy of the spectrophotometric flux (re)calibration is, by a good margin, sufficient to detect the UV bump signature or, equivalently, that any residual flux calibration uncertainties can have only a very small effect on the measurement of this feature.

In the remainder of the paper, the term ``observed VIMOS spectra'' will always refer to these accurately spectrophotometrically recalibrated spectra.

\subsection{Characterization of the UV Continua}
\label{sec:method_UVcontinua}

\begin{figure}[tbp]
\begin{center}
\includegraphics[width=3.2in]{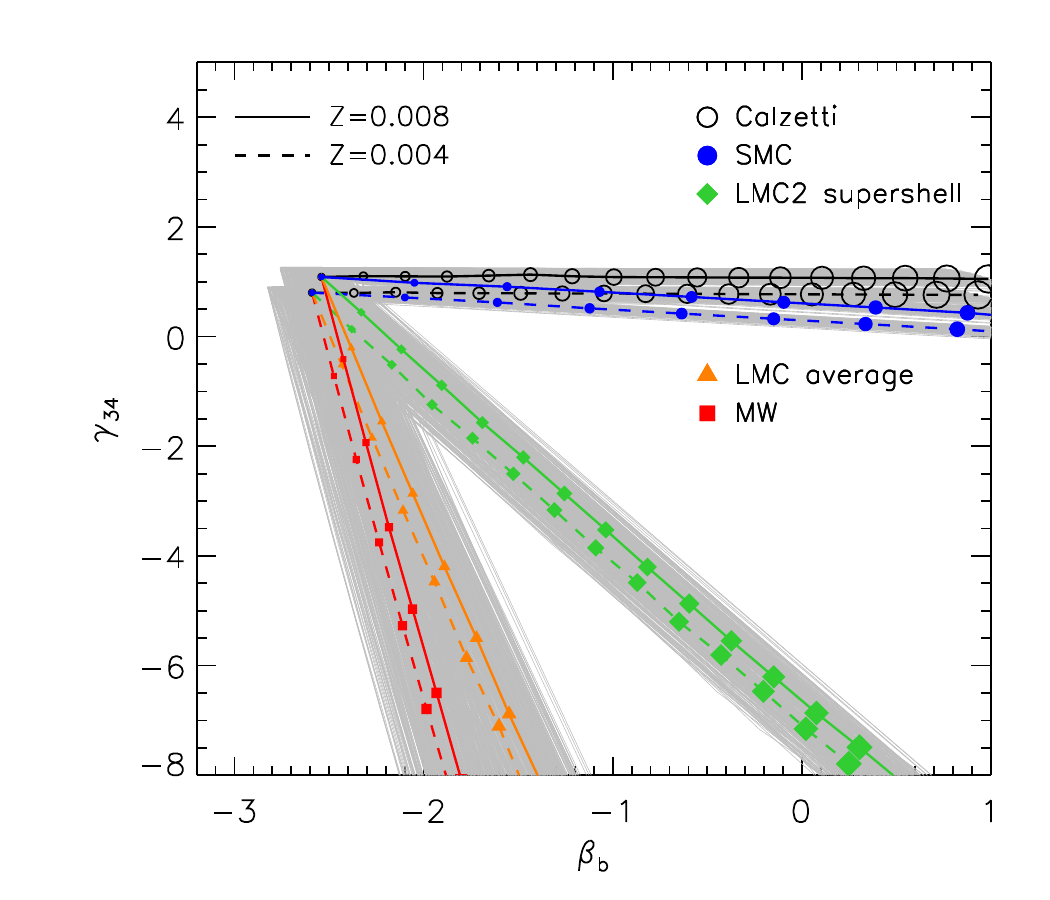}
\includegraphics[width=3.2in]{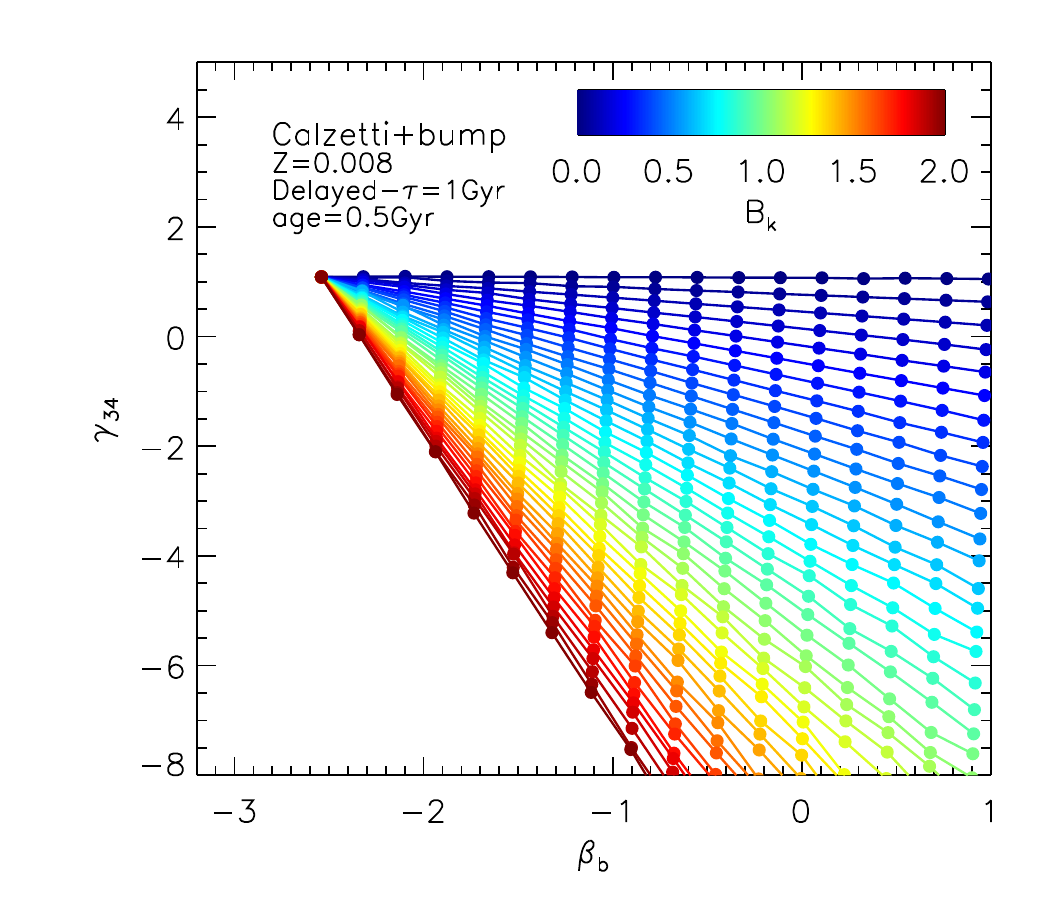}
\caption{
The $\beta_\mathrm{b}$ vs. $\gamma_{34}$ diagrams for model spectra.  Upper panel: colored symbols indicate $\beta_\mathrm{b}$ and $\gamma_{34}$ computed from representative \citet{2003MNRAS.344.1000B} models (a delayed SFH with $\tau=1~\mathrm{Gyr}$, age of 0.50~Gyr, and two stellar metallicities) in steps of $\Delta E(B-V)=0.05$.  The symbol size increases with $E(B-V)$.  Different colors and symbols correspond to different dust extinction/attenuation laws, and different line types are for different metallicities of the stellar synthesis models (see legend).
Gray lines are for models within plausible ranges of SFH and age, which indicate the expected scatters of the $\beta_\mathrm{b}$--$\gamma_{34}$ relations due to the galaxy stellar populations for each extinction law.
Lower panel: $\beta_\mathrm{b}$ vs. $\gamma_{34}$ for attenuation laws that consist of the baseline \citet{2000ApJ...533..682C} law and the bump excess of different peak amplitudes ($0\le B_k\le2$, in steps of 0.05).  The \citet{2003MNRAS.344.1000B} model with a delayed SFH with $\tau=1~\mathrm{Gyr}$, age of 0.50~Gyr, and $Z=0.008$ is used as a representative.  Circles mark $E(B-V)$ in steps of 0.05.
\label{fig:beta_b_vs_gamma34_sim}}
\end{center}
\end{figure}

In this subsection, we characterize the UV continua of the galaxy spectra in the region of the 2175~{\AA} feature.
We follow the parameterization introduced by \citet{2005A&A...444..137N} and used in their subsequent papers \citep{2007A&A...472..455N,2009A&A...499...69N}.
The first parameter $\gamma_{34}$ represents the difference ($\gamma_3-\gamma_4$) between the power-law slopes $\gamma_3$ and $\gamma_4$ measured in the wavelength ranges 1900--2175 and 2175--2500~{\AA}, respectively.  Thus, the parameter $\gamma_{34}$ is a measure of the strength of the 2175~{\AA} feature; a larger negative $\gamma_{34}$ corresponds to a stronger feature.
The second observational parameter is the overall slope $\beta_\mathrm{b}$, an indicator of reddening of the UV continuum, measured across the UV bump using the rest-frame wavelength interval 1750--2600~{\AA} but excluding 1950--2400~{\AA}.  In measuring these from the actual observed spectra, we exclude any strong narrow spectral features.

These two parameters, $\gamma_{34}$ and $\beta_\mathrm{b}$, will vary together as a function of reddening following a relation that is determined by the dust extinction/attenuation law.  
We simulated these quantities using a range of \citet{2003MNRAS.344.1000B} stellar synthesis templates with typical ranges of SFHs and ages and for two metallicities ($Z=0.004$ and 0.008).  We modified these artificial spectra by applying dust attenuation with four different dust extinction/attenuation laws: 
the \citet{2000ApJ...533..682C} law, which is used in our SED fitting (Section \ref{sec:SEDfitting}); the MW extinction law from \citet{2007ApJ...663..320F}; the LMC2 supershell curve; and the SMC one \citep{2003ApJ...594..279G}.  

Figure \ref{fig:beta_b_vs_gamma34_sim} shows that the relationship between $\beta_\mathrm{b}$ and $\gamma_{34}$ depends significantly on the dust attenuation law, with some scatter caused by the variation in the input intrinsic UV continuum spectra, as shown by the thin gray lines.  The $\beta_\mathrm{b}$--$\gamma_{34}$ relations show variations of $\Delta \gamma_\mathrm{34})\lesssim 0.5$ and $\Delta \beta_\mathrm{b}\lesssim 0.2$ at fixed $E(B-V)$.
For reference, we highlight the loci for the intrinsic UV continuum spectra of a delayed SFH with $\tau=1$~Gyr at an age of $0.50~\mathrm{Gyr}$ for each constant stellar metallicity. 

Figure \ref{fig:beta_b_vs_gamma34_sim} demonstrates that the presence of the 2175~{\AA} feature could, in principle, be determined by measuring any $\gamma_{34}\lesssim -1$. The figure also shows that, if all galaxies in a given sample follow a single universal attenuation law, then measurements of $(\beta_\mathrm{b}, \gamma_{34})$ for the sample enable us to constrain the strength of the 2175~{\AA} feature.

Similarly, we simulated $\gamma_{34}$ and $\beta_\mathrm{b}$ by applying synthetic attenuation laws that consist of the \citet{2000ApJ...533..682C} curve as a baseline, to which is added a bump component with different amplitudes (Equation \ref{eq:kbump}; the Drude profile width parameter $\gamma$ is fixed to $0.64~\mathrm{\mu m^{-1}}$).  The lower panel of Figure \ref{fig:beta_b_vs_gamma34_sim} shows that the gradient of the $\beta_\mathrm{b}$--$\gamma_{34}$ relation becomes steeper for higher $B_k$, as expected.  The relations of $B_k\approx1.2$ yield a $\beta_\mathrm{b}$--$\gamma_{34}$ relation that is similar to the prediction for the empirical LMC2 supershell extinction law.

\subsection{Stacking Analysis and Detection of the 2175 ~{\AA} Bump Feature}
\label{sec:stacking}

\begin{figure*}[tbp]
\begin{center}
\includegraphics[width=7.0in]{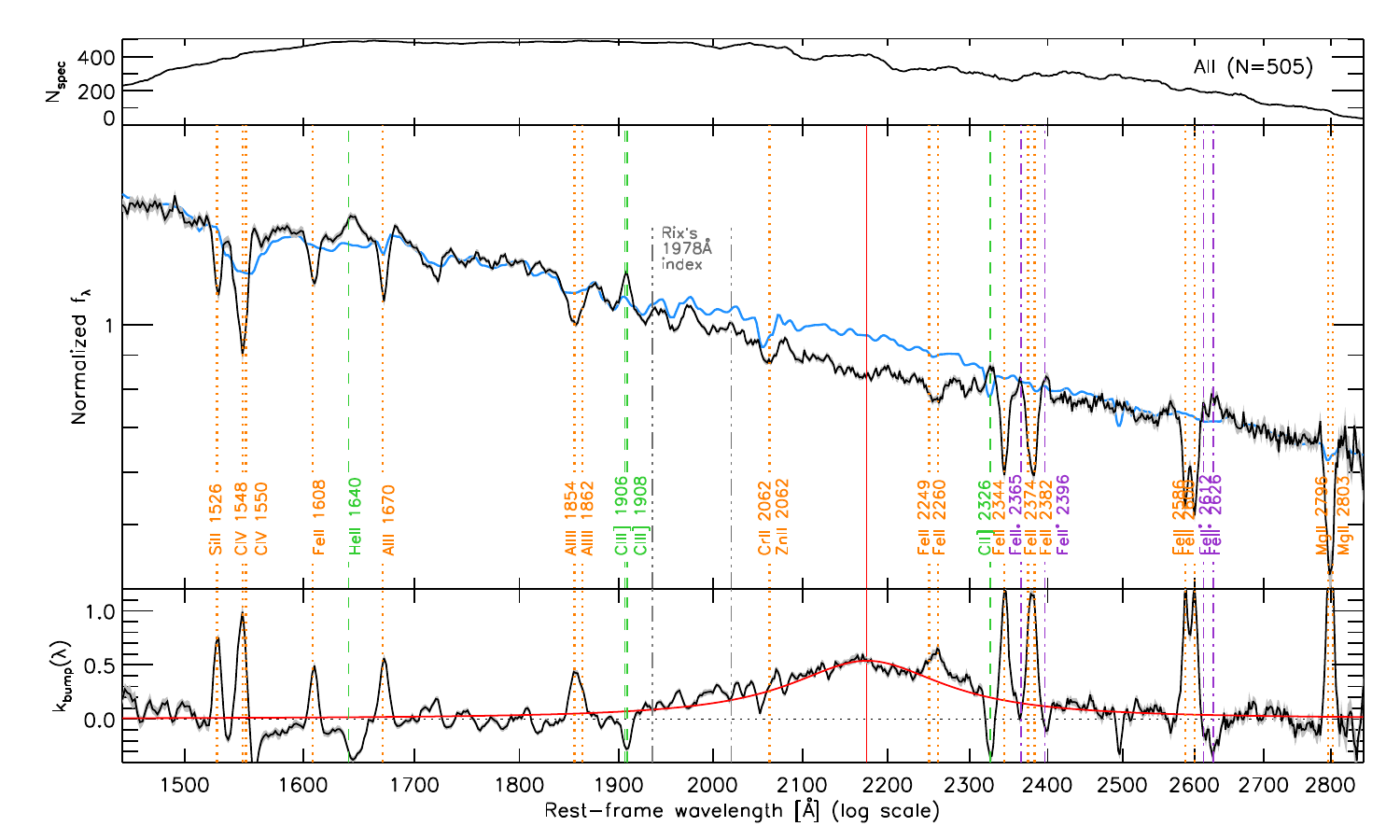}
\caption{Composite VIMOS spectrum of the entire sample of 505 galaxies at $1.3\le z\le 1.8$. 
Top panel: number of spectra that have been stacked at each wavelength grid. 
Middle panel: stacked spectrum (black line) in which some prominent absorption and emission features are identified as marked by color-coded labels: interstellar absorption features (orange), nebular emission lines (green), and fluorescence Fe\,\textsc{ii}$^\ast$ lines (purple).  The spectral region of the \citet{2004ApJ...615...98R} 1978~{\AA} index is marked by triple-dotted-dashed lines.  The narrow gray region along the spectrum indicates the noise level in the composite spectrum estimated by bootstrap resampling.
The blue line indicates the composite of the PDF-weighted median spectra obtained from SED fitting in which dust extinction following the bump-free \citet{2000ApJ...533..682C} attenuation law is applied.
Bottom panel: 2175~{\AA} bump feature in the attenuation curve.  The black line indicates the stack of the individual $k_\mathrm{bump}(\lambda)$ computed by Equation \ref{eq:kbump_obs} for the whole sample.
The red line indicates the best-fit Drude profile fit of the bump component shown.
\label{fig:stack_all}}
\end{center}
\end{figure*}

To accurately determine the shape of the 2175~{\AA} bump, we rely on a stacking analysis of subsamples of galaxies selected by various intrinsic properties, specifically stellar mass and SFR.  We stack the observed VIMOS spectra, as well as the PDF-weighted median model spectra derived from SED fitting, as follows.
We first transform all of the individual VIMOS and model spectra to the rest-frame wavelength based on their spectroscopic redshift. Both the VIMOS and model spectra are resampled to a common wavelength grid with a spacing of 2.0~{\AA}~pixel$^{-1}$.
Each spectrum is then normalized by the average flux density within the rest-frame 1950--2200~{\AA} region.  
The composite stacked spectra are then produced by taking the median value of the individual spectra at each wavelength grid while ignoring any spectral regions that are missing and/or contaminated by, for example, zeroth-order contamination.  The associated errors are estimated using bootstrap resampling. 

Figure \ref{fig:stack_all} shows the composite stacked spectrum (middle panel) of the whole sample.  Some prominent narrow spectral features are clearly identified in the stacked spectra, as marked by vertical lines.  Note that, given the redshift range of our sample, the rest-frame wavelength range of 1550--2400~{\AA} is covered by nearly all of the sources, while shorter and longer wavelengths are less fully represented within the sample. The top panel of the figure shows the number of spectra that have contributed to the stack at each wavelength. This begins to decrease at $\lambda>2000$~{\AA} because of the removal of a significant number of spectra that are contaminated at a given wavelength grid by sky lines that fall on the red side of the observing window (see Section \ref{sec:observations}).  

In Figure \ref{fig:stack_all}, the stacked spectrum is in good agreement with the stacked model spectrum at both the blue and red ends of the spectrum, either side of the wavelength interval of the bump feature ($\sim 1900\textrm{--}2500$~{\AA}).  To be clear, the stacked VIMOS spectra are independent of SED fitting.  On the other hand, the SED fitting does not include any information from the observed spectra except the minor corrections of the photometry for the narrow spectral features (see Section \ref{sec:SEDfitting}).  Therefore, the agreement is a consequence of the precise flux calibration constructed in Section \ref{sec:flux_calibration}.  Conversely, however, there is a clear gap between the stacked and model spectra across $\lambda\sim1900\textrm{--}2500~\textrm{\AA}$.  Recalling that the latter was constructed using a featureless \citet{2000ApJ...533..682C} attenuation law, we can conclude that this gap is the result of a broad absorption excess that is attributed to the usual UV bump feature.

The absolute excess absorption, $A_\mathrm{bump} (\lambda)$, can be measured as  
\begin{equation}
    A_\mathrm{bump}(\lambda)=2.5 \log \left( \frac{f^\mathrm{mod}_\lambda (\lambda)}{f^\mathrm{obs}_\lambda(\lambda)}\right),
    \label{eq:Abump_obs}
\end{equation}
where $f^\mathrm{obs}_\lambda$ and $f^\mathrm{mod}_\lambda$ are, respectively, the observed and model spectra constructed using the featureless attenuation curve.  Correspondingly, the bump component in the actual observed attenuation curve is given as 
\begin{equation}
    k_\mathrm{bump}(\lambda)=A_\mathrm{bump}(\lambda)/E(B-V),
    \label{eq:kbump_obs}
\end{equation}
where $E(B-V)$ is the reddening value derived from the SED fitting.

To measure the bump component, we therefore first calculate $A_\mathrm{bump}(\lambda)$ or $k_\mathrm{bump}(\lambda)$ for the individual galaxies and then coadd them, instead of computing it directly from the stacked spectra and, for $k_\mathrm{bump}(\lambda)$, a representative value of $E(B-V)$ in stacking samples.  By doing so, we mitigate the possible effects of the variation in $E(B-V)$ from one galaxy to another, though the conclusion does not depend on the detailed method of stacking.  The associated errors in the stacks are estimated via bootstrap resampling.  See Section \ref{sec:uncertainties} for error estimation including the uncertainties in the model spectra.

The bottom panel of Figure \ref{fig:stack_all} shows a Drude profile fitted to the stacked $k_\mathrm{bump}(\lambda)$.  For fitting, we excluded the known strong spectral lines that are marked in the figure and the region of the \citet{2004ApJ...615...98R} 1978~{\AA} index (1935--2020~{\AA}) where significant absorption features arise from the blending of numerous Fe\,{\sc iii} transitions.  We fixed the central wavelength of the Drude profile, $\lambda_0$, to 2175~{\AA}.  We will come back to this result in detail in Section \ref{sec:results_bumpProfiles}. 

It is worth commenting on the small disagreement between the composite observed and synthesized spectra (Figure \ref{fig:stack_all}, middle panel) and, correspondingly, the deviation in $k_\mathrm{bump}(\lambda)$ from zero at $\lambda_\mathrm{rest}<1500$~{\AA} (bottom panel).  This deviation would plausibly be due to the relatively large uncertainties in the secondary flux calibration at the bluest wavelengths and thus hardly affects the bump measurements as mentioned in Section \ref{sec:flux_calibration}.

\subsection{Uncertainties in the Bump Measurements}
\label{sec:uncertainties}

The errors in the measurements of the bump amplitudes, $B_A$ and $B_k$, are subject not only to the errors in the observed VIMOS spectra but also to the uncertainties in the models.  For simplicity, we approximately estimate the errors on $B_A$ and $B_k$ by assuming that the squared errors can be expressed as the sum of two independent terms, which are attributed to either the errors in the observed spectra or the uncertainties in the models:
\begin{eqnarray}
    & \sigma_\mathrm{tot}^2 = \sigma_\mathrm{obs}^2 + \sigma_\mathrm{mod}^2
    \label{eq:dE_A}.
    \end{eqnarray}
The former term is estimated through the Drude profile fitting based on the stack and the associated error spectrum but ignoring the uncertainties in the models (i.e., $f^\mathrm{mod}_\lambda$ and $E(B-V)$).  The error spectrum is obtained via bootstrap resampling.

Turning to the latter term, it is the flux uncertainties in the model spectra around $\lambda_\mathrm{rest}\approx2175$~{\AA} that affect the measurements of the bump amplitude.  We therefore consider the mean flux value, $f^\mathrm{mod}_{2175}$, obtained by averaging $f_\lambda^\mathrm{mod}(\lambda)$ within $\lambda_\mathrm{rest} = 2075\textrm{--}2275$~{\AA}.  For each galaxy, we then calculate the marginal PDF of $r_{2175}=f^\mathrm{mod}_{2175}/\tilde{f}^\mathrm{mod}_{2175}$, where $\tilde{f}^\mathrm{mod}_{2175}$ is the value obtained from the PDF-weighted median spectra.  Given the marginal PDFs of $r_{2175}$ for all objects, we can calculate the model-origin uncertainties on $B_A$ as
\begin{equation}
    \sigma_\mathrm{mod}^2(B_A) = \mathrm{var}\left(~\mathrm{med}\left( 2.5\log r_{2175} \right) \right).
\end{equation}
Here $\mathrm{var}(x)$ denotes the expected variance of $x$, and $\mathrm{med}(x)$ denotes the median of $x$ taken over all objects in a subsample of a stack.  The value $r_{2175}$ of each object is a stochastic variable that follows the relevant PDF.

In the case of $B_k$, we need to include the PDF of $E(B-V)$, and can approximately express the value of $\sigma_\mathrm{mod}^2(B_k)$ as
\begin{equation}
    \sigma_\mathrm{mod}^2(E_k) = \mathrm{var}\left(~\mathrm{med}\left(~\frac{\left<B_A\right> + 2.5\log r_{2175}}{E(B-V)}\right)\right).
\end{equation}
Here $r_{2175}$ and $E(B-V)$ are both stochastic variables, which are assumed to be independent of each other, and $\left<B_A\right>$ is the best estimate obtained using the stack of the subsample in question.  
In practice, we estimated $\sigma_\mathrm{mod}^2(B_k)$ and $\sigma_\mathrm{mod}^2(B_A)$ by (1) generating random $r_{2175}$ and $E(B-V)$ values for each galaxy following the marginalized PDFs, (2) calculating the median values of $2.5\log r_{2175}$ and $(\left<E_A\right>+2.5\log r_{2175})/E(B-V)$) over the subsample, and (3) repeating this process $10^4$ times with different random values to estimate the variance of these quantities.  

We found that the contribution of the uncertainties in the models to the final errors on the $B_k$ measurements is roughly comparable to that from the errors in the observed VIMOS spectra, i.e., $\sigma_\mathrm{mod}\sim \sigma_\mathrm{obs}.$

\section{Results}
\label{sec:results}

In this section, we first show the characteristics of the UV continuum of individual galaxies and then present results on the detailed shapes of the 2175~{\AA} bump feature that are obtained when the spectra are stacked according to stellar mass and SFR.  

\subsection{The Shape of the UV Continua in Individual Galaxies}
\label{sec:results_UVcontinua}

\begin{figure}[tbp]
\begin{center}
\includegraphics[width=3.3in]{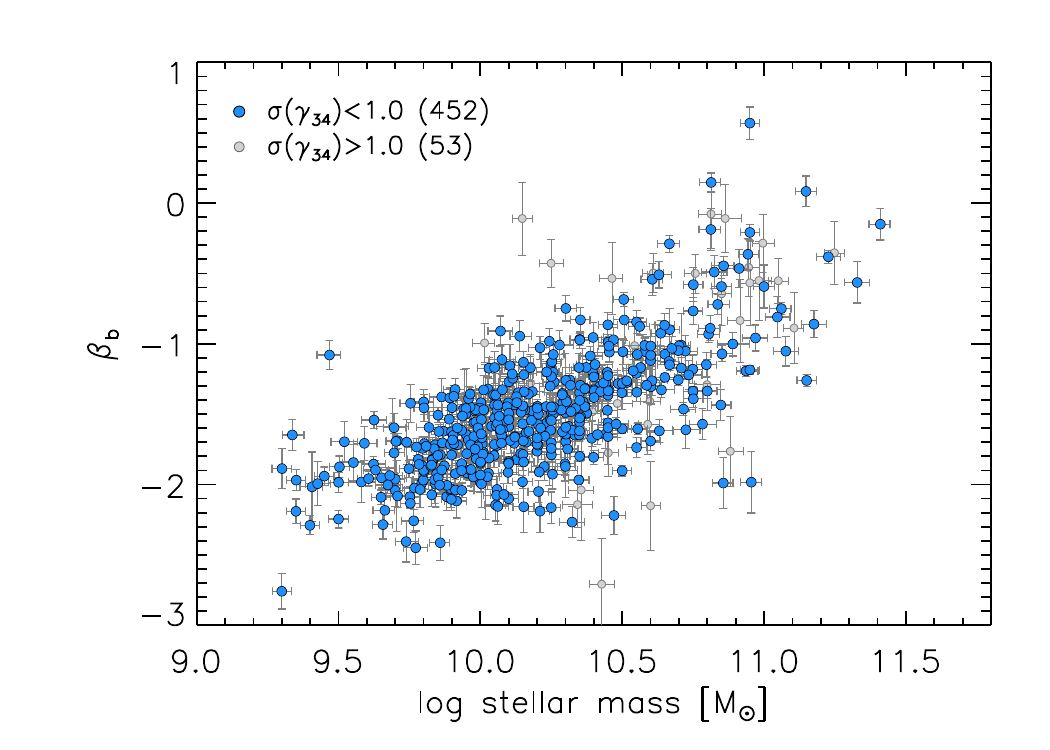}
\includegraphics[width=3.3in]{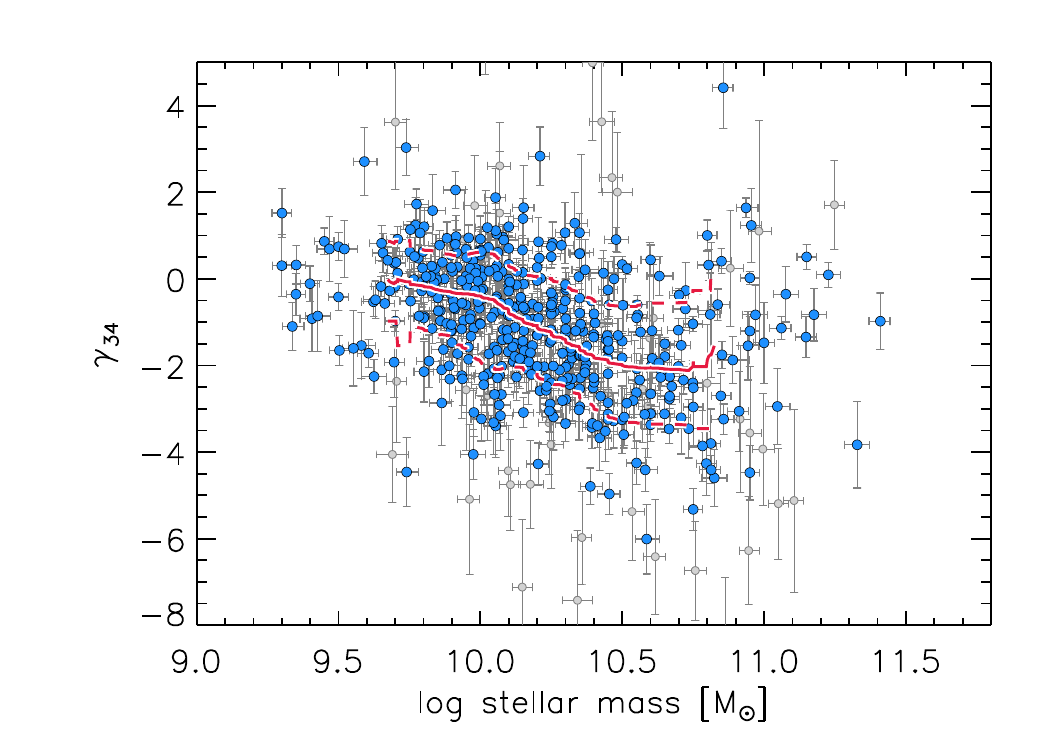}
\caption{Upper panel: stellar mass vs. $\beta_\mathrm{b}$.  Lower panel: stellar mass vs. $\gamma_{34}$.  Blue circles indicate 452 objects with $\sigma(\gamma_{34})<1.0$, and gray circles are for the remaining 53 objects.  Red solid and dashed lines indicate the running medians and 16th--84th percentiles of $\gamma_{34}$.\label{fig:Mstar_vs_beta_b}}
\end{center}
\end{figure}

\begin{figure}[tbp]
\begin{center}
\includegraphics[width=3.3in]{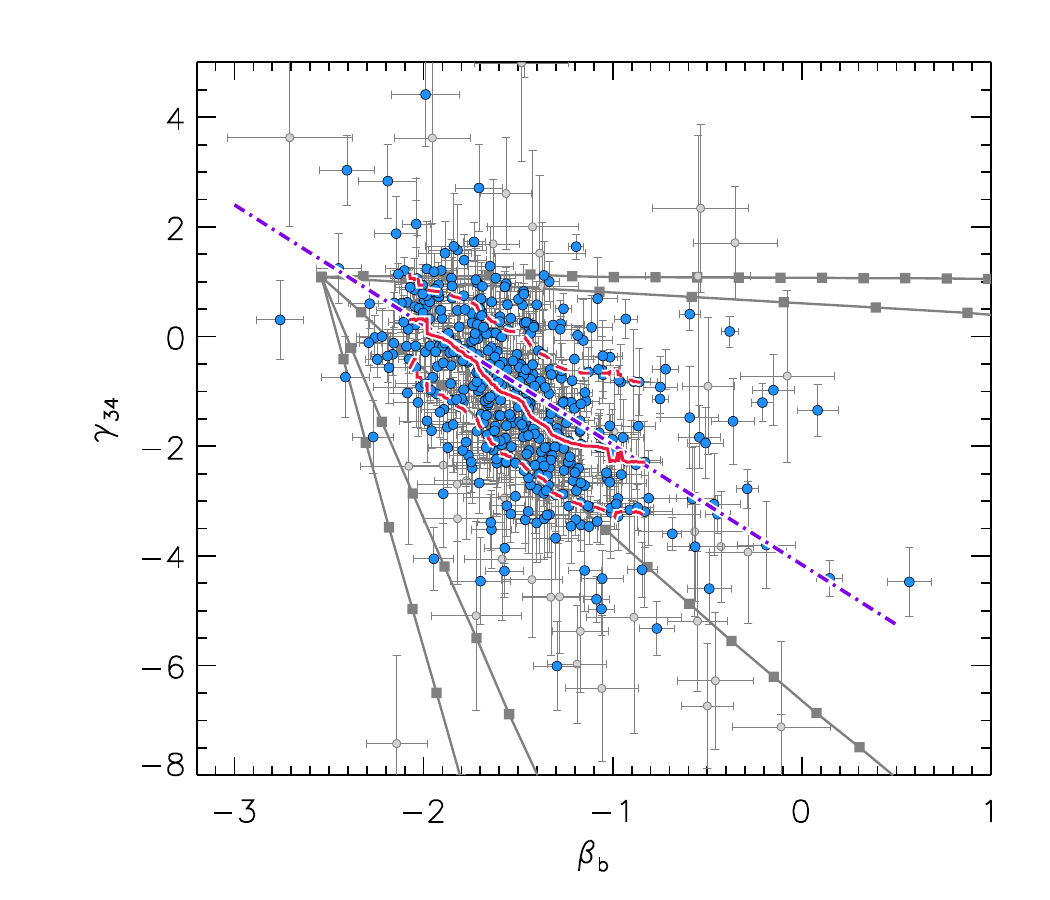}
\caption{
The $\beta_\mathrm{b}$ vs. $\gamma_{34}$ for the sample of 505 galaxies.  Blue circles indicate 452 objects with $\sigma(\gamma_{34})<1.0$, and gray circles are for the remaining 53 objects (same as in Figure \ref{fig:Mstar_vs_beta_b}).
Red solid and dashed lines indicate the running medians and 16th--84th percentiles of $\gamma_{34}$ in bins of $\beta_\mathrm{b}$. 
The purple dotted-dashed line indicates a linear fit (Equation \ref{eq:gamma34_beta_b}) to the secure data points (blue circles).
For comparison, the gray lines indicate the predictions for a representative model ($Z=0.008$, delayed SFH with $\tau=1~\mathrm{Gyr}$, and age of 0.50~Gyr) that are highlighted in Figure \ref{fig:beta_b_vs_gamma34_sim}.
\label{fig:beta_b_vs_gamma34}}
\end{center}
\end{figure}

Figure \ref{fig:Mstar_vs_beta_b} shows the measurements of $\beta_\mathrm{b}$ and $\gamma_{34}$ (see Section \ref{sec:method_UVcontinua}) from the individual spectra as a function of stellar mass.
It is clear that these parameters are both correlated with $M_\ast$. The positive correlation between $M_\ast$ and $\beta_\mathrm{b}$ is a reflection of the correlation between $M_\ast$ and the reddening as already reported in the literature \citep[e.g.,][]{2015ApJ...807..141P}.  
The interpretation of $\gamma_{34}$ is more complicated because the excess attenuation due to the UV bump depends on both the bump strength implemented in the attenuation curve and the absolute level of attenuation.  

Figure \ref{fig:Mstar_vs_beta_b} shows that $\gamma_{34}$ decreases with $M_\ast$, but the correlation becomes unclear at $\log M_\ast/M_\odot \gtrsim 10.5$ with a possible increase of the scatter.  The correlation also appears to vanish at lower masses ($\log M_\ast/M_\odot \lesssim 9.4$) as $\gamma_{34}$ gets close to the upper limit corresponding to the zero attenuation.  Note that the errors on $\gamma_{34}$ and $\beta_\mathrm{b}$ are subject just to the noise in the observed spectra and not to the uncertainties in the SED modeling because these values are purely measured on the observed individual spectra.

Figure \ref{fig:beta_b_vs_gamma34} shows $\beta_\mathrm{b}$ versus $\gamma_{34}$ together with the simulated relations for different extinction and attenuation laws (see Figure \ref{fig:beta_b_vs_gamma34_sim}).  Here we show as a reference a stellar population model with $Z=0.008$ and adopting a delayed SFH with $\tau=1~\mathrm{Gyr}$ at an age of 0.50~Gyr, which is the representative case shown by solid lines in Figure \ref{fig:beta_b_vs_gamma34_sim}.
The majority of the sample is located between the ones for the bump-free (i.e., the \citet{2000ApJ...533..682C} and SMC) dust laws and the LMC2 supershell extinction curve, with some outliers that are presumably due to measurement errors.  The location occupied by the sample is very consistent with that found by \citet[][see their Figures 3 and 5]{2009A&A...499...69N}.

There is a clear correlation between $\beta_\mathrm{b}$ and $\gamma$.  A linear fit to the relatively secure data points ($\sigma(\gamma_{34})<1.0$) yields 
\begin{equation}
    \gamma_{34}=-2.19\beta_\mathrm{b}-4.16.
\label{eq:gamma34_beta_b}
\end{equation}
The best-fit relation closely passes the location of zero attenuation.  The slope of the fit is between the relations for the SMC and LMC2 supershell extinction curves.  Thus, we can state that, to first order, the dust attenuation of the sample can be represented by an attenuation curve with a moderate UV bump that can nearly reproduce the best-fit $\beta_\mathrm{b}$--$\gamma_{34}$ relation.  However, the substantial scatter remains ($\mathrm{rms}\approx1.15$ in $\gamma_{34}$) after accounting for the measurement errors, which clearly indicates the presence of intrinsic variations in the attenuation curves across the sample.

Given the threshold of $\gamma_\mathrm{34}=-2$ adopted by \citet{2009A&A...499...69N}, we find that the fraction of the sample having a nominal $\gamma_\mathrm{34}<-2$ to be 30\% (27\% if limited to those with $\sigma(\gamma_{34})<1.0$).  This fraction is in very good agreement with that found by \citet{2009A&A...499...69N} at $1<z<2.5$.

\subsection{The UV Bump Profiles in Stacked Spectra}
\label{sec:results_bumpProfiles}

\begin{figure*}[htbp]
\begin{center}
\includegraphics[width=6.0in]{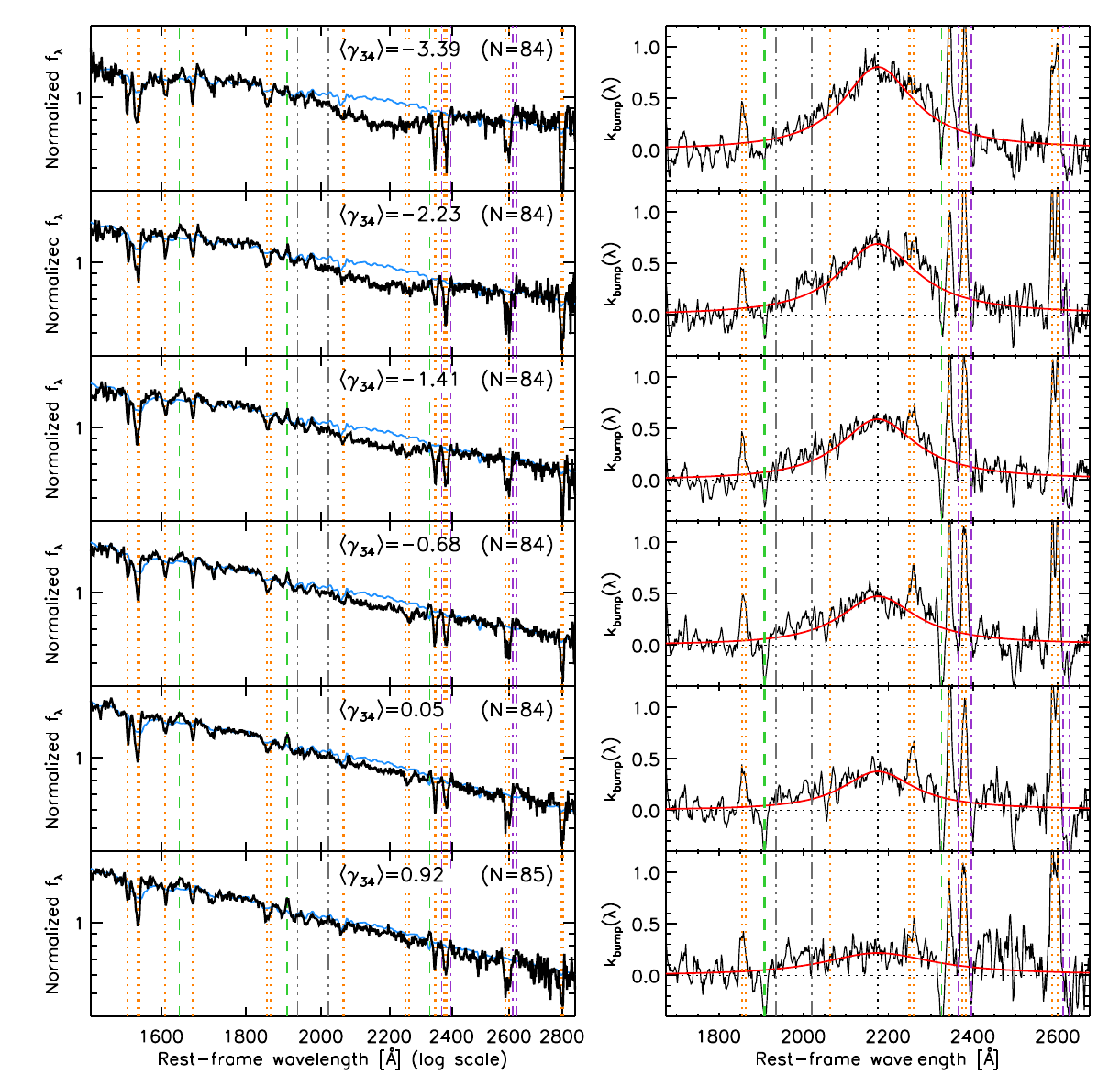}
\caption{Composite spectra in bins of the individual $\gamma_{34}$ measurements.  Prominent narrow spectral features are marked as in Figure \ref{fig:stack_all}.
Left panels: stacked VIMOS spectra (black lines) and coadded model spectra (blue lines).  The median $\gamma_{34}$ is indicated with the number of galaxies in each panel.
Right panels: coadded excess in the extinction curves, $k_\mathrm{bump}(\lambda)$, in the same subset as the left panels. The best-fit Drude profile is overplotted in each panel (red curve). \label{fig:stacks}}
\end{center}
\end{figure*}

\begin{deluxetable*}{lccccccccc}
\tablecaption{UV Bump Parameters\tablenotemark{a} \label{tb:params}}
\tablehead{
    \colhead{Sample}&
    \colhead{$N$}&
    \colhead{$\left< \gamma_{34}\right>_\mathrm{med}$\tablenotemark{b}}&
    \colhead{$\left< M_\ast \right>_\mathrm{med}$\tablenotemark{c}}&
    \colhead{$\left< \mathrm{sSFR} \right>_\mathrm{med}$\tablenotemark{d}}&
    \colhead{$\left< E(B-V) \right>_\mathrm{med}$\tablenotemark{e}}&
    \colhead{$\gamma~\mathrm{(\mu m^{-1})}$}&
    \colhead{$w_\lambda$~({\AA})\tablenotemark{f}}&
    \colhead{$B_k$\tablenotemark{g}}&
    \colhead{$B_A$~(mag)\tablenotemark{g}}}
\startdata
       All &  505 & $   -1.020 $ & $ 10.177 $ & $   -8.782 $ & $ 0.285 $ & $ 0.473 \pm 0.007 $ & $ 224 \pm   3 $ & $ 0.538 \pm 0.008 $ & $ 0.149 \pm 0.002 $ \\
\hline
$\gamma_{34}$-bin\_1 & 84 & $   -3.389 $ & $ 10.499 $ & $   -8.838 $ & $ 0.361 $ & $ 0.436 \pm 0.012 $ & $ 206 \pm   5 $ & $ 0.802 \pm 0.021 $ & $ 0.290 \pm 0.007 $ \\
$\gamma_{34}$-bin\_2 & 84 & $   -2.234 $ & $ 10.300 $ & $   -8.783 $ & $ 0.309 $ & $ 0.472 \pm 0.014 $ & $ 224 \pm   6 $ & $ 0.688 \pm 0.019 $ & $ 0.218 \pm 0.006 $ \\
$\gamma_{34}$-bin\_3 & 84 & $   -1.407 $ & $ 10.248 $ & $   -8.791 $ & $ 0.286 $ & $ 0.465 \pm 0.014 $ & $ 220 \pm   6 $ & $ 0.590 \pm 0.019 $ & $ 0.172 \pm 0.005 $ \\
$\gamma_{34}$-bin\_4 & 84 & $   -0.678 $ & $ 10.121 $ & $   -8.744 $ & $ 0.263 $ & $ 0.461 \pm 0.018 $ & $ 218 \pm   8 $ & $ 0.473 \pm 0.019 $ & $ 0.124 \pm 0.005 $ \\
$\gamma_{34}$-bin\_5 & 84 & $    0.047 $ & $  9.984 $ & $   -8.783 $ & $ 0.237 $ & $ 0.436 \pm 0.023 $ & $ 206 \pm  11 $ & $ 0.375 \pm 0.021 $ & $ 0.088 \pm 0.005 $ \\
$\gamma_{34}$-bin\_6 & 85 & $    0.918 $ & $ 10.019 $ & $   -8.745 $ & $ 0.256 $ & $ 0.683 \pm 0.068 $ & $ 325 \pm  32 $ & $ 0.216 \pm 0.021 $ & $ 0.050 \pm 0.005 $ \\
\hline
$M_\ast$-bin\_1 & 84 & $   -0.118 $ & $  9.754 $ & $   -8.768 $ & $ 0.190 $ & $ 0.360 \pm 0.023 $ & $ 170 \pm  11 $ & $ 0.472 \pm 0.034 $ & $ 0.088 \pm 0.006 $ \\ 
$M_\ast$-bin\_2 & 84 & $   -0.285 $ & $  9.964 $ & $   -8.721 $ & $ 0.242 $ & $ 0.425 \pm 0.023 $ & $ 201 \pm  11 $ & $ 0.434 \pm 0.023 $ & $ 0.103 \pm 0.005 $ \\ 
$M_\ast$-bin\_3 & 84 & $   -0.805 $ & $ 10.100 $ & $   -8.790 $ & $ 0.262 $ & $ 0.385 \pm 0.017 $ & $ 182 \pm   8 $ & $ 0.496 \pm 0.023 $ & $ 0.129 \pm 0.006 $ \\ 
$M_\ast$-bin\_4 & 84 & $   -1.302 $ & $ 10.255 $ & $   -8.780 $ & $ 0.283 $ & $ 0.497 \pm 0.015 $ & $ 235 \pm   7 $ & $ 0.574 \pm 0.018 $ & $ 0.161 \pm 0.005 $ \\ 
$M_\ast$-bin\_5 & 84 & $   -1.806 $ & $ 10.450 $ & $   -8.790 $ & $ 0.326 $ & $ 0.502 \pm 0.014 $ & $ 238 \pm   6 $ & $ 0.625 \pm 0.017 $ & $ 0.206 \pm 0.005 $ \\ 
$M_\ast$-bin\_6 & 85 & $   -2.013 $ & $ 10.800 $ & $   -8.853 $ & $ 0.408 $ & $ 0.541 \pm 0.016 $ & $ 257 \pm   7 $ & $ 0.567 \pm 0.015 $ & $ 0.226 \pm 0.006 $ \\ 
\enddata
\tablenotetext{a}{The central wavelength ($\lambda_0=1/x_0$) was fixed to 2175~{\AA} ($x_0=4.598~\mathrm{\mu m^{-1}}$).}
\tablenotetext{b}{Median values of the individual $\gamma_{34}$ measurements.}
\tablenotetext{c}{Median values of the individual PDF-weighted median $\log(M_\ast/M_\odot$) values derived from SED fitting.}
\tablenotetext{d}{Median values of the individual $\log(\mathrm{sSFR}_\mathrm{UV,corr}/\mathrm{yr}^{-1})$.}
\tablenotetext{e}{Median values of the individual PDF-weighted median $E(B-V)$ values derived from SED fitting.}
\tablenotetext{f}{The FWHM in wavelength ($w_\lambda\approx \gamma \lambda_0^2$).  }
\tablenotetext{g}{The errors include the effects of the uncertainties in the SED models (see Section \ref{sec:uncertainties}).  }
\end{deluxetable*}

In this subsection, we show the direct measurements of the UV bump profiles based on different stacked spectra. 
As already demonstrated in Figure \ref{fig:stack_all}, we found that the stacked $k_\mathrm{bump}(\lambda)$ of the entire sample of 505 galaxies is well described by a Drude profile.  The UV bump parameters of the fit that minimize the $\chi^2$ value are given in Table \ref{tb:params}.  The amplitude is found to be $B_k=\Bkavg \pm 0.008$,\footnote{The errors on the amplitude parameters ($B_k$ and $B_A$) denote the $1\sigma$ uncertainties, including the contribution of the uncertainties in the SED models.} which corresponds to a fractional bump absorption of $f_\mathrm{bump}=B_k/(B_k+k_\mathrm{Cal}(2175~\textrm{\AA}))=0.060$. 
The width $\gamma=0.473\pm 0.007~\mathrm{\mu m^{-1}}$ (or $224\pm 3$~{\AA}) is in agreement with that measured for $1<z<2.5$ galaxies with a similar approach by \citet{2009A&A...499...69N}.  The fiducial results have been obtained by fixing the central wavelength $\lambda_0$ to 2175~{\AA}. If it is treated as a free parameter, we found $\lambda_0=2167$~{\AA}, while the amplitude $B_k$ and width $\gamma$ change very little.

\begin{figure}[tbp]
\begin{center}
\includegraphics[width=3.3in]{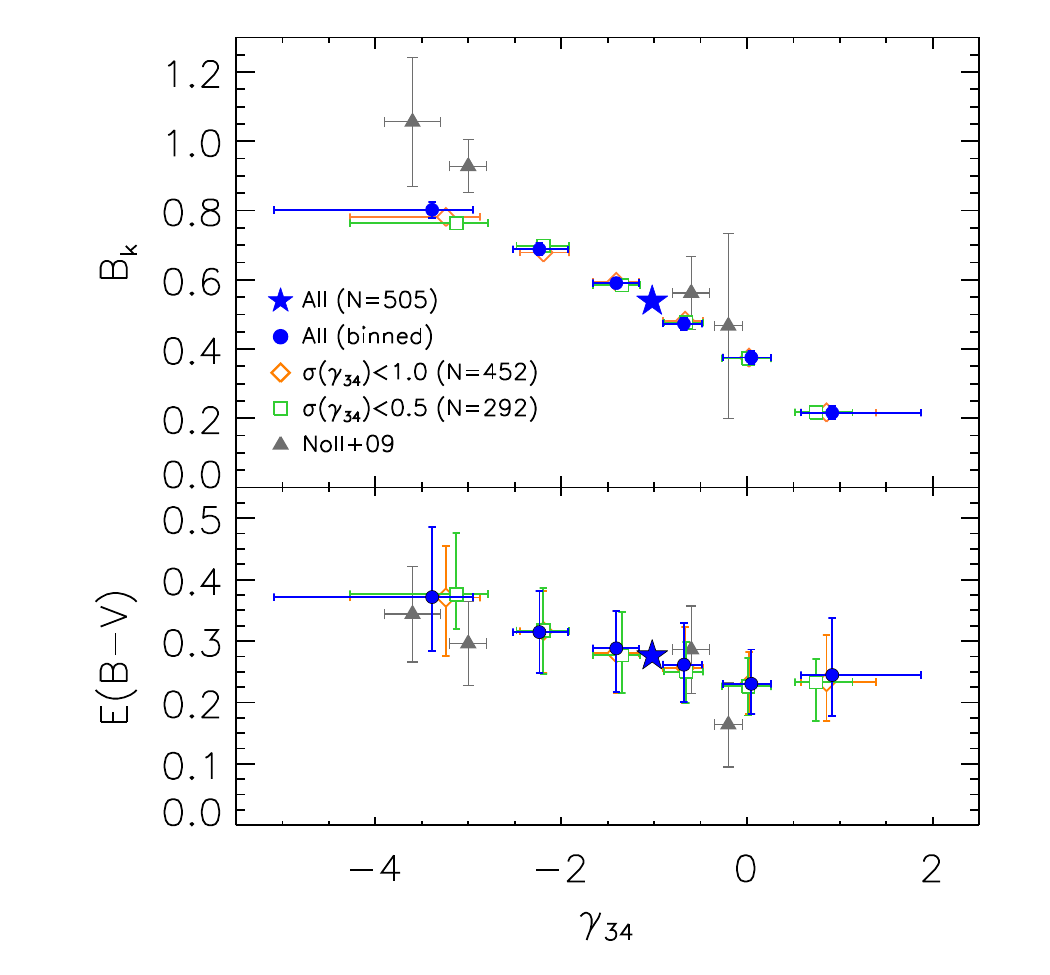}
\caption{
Upper panel: amplitude of the UV bump, $B_k$, vs. median $\gamma_{34}$ measured in the stacked spectra in six bins of the individual $\gamma_{34}$ measurements. The blue circles indicate the binned stacks using the full sample, while the open diamonds and squares are limited, respectively, to those with $\sigma(\gamma_{34})<1.0$ and $<0.5$.  The horizontal error bars denote the 16th--84th percentiles in $\gamma_{34}$.  The star indicates the stack of the entire sample.  The gray triangles indicate the measurements from \citet{2009A&A...499...69N}.
Lower panel: median values of the individual PDF-weighted median $E(B-V)$ in the same bins as above.  The vertical error bars present the 16th--84th percentiles in $E(B-V)$. \label{fig:gamma34_vs_Ebump}}
\end{center}
\end{figure}

We show in Figure \ref{fig:stacks} the stacked spectra and $k_\mathrm{bump}(\lambda)$ in bins of the nominal value of $\gamma_{34}$.  As expected, the stacked spectra appear to be more bent at larger negative $\gamma_{34}$.  The right panels show that the amplitude of the UV bump, $B_k$, decreases with increasing $\gamma_{34}$.  
The best-fit Drude parameters are given in Table \ref{tb:params}.  
If $x_0=1/\lambda_0$ is treated as a free parameter, we found highly consistent values ($\Delta \lambda_0 = |\lambda_0-2175~\textrm{\AA}| <20$~{\AA}; i.e., within 1\%) when the bump is strong, as in the upper panels of Figure \ref{fig:stacks}, while we measured $\lambda_0=2219\pm4$~{\AA} in the highest-$\gamma_{34}$ bin with the weakest bump. 
The amplitude $B_k$ and the width $\gamma$ hardly depend on whether or not the peak wavelength is fixed.  Note that, in the highest $\gamma_{34}$ bin, we measured the width parameter $\gamma$ ($=0.68\pm0.07~\mathrm{\mu~m^{-1}}$; or $w_\lambda=325\pm32$~{\AA}) to be larger than the rest, with a relatively large error.  This is presumably due to noise and the low $B_k$, rather than reflecting such a real wide bump.  
However, the bump amplitude $B_k$ itself is more or less accurate.
Among these subsamples, the median $M_\ast$ decreases weakly with $\gamma_{34}$, and the median sSFR hardly varies (see Table \ref{tb:params}).  This means that the galaxies in a single $\gamma_{34}$ bin have a range of $M_\ast$ and sSFR.  Even at a given $\gamma_{34}$, the variations in the intrinsic galaxy properties may bring a variation in the bump amplitude.  We will see this later in Section \ref{sec:results_bump_vs_gal}.

Figure \ref{fig:gamma34_vs_Ebump} indicates a tight negative correlation between $B_k$ and $\gamma_{34}$.  The bump amplitude reaches to $B_k \approx 0.8$ in the lowest $\gamma_{34}$ bin.  A concern with this approach, however, is that stacking those binned by the observed $\gamma_{34}$ may induce an artificial magnification of the bump strength, since we are constructing the stack on the basis of the quantity of interest itself. 
We thus repeated the analysis excluding those spectra with larger uncertainties in $\gamma_{34}$.  We attempted two thresholds of $\sigma(\gamma_{34})=0.5$ and 1.0 while adopting the same binning grid as the original one; the numbers of galaxies in the bins are thus no longer equal.  The results are shown together in Figure \ref{fig:gamma34_vs_Ebump}, demonstrating that the measurements of $B_k$ appear to converge in all but the lowest $\gamma_{34}$ bin, where the exclusion of lower-quality measurements results in a slightly lower $B_k$. The presence of a tight $\gamma_{34}$--$B_k$ correlation is robust.  Our result thus confirms the result of \citet[][see their Figure 8]{2009A&A...499...69N}, whose $\gamma_{34}$--$B_k$ data points are plotted in the figure, and strengthens their statement that the more negative $\gamma_{34}$ could be associated with more prominent UV bump in the attenuation curves.

\begin{figure*}[tbp]
\begin{center}
\includegraphics[width=2.3in]{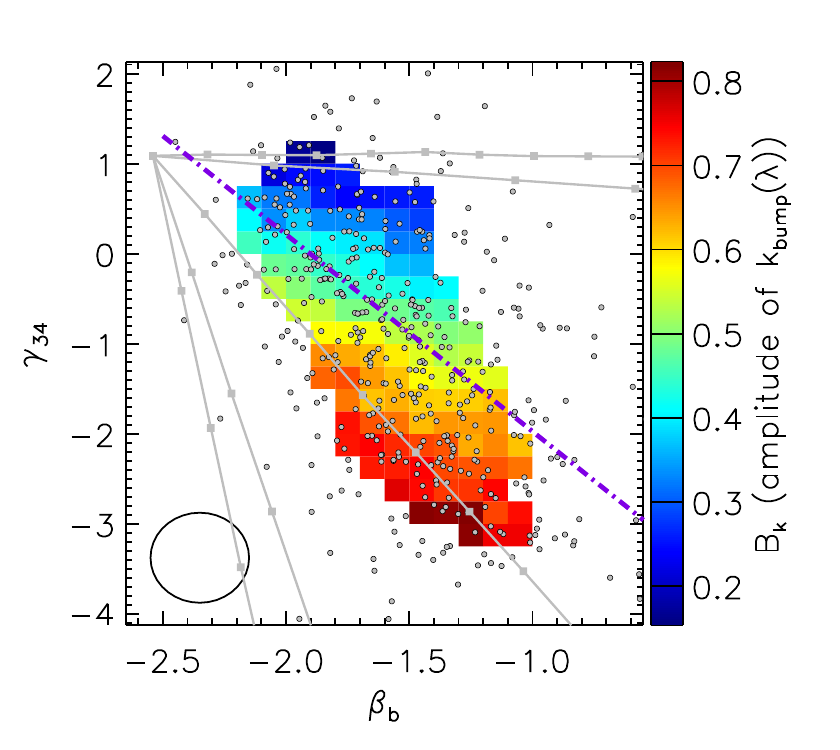}
\includegraphics[width=2.3in]{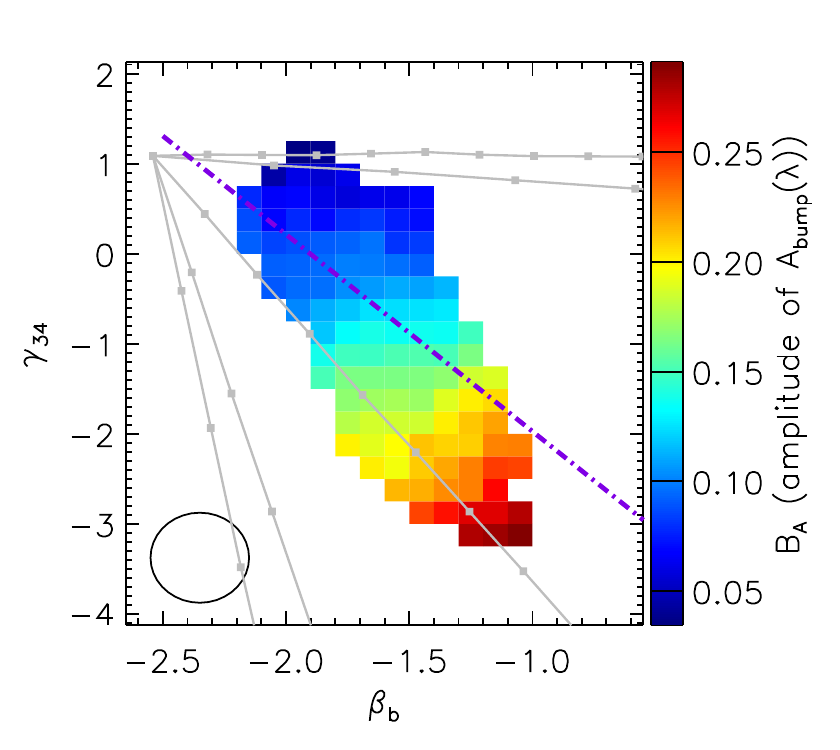}
\includegraphics[width=2.3in]{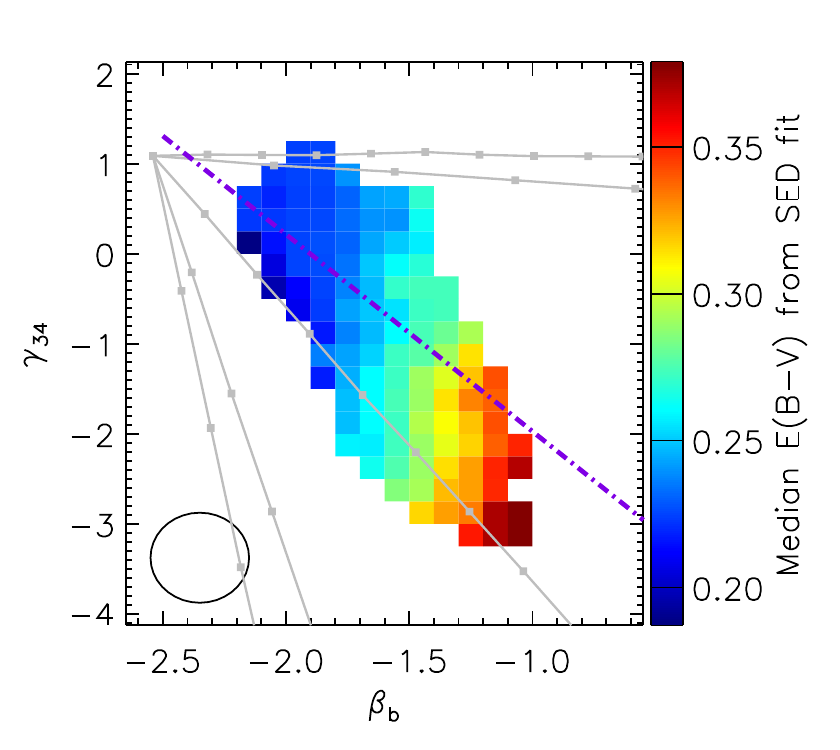}
\caption{
Left panel: stacked measurements of $B_k$ as a function of $\beta_\mathrm{b}$ and $\gamma_{34}$.  The ellipse in the lower left corner indicates the bin size.
The gray lines indicate the predictions for a representative model ($Z=0.008$, delayed SFH with $\tau=1~\mathrm{Gyr}$, and age of 0.50~Gyr), and the purple dotted-dashed line indicates the linear fit to the individual measurements (Equation \ref{eq:gamma34_beta_b}; same as in Figure \ref{fig:beta_b_vs_gamma34}).
Middle panel: stacked measurements of $B_A$ in the same stacking bins.
Right panel: median $E(B-V)$ in the same stacking bins.
\label{fig:beta_b_vs_gamma34_2Dbin}}
\end{center}
\end{figure*}

We next measure the bump strengths as a function of $\beta_\mathrm{b}$ and $\gamma_{34}$ by stacking galaxies at grid points in the $\beta_\mathrm{b}$--$\gamma_{34}$ plane in steps of 0.1 and 0.25, respectively, in $\Delta \beta_\mathrm{b}$ and $\Delta \gamma_{34}$.
Note that the size of the elliptic bin ($\Delta \beta_\mathrm{b}=0.2$ and $\Delta \gamma_{34}=0.5$ in each radius) is larger than the grid separations to achieve a reasonable S/N in the stacked spectra.  The left panel of Figure \ref{fig:beta_b_vs_gamma34_2Dbin} shows a clear trend of the stacked $B_k$ that is similar to the simulations that are shown in the lower panel of Figure \ref{fig:beta_b_vs_gamma34_sim}.  

We also show in Figure \ref{fig:beta_b_vs_gamma34_2Dbin} (middle panel) the observed amplitudes of the absolute excess absorption, $B_A$.  The excess absorption reaches $\sim0.3$~mag at 2175~{\AA} in the population of lowest $\gamma_{34}$ and highest $\beta_\mathrm{b}$. The observed $B_A$ increases with the distance from the locus that corresponds to $E(B-V)=0$ in the upper left corner.

Lastly, the right panel of Figure \ref{fig:beta_b_vs_gamma34_2Dbin} shows the median $E(B-V)$ in the corresponding bins.  As expected, the $E(B-V)$ is tightly correlated with $\beta_\mathrm{b}$ but almost independent of $\gamma_{34}$.  Thus, we can conclude that the variations in $\gamma_{34}$ at a given $\beta_\mathrm{b}$ reflect a real diversity of the bump strengths in the attenuation curves rather than a variation in the amount of overall reddening of the spectra.

\subsection{UV Bump Strength versus Galaxy Stellar Mass and sSFR}
\label{sec:results_bump_vs_gal}

\begin{figure}[tbp]
\begin{center}
\includegraphics[width=3.3in]{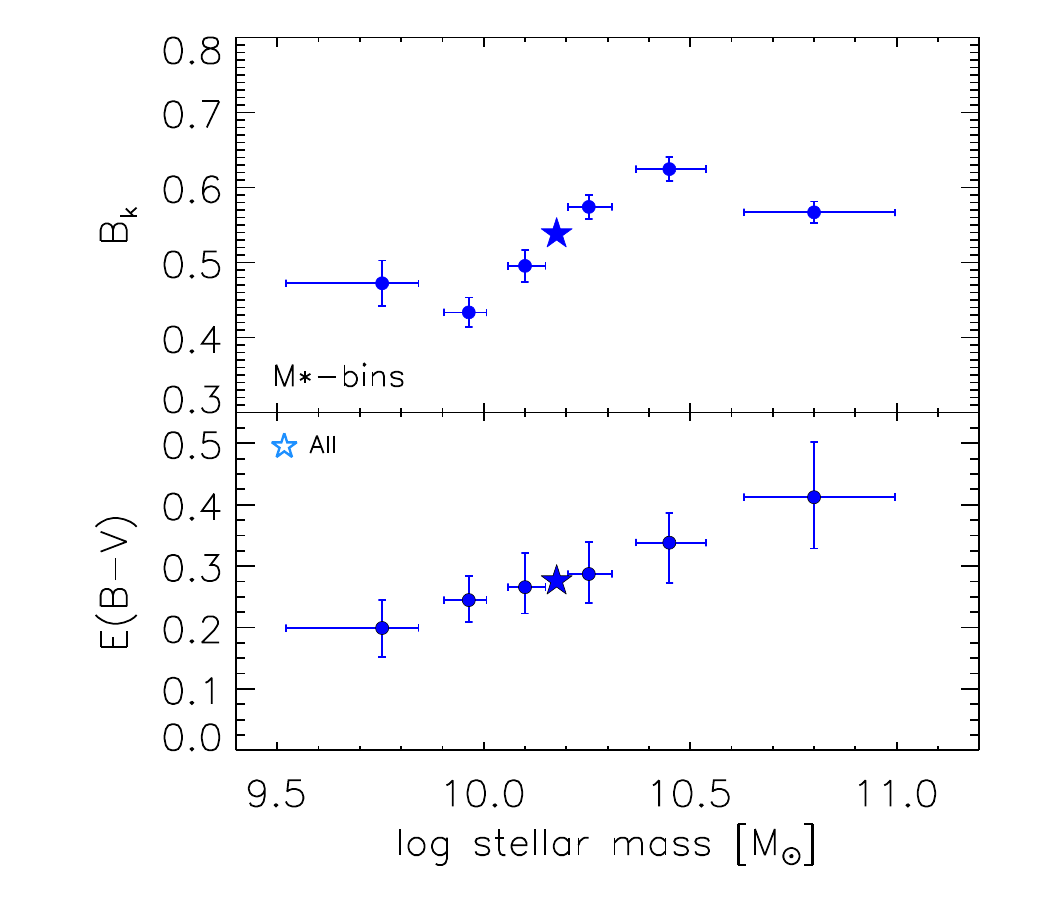}
\caption{
Upper panel: $B_k$ vs. median $M_\ast$ in six bins of the individual $M_\ast$ estimates.  The horizontal error bars present the 16th--84th percentiles in $M_\ast$.  
Lower panel: median values of the individual PDF-weighted median $E(B-V)$ in the same bins as above.  The vertical error bars present the 16th--84th percentiles in $E(B-V)$. \label{fig:mass_vs_Ebump}}
\end{center}
\end{figure}

In this subsection, we correlate the amplitude of $k_\mathrm{bump}(\lambda)$ with galaxy properties.  We first look at the global trend along the MS by stacking the individual spectra in six bins of the nominal stellar mass of the galaxies.  Table \ref{tb:params} summarizes the bump parameters obtained from the fits, and Figure \ref{fig:mass_vs_Ebump} shows the measured $B_k$ as a function of $M_\ast$.  We found a tight positive correlation across $9.8\lesssim \log M_\ast/M_\odot \lesssim 10.5$, though the correlation may not hold at the lowest- and highest-mass ends.

We next focus on both stellar mass and sSFR.  In order to understand the dependence of the bump strength on $M_\ast$ and sSFR separately, we measure the $k_\mathrm{bump}(\lambda)$ in stacked spectra of galaxies constructed at grid points in the $\log M_\ast$--$\log$~sSFR plane in steps of 0.1~dex in $M_\ast$ and 0.05~dex in sSFR. 
At each grid point, objects within an elliptic bin of $\Delta \log M_\ast=0.2~\mathrm{dex}$ and $\Delta \log \mathrm{sSFR}=0.1~\mathrm{dex}$ in each radius were stacked.  Therefore, the adjacent grid points partially share the same galaxies, and thus the measurements will be correlated.  We limited bins to those containing $\ge 20$ galaxies so as to achieve a reasonable S/N in the stacked spectra.

\begin{figure*}[tbp]
\begin{center}
\includegraphics[width=2.3in]{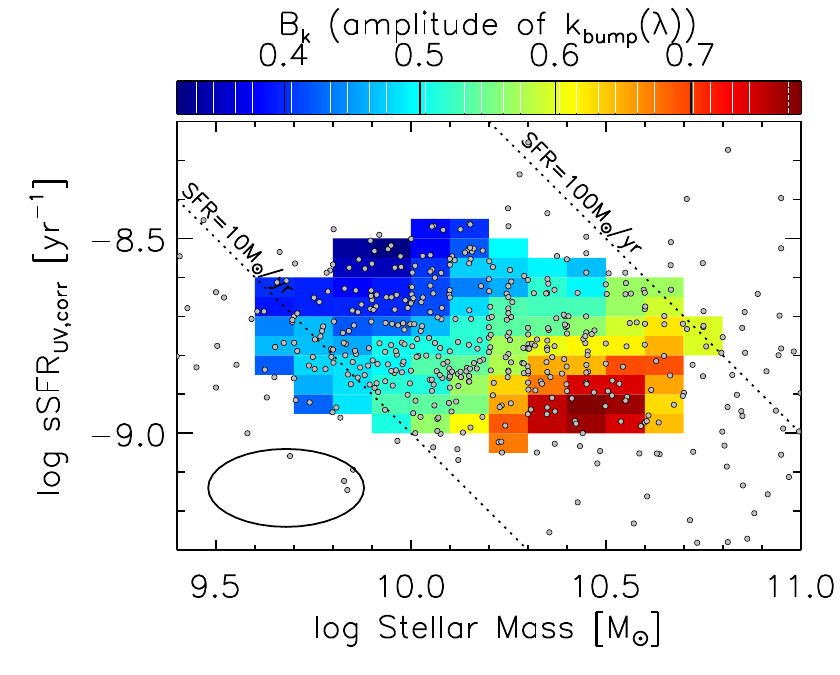}
\includegraphics[width=2.3in]{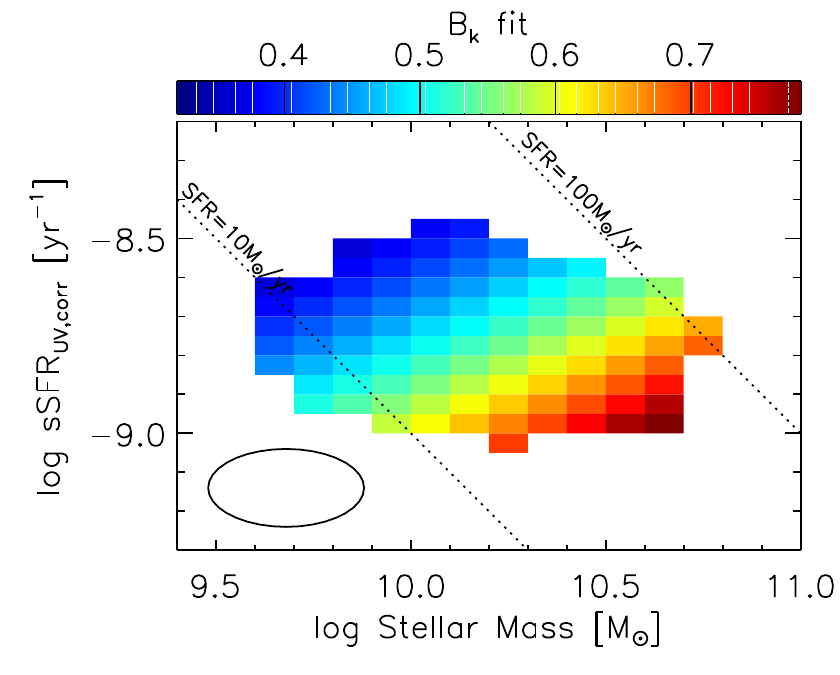}
\includegraphics[width=2.3in]{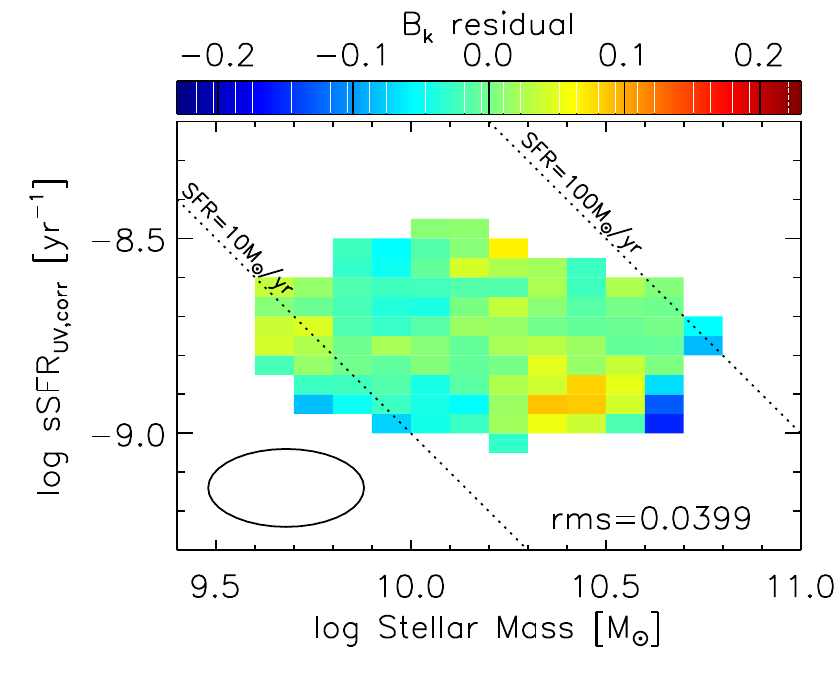}
\includegraphics[width=2.3in]{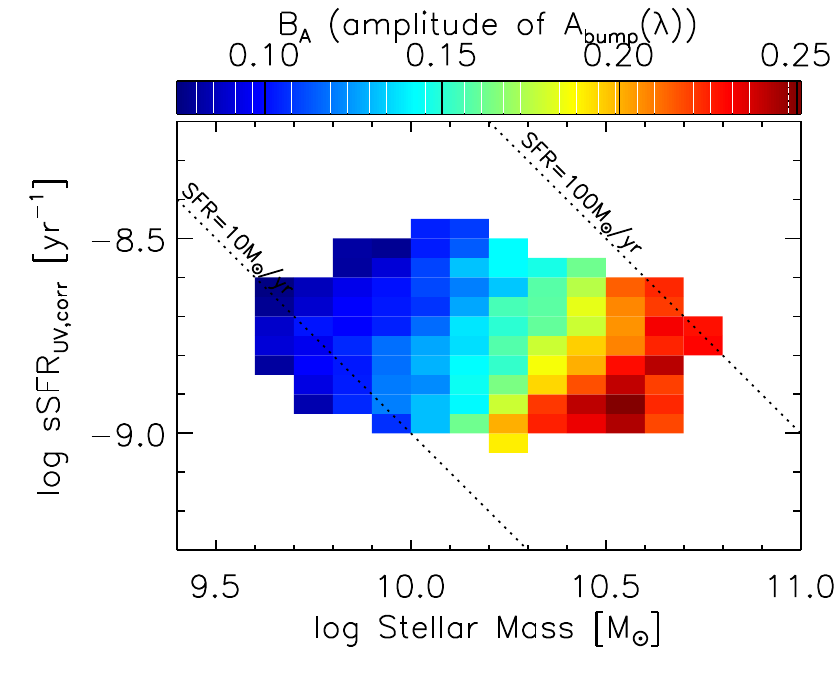}
\includegraphics[width=2.3in]{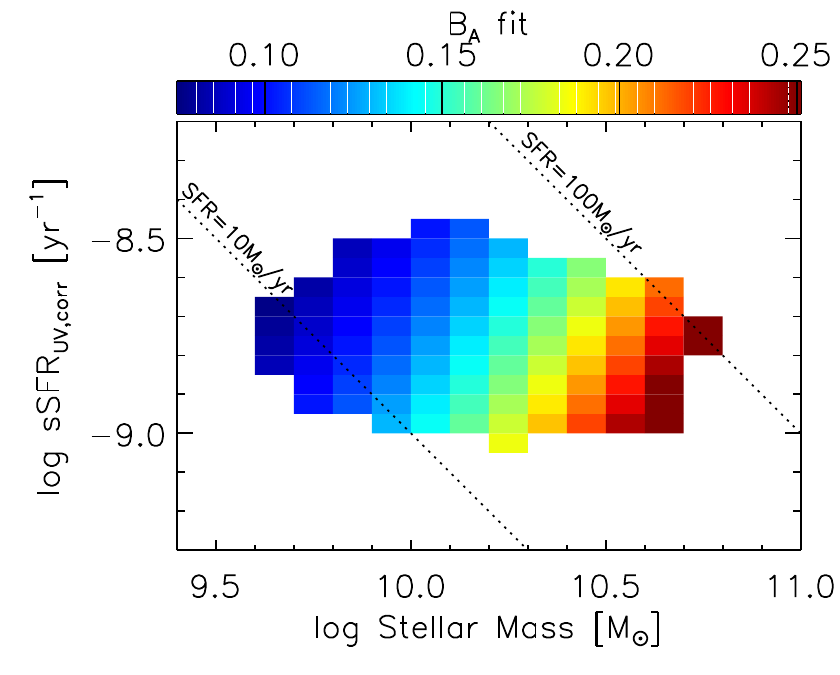}
\includegraphics[width=2.3in]{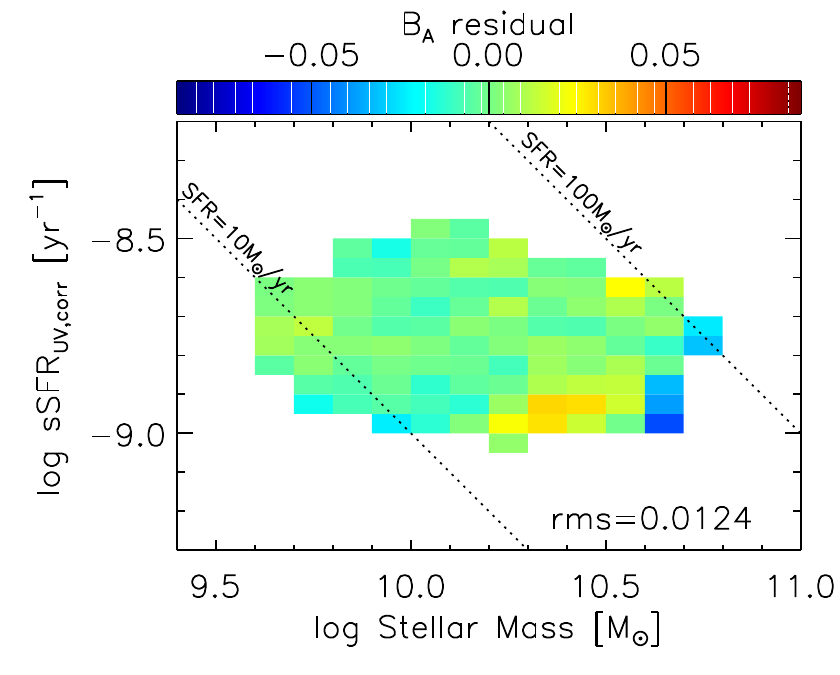}
\includegraphics[width=2.3in]{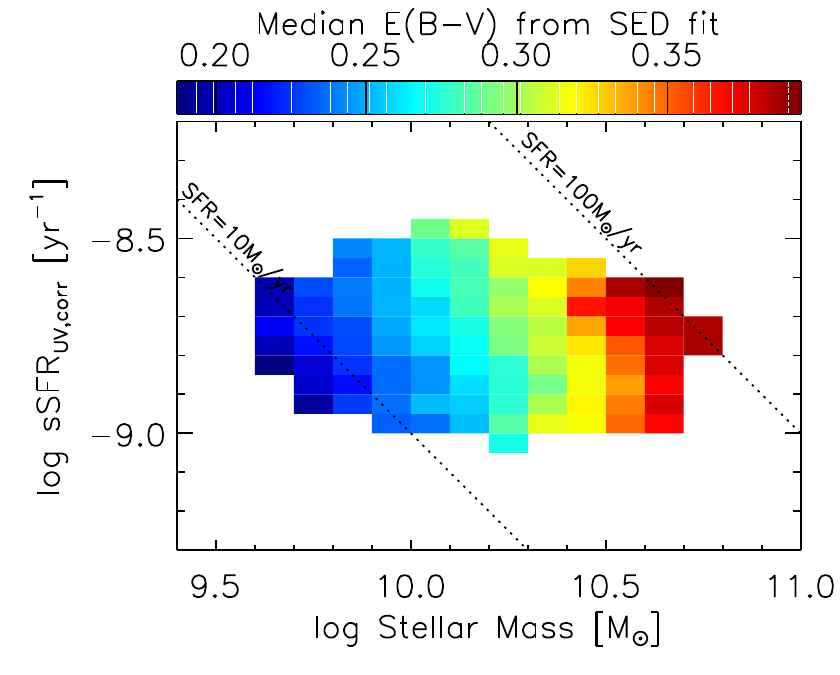}
\includegraphics[width=2.3in]{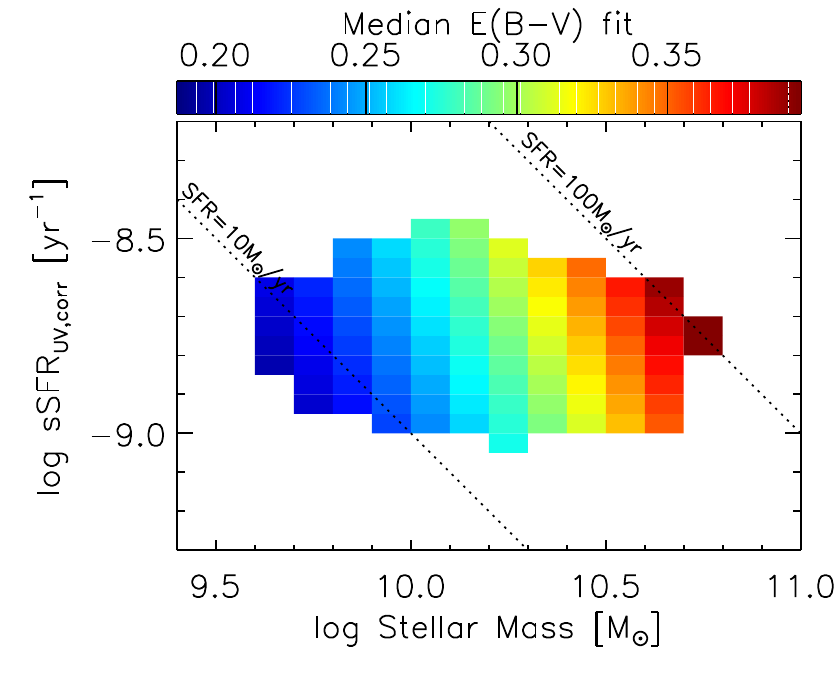}
\includegraphics[width=2.3in]{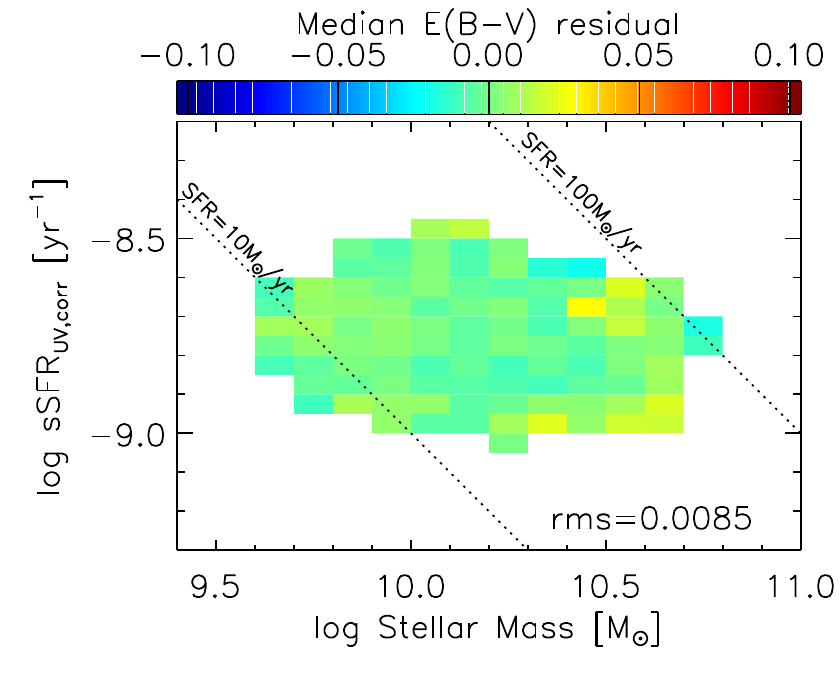}
\caption{
Top row: 
stacked measurements of $B_k$ as a function of $M_\ast$ and sSFR (left panel), the linear fit to $B_k$ (Equation \ref{eq:B_k=M,sSFR}; middle panel), and residuals from the fit (right panel).
The large ellipse indicates the bin size with radii of $0.2~\mathrm{dex}$ in $M_\ast$ and 0.1~dex in sSFR.  
Small circles indicate the individual galaxies.
For the residuals, the color range matches that of the left panel, but the middle value is shifted to zero.
Middle row: same as the top row but for the stacked measurements of $B_A$ in the same stacking bins.
Bottom row: for the median $E(B-V)$ in the same stacking bins.
\label{fig:M_vs_sSFR_2Dbin}}
\end{center}
\end{figure*}

Figure \ref{fig:M_vs_sSFR_2Dbin} shows in the top row the stacked $B_k$ as a function of $M_\ast$ and sSFR.
Across the accessible range, $B_k$ depends on both $M_\ast$ and sSFR, with a stronger dependence on sSFR.  We find that $\log B_k$ can be expressed by a linear function of $\log~M_\ast$ and $\log~\mathrm{sSFR}$.  The best fit is given by
\begin{eqnarray}
    \log B_k = && -0.212+0.191\log (M_\ast/10^{10}M_\odot) \nonumber \\ 
    && -0.427 \log (\mathrm{sSFR}/\mathrm{Gyr^{-1}}),
    \label{eq:B_k=M,sSFR}
\end{eqnarray}
which is shown in the middle panel in the top row.  The rms of the residuals in $B_k$ from this fitted surface is 0.0399 (right panel).  The result indicates that $B_k$ decreases with increasing sSFR at fixed $M_\ast$ and increases weakly with $M_\ast$ at fixed sSFR.  
Note that this empirical functional form has been obtained from a limited region in the $M_\ast$ versus sSFR space; thus, extrapolation of this fitted relation for galaxies lying outside the region may not be justified.

Figure \ref{fig:M_vs_sSFR_2Dbin} also shows in the middle row the amplitude of the absolute excess absorption, $B_A$, in the same stacking bins.  The amplitude $B_A$ ranges from $\approx0.07$ to 0.25~mag, increases with increasing $M_\ast$, and decreases with sSFR.  A linear fit to $\log B_A$ yields
\begin{eqnarray}
    \log B_A = && -0.857+0.457\log (M_\ast/10^{10}M_\odot) \nonumber \\ 
    && -0.303\log (\mathrm{sSFR}/\mathrm{Gyr^{-1}}),
    \label{eq:B_A=M,sSFR}
\end{eqnarray}
with the rms residual of $B_A$ of 0.0124~{mag}.  

The bottom panels of Figure \ref{fig:M_vs_sSFR_2Dbin} show that the median $E(B-V)$ increases with both $M_\ast$ and sSFR.  A linear fit yields
\begin{eqnarray}
    \log E(B-V) = && -0.602+0.251\log (M_\ast/10^{10}M_\odot) \nonumber \\ 
    && +0.150\log (\mathrm{sSFR}/\mathrm{Gyr^{-1}}),
    \label{eq:EBV=M,sSFR}
\end{eqnarray}
where the rms residual of $E(B-V)$ is 0.0092.  Both positive coefficients on $M_\ast$ and sSFR indicate that the $E(B-V)$ is rather correlated with the absolute SFR. The presence of a tight positive correlation between the reddening and SFR (or the UV luminosity) is consistent with those found at $z\sim2\textrm{--}3$ in previous studies \citep[e.g.,][]{1999ApJ...521...64M,2009ApJ...705..936B}.

It may be noted that the dependence of $B_A$ (Equation \ref{eq:B_A=M,sSFR}) can be reproduced trivially from a combination of those of $B_k$ and $E(B-V)$ (Equations \ref{eq:B_k=M,sSFR} and \ref{eq:EBV=M,sSFR}) and vice versa, as expected from the definition.

\section{Discussion}
\label{sec:discussion}

\subsection{Comparison with Local Starbursts}
\label{sec:discussion:comparison}

We have detected the broad excess absorption due to the UV bump in the attenuation curve of $1.3 \le z \le 1.8$ galaxies and mapped the dependence of this on $M_\ast$ and sSFR within the accessible region.  The impact of the UV bump on the spectra is intermediate between that of the LMC2 supershell extinction law and that of a featureless attenuation law, like the \citet{2000ApJ...533..682C} curve or the SMC extinction law, without a bump.  Our result is thus at face value different from what has been found for local starburst galaxies in the series of papers that led to the so-called Calzetti law \citep{1994ApJ...429..582C,1997AJ....113..162C,2000ApJ...533..682C}.
We therefore here attempt to directly compare our high-$z$ sample of MS galaxies and the local starbursts from which the featureless attenuation curve has been derived. We refer back to Section \ref{sec:local} for the selection of the local starbursts.  

The measurements of the absolute excess attenuation due to the UV bump are available for 23 starbursts (for 17 of them, the estimates of $M_\ast$ and sSFR are also available) in \citet{1994ApJ...429..582C} and thus can be compared with our $B_A$ measurements or $B_k$ by dividing by the reddening measurements of the sources, although there are some methodological differences in detail.\footnote{\citet{1994ApJ...429..582C} parameterized the absorbed flux as $\eta\equiv\Delta \log f_\lambda$ at $\lambda_\mathrm{rest}=2175$~{\AA}. This can be converted to our $B_A$ as $B_A=-2.5\eta$.}  An absorption excess, i.e., a positive bump amplitude, has been measured for 14/23 of the galaxies, or 9/17 if limited to those with an available $M_\ast$ and SFR. For the reminder, a negative amplitude has been obtained, presumably due to uncertainties in the spectral data.

\begin{figure}[tbp]
\begin{center}
\includegraphics[width=3.35in]{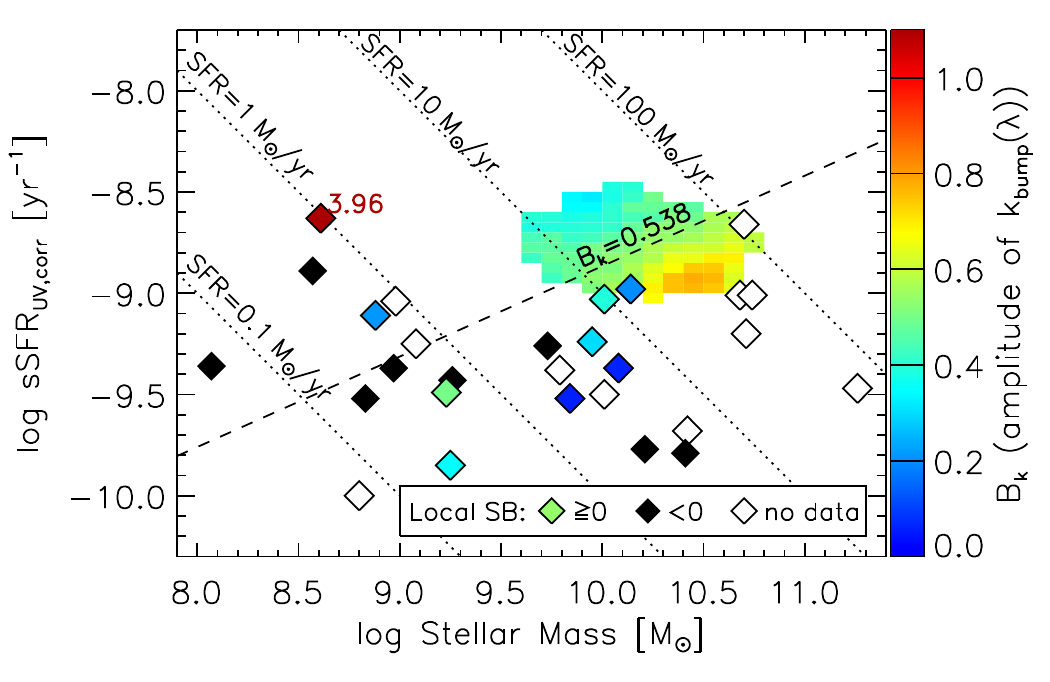}
\caption{
Bump amplitude, $B_k$, of the local starbursts from the Calzetti et al. papers (Section \ref{sec:local}) in the $M_\ast$ vs. sSFR plane compared with our results (colored grid).  The filled diamonds indicate the local starbursts and are color-coded by the bump amplitude, but the black filled diamonds correspond to the negative measurements of the bump.  The white diamonds indicate the local sources for which no individual bump measurements are available.
The color scale has been changed from Figure \ref{fig:M_vs_sSFR_2Dbin} to cover the measurements of the local sources down to zero.  
The dashed line indicates Equation (\ref{eq:B_k=M,sSFR}) at $B_k=\Bkavg$, which is obtained for our stack of the entire sample.
Note that a single red diamond has $B_k=3.96$, as labeled, which is largely exceeding the upper limit of the color range.
\label{fig:M_vs_sSFR_wLocal}}
\end{center}
\end{figure}

In Figure \ref{fig:M_vs_sSFR_wLocal}, we show the positions of the local starburst galaxies in the $M_\ast$ versus sSFR diagram relative to the region in which our stacked spectra have yielded $B_k$ measurements.  
As noted earlier, the local ``starburst'' sources have sSFRs that are, on average, $\sim0.5~\mathrm{dex}$ lower than our high-$z$ MS galaxy sample while spanning a wider range of stellar mass of $8\lesssim \log M_\ast/M_\odot \lesssim11$.  
The dependence of $B_k$ that we have observed within our high-redshift sample indicates that lower sSFR galaxies tend to have a stronger bump in the attenuation curves, and the extrapolation of this observed trend would clearly predict that the local sources should generally have much larger UV bumps.
Since the majority of the local sample are located below the dashed line in Figure \ref{fig:M_vs_sSFR_wLocal}, we would definitely expect that the $B_k$ of the local sample should exceed the value observed in the stacked spectrum of our entire sample ($B_k=\Bkavg$). Considering the then-current data quality, it should have been easy to detect such a bump signature in the average attenuation law (see, e.g., Figure 20 of \citealt{1994ApJ...429..582C}).  

The local starbursts, however, show on average much weaker bump strengths with the median amplitude $\left<B_k\right>_\mathrm{med}=0.20$ ($\left<B_A\right>_\mathrm{med}=0.028$~mag) for all 23 local sources and $\left<B_k\right>_\mathrm{med}=0.06$ ($\left<B_A\right>_\mathrm{med}=0.025$~mag) for the plotted 17 sources.  These average excess absorptions are significantly smaller than the $B_k=\Bkavg$ ($B_A=\BAavg~\mathrm{mag}$) measured for our stack of the entire sample. 

It is clear from the color-coding of Figure \ref{fig:M_vs_sSFR_wLocal} that the $B_k$ values of the local starbursts are completely inconsistent with the trend established by the higher-redshift sample.  We can thus conclude that local starbursts, or at least those used in the Calzetti et al. papers, have an attenuation curve in which the UV bump is, on average, substantially weaker than what is seen in $1.3 \le z\le 1.8$ star-forming galaxies, despite the latter's higher sSFR.

We here recall, however, that the local ``starbursts'' are, as their name implies, almost certainly undergoing a substantial, recent, and short-lived elevation of their SFR.  In contrast, the galaxies in the higher-redshift sample are close to the MS and have probably been forming stars at a more or less steady rate.

To gain further insights into how the bump feature depends on the star formation activity, we now compare the high-$z$ sample and the local starbursts by renormalizing their sSFRs to that of the MS at the appropriate epoch; i.e., we consider the 
$\Delta \log \mathrm{sSFR} = \log (\mathrm{sSFR}/\left<\mathrm{sSFR}\right>_\mathrm{MS})$ relative to the appropriate MS.
We utilize the MS relation at $z=1.5\textrm{--}2.0$ derived by \citet{2014ApJ...795..104W} to compute $\Delta \log \mathrm{sSFR}$ for our high-$z$ sample and a relation combining the \citet{2014ApJS..214...15S} relation at $z=0.01$ ($M_\ast \ge10^{9.66}~M_\odot$) and the one from \citet[][$<10^{9.66}~M_\odot$]{2015ApJ...801L..29R} for the local sources (see Figure \ref{fig:Mstar_vs_SFRUV}).
We then rebin our sample in the $\log M_\ast$ versus $\Delta \log \mathrm{sSFR}$ plane in steps of, respectively, 0.1 and 0.05~dex using elliptic apertures with radii of 0.2 and 0.1~dex in each axis.  

\begin{figure}[tbp]
\begin{center}
\includegraphics[width=3.35in]{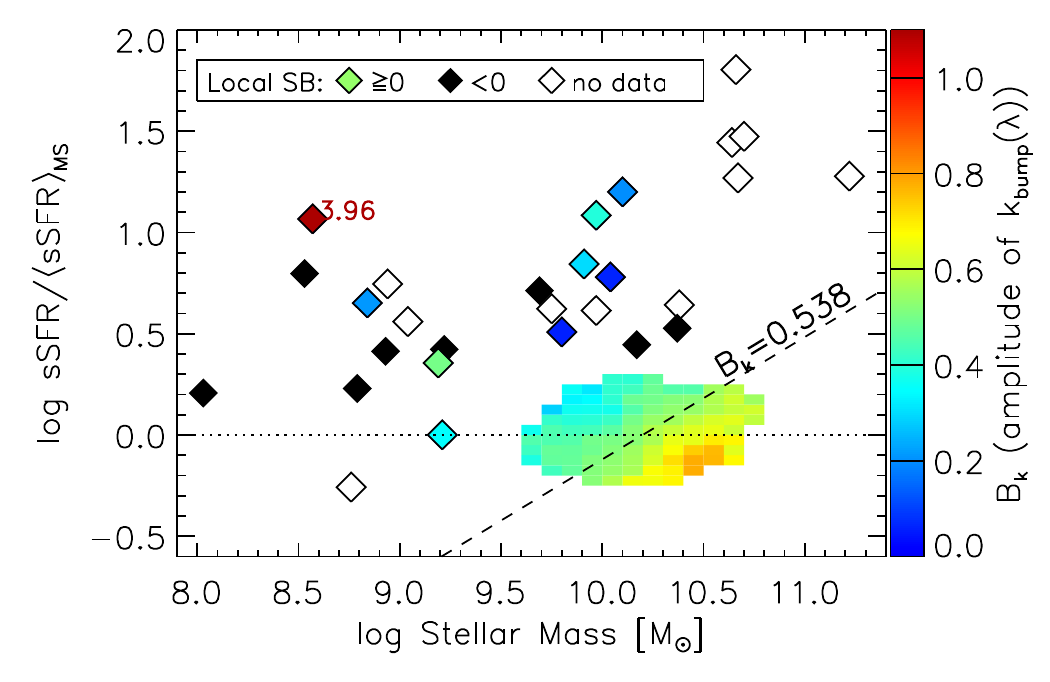}
\includegraphics[width=3.35in]{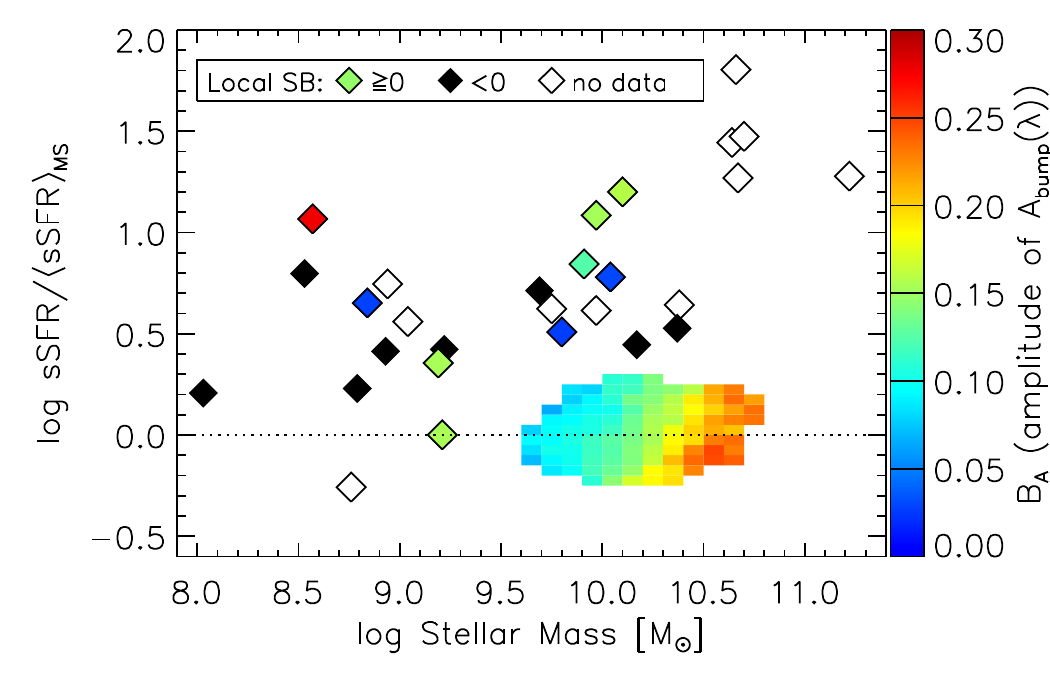}
\includegraphics[width=3.35in]{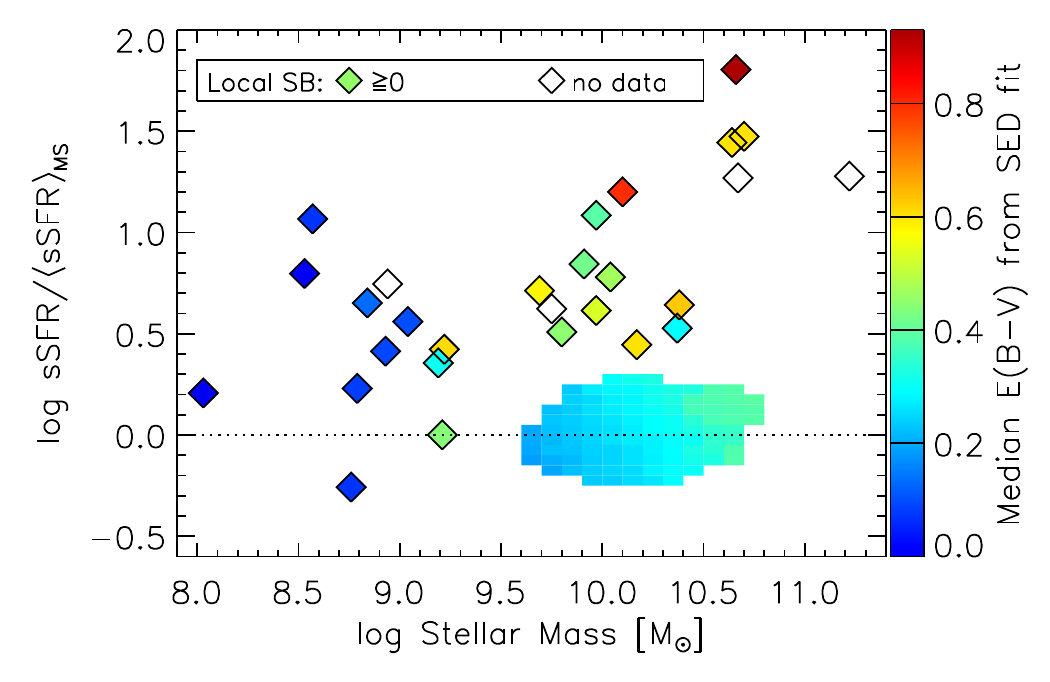}
\caption{
Bump amplitude, $B_k$ (top) and $B_A$ (middle), and $E(B-V)$ (bottom) as a function of $M_\ast$ and $\Delta \log \mathrm{sSFR}=\log (\mathrm{sSFR}/\left<\mathrm{sSFR}\right>_\mathrm{MS})$ for our sample (colored grid) and the local starbursts (diamonds).  The horizontal dotted lines indicate $\Delta \log \mathrm{sSFR}=0$.  The dashed line in the top panel indicates Equation (\ref{eq:B_k=M,dMS}) at $B_k=\Bkavg$, which is obtained for our stack of the entire sample.  
\label{fig:M_vs_dMS_wLocal}}
\end{center}
\end{figure}

Figure \ref{fig:M_vs_dMS_wLocal} summarizes the new bump measurements, $B_k$ and $B_A$, and the reddening $E(B-V)$ for our high-$z$ MS galaxies, together with the local starbursts, in the $\log M_\ast$ versus $\Delta\log~\mathrm{sSFR}$ plane.
Our sample covers $\pm0.3~\mathrm{dex}$ with respect to the MS.  It may be worth providing the empirical fits to these measurements as a function of $M_\ast$ and $\Delta\log~\mathrm{sSFR}$.  The best linear fits for our sample are given as follows:
\begin{eqnarray}
    \log B_k = && -0.319+0.245\log (M_\ast/10^{10}M_\odot) \nonumber \\ 
    && -0.413~\Delta\log\mathrm{sSFR},\\
    \label{eq:B_k=M,dMS}
    \log B_A = && -0.933+0.497\log (M_\ast/10^{10}M_\odot) \nonumber \\ 
    && -0.268~\Delta\log\mathrm{sSFR},\\
    \label{eq:B_A=M,dMS}
    \log E(B-V) = && -0.602+0.251\log (M_\ast/10^{10}M_\odot) \nonumber \\ 
    && +0.150~\Delta\log\mathrm{sSFR}.
    \label{eq:EBV=M,dMS}
\end{eqnarray} 
The rms residuals in $B_k$, $B_A$, and $E(B-V)$ are, respectively, 0.0378, 0.0113, and 0.0092.

It is obvious in the top panel of Figure \ref{fig:M_vs_dMS_wLocal} that the majority of the local starbursts differ from the local MS, lying well above the commonly used threshold of $\Delta \log~\mathrm{sSFR} >4$ (0.6~dex) for starbursts.  
Interestingly, the observed UV bump strengths, $B_k$, in the local starbursts are now in much better agreement, at least qualitatively, with the extrapolation of the trends in our high-$z$ sample that predicts weaker bumps for higher $\Delta \log~\mathrm{sSFR}$.  This suggests that the weak or absent UV bump in the local starbursts is linked to their large positive offset from the MS, i.e., their $\Delta\log~\mathrm{sSFR}$, rather than the their absolute SFRs or sSFRs.  In the following subsection, we will come back to this point in interpreting the observed behavior of the bump feature.

\subsection{Interpretation of the Dependence of $B_k$}
\label{sec:discussion:interpretation}

The variations in bump strength can be produced by the changes in the nature of dust grains, such as size distribution and/or composition, or the radiative transfer effects in different dust-to-star geometries.  Here we attempt to interpret the observed trends of $B_k$, first in terms of the underlying grain properties and later in terms of geometrical effects in the subsequent subsection.

The identification of carriers of the UV bump is a long-standing topic of controversy (see \citealt{1989IAUS..135..313D} for a review).  
Carbonaceous grains containing sp$^2$-bonded structures are the most widely accepted materials because they exhibit a broad excess in the absorption cross section at $\lambda\sim2200$~{\AA} due to resonant absorption in the Rayleigh limit (i.e., grain size $a \lesssim 0.01~\mathrm{\mu m}\ll \lambda$; e.g., \citealt{1971Natur.229..237G}).  Multiple possible forms of the carbonaceous bump carrier candidates have been considered, including (partially) graphitized particles, a random assembly of microscopic sp$^2$ carbon chips \citep{2009MNRAS.394.2175P}, and polycyclic aromatic hydrocarbons (PAHs; e.g., \citealt{1994ApJ...422..176M,2010ApJ...712L..16S,2020MNRAS.492.3779H}).
In the following discussion, we assume that the UV bump is attributed to small ($a\lesssim0.01~\mathrm{\mu m}$) carbonaceous grains, whatever their form.

The strength of the bump feature will reflect the destruction and creation of the bump carriers.  The destruction of small-sized ($a\lesssim0.01~\mathrm{\mu m}$) grains that may be able to give rise to the bump feature, whatever the chemical composition, is accelerated by an enhanced frequency of supernovae (SNe) and/or intense hard radiation fields \citep{1998ApJ...500..816G,2008ApJ...686.1056S}.  Harder and more intense radiation fields may arise due to the lower metallicity in less massive and/or higher sSFR galaxies \citep[e.g.,][]{2014ApJ...791..130Z,2018ApJ...858...99S}, and these may more efficiently destroy small grains \citep{2008ApJ...686.1056S}.

The production rate of small grains is also a factor.  It is thought that dust grains of size $a\sim0.1~\mathrm{\mu m}$ are produced by SNe and asymptotic giant branch (AGB) stars.  Carbonaceous grains that could be the precursors of bump carriers would thus be attributed mainly to carbon-rich AGB stars \citep{2011A&ARv..19...43G}.
Because only moderate-mass stars ($\lesssim 3M_\odot$) can be carbon stars \citep{1981A&A....94..175R}, the dust production in the AGB phase achieves its peak efficiency $\sim1$~Gyr (the lifetime of $\approx2M_\odot$ stars) after star formation \citep{2008A&A...479..453Z}.  
Contrarily, the time delays in the appearance of Type II SNe and oxygen-rich AGB stars, whose progenitor masses are $>8~M_\odot$ and $\sim (3\textrm{--}8)\times M_\odot$, respectively, are significantly shorter than 1~Gyr.
The ejected carbonaceous grains then need to be shattered into small grains and, if amorphous, at least partially graphitized and/or aromatized to be bump carriers.  The timescales of these processes, however, are both thought to be shorter ($\sim100~\mathrm{Myr}$) than the time delay in the appearance of the carbon-rich AGB stars \citep{1990MNRAS.243..570S,2015MNRAS.447.2937H}.

We attempt to connect the observed dependence of $B_k$ to possible processes relevant to the evolution of dust grains.  In the Local Group (i.e., among the MW, LMC, and SMC), more massive and/or metal-rich galaxies tend to have extinction curves that exhibit stronger bumps. 
At $z\sim2$, \citet{2020ApJ...899..117S} reported that attenuation curves of higher-metallicity galaxies exhibit stronger bumps.  To first order, the observed $M_\ast$--$B_k$ correlation (Figure \ref{fig:mass_vs_Ebump}) appears to be consistent with these trends, given the metallicity being tightly correlated with $M_\ast$ \citep[e.g.,][]{2014ApJ...791..130Z,2017ApJ...835...88K}.
The abundance of PAHs, estimated from the mid-IR luminosity, is known to depend strongly on the metallicity of the galaxies, which may be linked to the possible metallicity dependence of the bump strength given PAHs being a plausible candidate of the bump carrier \citep[e.g.,][]{2005ApJ...628L..29E,2008ApJ...672..214G,2017ApJ...837..157S}.  
The PAH-to-dust abundance ratio increasing with metallicity is also predicted by theoretical models in terms of the changes in the efficiency of shattering and coagulation processes, which both play roles in producing small grains \citep{2014MNRAS.439.2186S,2019MNRAS.489.5218R}.

Moreover, our results, obtained from binning the sample onto the $M_\ast$--sSFR plane, indicate that the $M_\ast$ dependence is only moderate when fixing sSFR.  This implies that an apparently stronger $M_\ast$ dependence may be derived if measuring $B_k$ only along the $M_\ast$ axis due to the anticorrelation between $M_\ast$ and sSFR, as well as between $B_k$ and sSFR within a representative sample of MS galaxies.  From now on, we focus on the dependence of $B_k$ on sSFR at fixed $M_\ast$ and will propose a simple approach to simultaneously explain our findings and the absence of the bump in local starbursts.

Knowledge of the physical processes playing on dust grains, as noted above, suggests that a key quantity in the balance between production and destruction of the bump carriers may be the recent SFR as measured at $\sim10^{7\textrm{--}8}$~yr compared with the SFR of order 1~Gyr ago (i.e., the time delay in the appearance of carbon-rich AGB stars).  For simplicity, we may denote, following \citet{2020ApJ...892...87W}, the ratio of the current instantaneous SFR (measured on $\sim10^{7\textrm{--}8}~\mathrm{yr}$ timescales) to that 1~Gyr ago as SF$_{79}$.   It should be noted that this definition is actually slightly different from that in \citet{2020ApJ...892...87W} who defined SF$_{79}$ to be the ratio of the instantaneous SFR to that averaged over the previous 1~Gyr.

We may then look at the behavior of SF$_{79}$ in two idealized cases.  First, for galaxies with a constant sSFR, both the stellar mass\footnote{Here we are referring to an sSFR computed using the stellar mass as the integral of the past SFH, so that the sSFR$^{-1}$ is the mass doubling timescale. This is different from the definition of the stellar masses estimated for the individual galaxies in our sample, which denotes the mass in stars that have survived to the time of observation.} and the SFR increase exponentially with an e-folding timescale given by the inverse sSFR$^{-1}$.  The change in SFR will therefore depend on the sSFR multiplied by the time interval of interest, which in this case is of order 1~Gyr:
\begin{equation}
    \ln \mathrm{SF}_{79} \sim \mathrm{sSFR} \times 1~\mathrm{Gyr}.
    \label{eq:SF79_1}
\end{equation}
It is then easy to see that different galaxies with different constant sSFRs will have different SF$_{79}$.  We can approximately write these in terms of some average fiducial value as 
\begin{equation}
    \Delta \log \mathrm{SF}_{79} \sim \Delta \log \mathrm{sSFR} \times \mathrm{sSFR}_\mathrm{Gyr}
\label{eq:SF79_2}
\end{equation}
if the offsets are small ($\ll 1~\mathrm{dex}$).  Here $\mathrm{sSFR}_\mathrm{Gyr}=\mathrm{sSFR}\times1~\mathrm{Gyr}$, i.e., the sSFR in units of Gyr$^{-1}$.   When the inverse sSFR$^{-1}$ timescale becomes comparable to the gigayear timescales of interest, i.e., $\mathrm{sSFR_{Gyr}} \sim$ unity, as is certainly the case at high redshift $z \sim 2$, variations in the (steady) sSFR will cause significant variations in SF$_{79}$, with a consequent effect expected on the amplitude of the absorption bump.  This effect alone could conceivably account for the observed trend of decreasing bump strength with sSFR that is seen within our high-redshift sample of MS galaxies.  Note, however, that at low redshift, where $\mathrm{sSFR}\ll 1~\mathrm{Gyr^{-1}}$, it can be seen that variations in the steady sSFR will have a much smaller effect on SF$_{79}$. 

However, a potentially much larger effect on SF$_{79}$ will be produced by any rapid temporal changes in the sSFR. In particular, even at low redshift, when most galaxies have $\mathrm{sSFR} \ll 1~\mathrm{Gyr}^{-1}$, a short sharp increase in the SFR, as in a ``starburst,'' could lead to a corresponding increase in SF$_{79}$.  If the burst is of short duration, much less than 1 Gyr, then it simply follows (independent of the value of the fiducial sSFR) that
\begin{equation}
   \Delta \log {\rm SF_{79}} \sim \Delta \log \mathrm{sSFR}.
\label{eq:SF79_3}
\end{equation}
This second effect is almost certainly the one that is relevant for the low-redshift sample of starburst galaxies.
Remarkably, it can be seen that Equations (\ref{eq:SF79_2}) and (\ref{eq:SF79_3}) have exactly the same form if Equation (\ref{eq:SF79_2}) is evaluated in the regime where $\mathrm{sSFR} \sim 1~\mathrm{Gyr}^{-1}$, i.e., at $z \sim 2$. 

This undoubtedly simplified picture therefore provides a natural explanation of the effects in Figure \ref{fig:M_vs_dMS_wLocal}, in which high-redshift MS galaxies and local starburst galaxies are combined; high-redshift MS galaxies in a quasi-steady state and local starburst galaxies undergoing a short temporal elevation in their sSFR happen to display the same relation when the amplitude of the bump feature is plotted against the $\Delta \log \mathrm{sSFR}$ relative to the appropriate MS but quite different relations when plotted against the sSFR.  
The success of this simple approach in explaining the observations presented in this paper adds support to the idea that the abundance of the carrier grains of the 2175~{\AA} excess absorption is strongly linked to variations in the recent SFH of the galaxies over the last billion years or so.

This picture, however, appears not to apply, at least immediately, to local MS galaxies; a prediction from Equation (\ref{eq:SF79_2}) is that we would expect to see a very much weaker or absent trend of bump strength with $\Delta \log$~sSFR for MS galaxies at low redshift with sSFR$_\mathrm{Gyr} \ll 1$.  On the other hand, \citet{2018ApJ...859...11S} found that more massive galaxies tend to have attenuation curves that are steeper and exhibit weaker bumps, and as a secondary dependence, the slope and the bump strength become, respectively, steeper and stronger toward both the upper and lower envelopes of the MS at fixed $M_\ast$.  In parallel, no dependence on the gas-phase metallicity was found in both the slope and the bump strength at fixed $M_\ast$.  These findings appear to conflict with the trend seen in the extinction curves in the Local Group galaxies and $M_\ast$ and metallicity trends seen at high redshifts (this paper and \citealt{2020ApJ...899..117S}).  
As discussed in \citet{2018ApJ...859...11S}, their results suggest that the dust-to-star geometries may play an important role in regulating the attenuation curves in these local galaxies.  We will discuss the the effects of geometry in the  next subsection.

We should also mention a caveat in the assumption that is a base of our interpretation.  Here the bump carriers are assumed to be small carbonaceous grains that were initially supplied by carbon-rich AGB stars and then processed in the ISM to be able to give rise to the bump feature.  Although many theoretical models employ this general picture \citep[e.g.,][]{2008A&A...479..453Z,2013MNRAS.432..637A,2020MNRAS.492.3779H}, on the other hand, there is a known paradoxical problem that the bump strength appears to be anticorrelated with the abundance ratio of carbon-rich to oxygen-rich stars (the C/M ratio); the SMC (with no bump) presents the highest C/M, while the MW (with a strong bump) presents the lowest among the Local Group objects \citep[e.g.,][]{1986ApJ...305..634C,1999IAUS..191..535G,2003MNRAS.338..572M}.  
This established observational fact appears to disfavor carbonaceous grains being the bump carriers, or at least scenarios in which the supply of bump carriers is owed to carbon-rich AGB stars.  There thus remains a tension between physical arguments favoring carbon grains and astrophysical arguments disfavoring them.

\subsection{Effects of Dust-to-star Geometry}
\label{sec:discussion:geometry}

In the last subsection, we proposed an interpretation of the sSFR dependence of $B_k$ in terms of the production and destruction of the bump carriers, or, in other words, the actual variations in the underlying extinction curves.  However, even if the extinction curve is constant, the radiative transfer effect in different dust-to-star geometries can produce a wide variation in the shape of the attenuation curves.  Radiative transfer calculations have demonstrated that, in general, the attenuation curve tends to be shallower (or flatter) toward the FUV part and exhibit a weaker bump for a higher dust column density (or higher overall attenuation), a clumpier ISM, and a geometry in which the stars and dust are mixed (instead of a geometry with a screen or shell of dust; e.g., \citealt{1996ApJ...463..681W,2000ApJ...528..799W,2016ApJ...833..201S,2018ApJ...869...70N}).
The radiative transfer effects have also been indicated in observations, especially through detecting the correlation between the slope and the bump strength (shallower slope $\Leftrightarrow$ weaker bump) and/or between the slope and the optical depth (higher $A_V \Leftrightarrow $ shallower slope;  \citealt{2013ApJ...775L..16K,2013MNRAS.432.2061C,2016ApJ...827...20S,2018ApJ...853...56R,2018ApJ...859...11S,2018MNRAS.475.2363T,2020ApJ...888..108B}).

It may be natural to expect that galaxies with higher sSFRs tend to have more complex, clumpy geometries of stars and dust.  Indeed, \citet{2013ApJ...775L..16K} showed that galaxies having a higher H$\alpha$ equivalent width tend to have attenuation curves that are flatter and exhibit weaker bumps. \citet{2020ApJ...888..108B} found a similar correlation with sSFR for $z\sim0.1$ galaxies.  Furthermore, \citet{2016ApJ...821...72S} found that the fraction of clumpy galaxies increases with sSFR across $0\lesssim z\lesssim3$.  The effects of dust-to-star geometry are therefore to be considered as an alternative explanation of the observed sSFR--$B_k$ anticorrelation, which does not require any changes in the underlying extinction curve, i.e., the nature of dust grains.
Distinguishing the effects of different geometries and grain compositions needs more detailed investigation.  

\subsection{The Assumed Baseline Attenuation Curve}
\label{sec:discussion:slope}

In this work, we adopted the \citet{2000ApJ...533..682C} law as the baseline of the attenuation curves for all galaxies in the sample, which is the same approach as employed by \citet{2009A&A...499...69N}.  The evolution of the dust population, however, implies that not only the bump profile but also the overall shape of the attenuation curve may be different at high redshift, as is the case locally.  As already mentioned above, the global shape of the attenuation curve also depends on the geometrical configuration of dust and stars \citep[e.g.,][]{2000ApJ...528..799W,2018ApJ...869...70N}.

For a given UV continuum, the measurement of $B_k$ depends on the overall slope of the baseline attenuation curve.  Application of a baseline curve that has a steeper rise toward the FUV results in a lower $E(B-V)$ for a given observed UV slope and thus a higher $B_k$ for a given level of absolute excess absorption due to the UV bump.  Possible variations in the underlying attenuation curve also have impacts on the SFR derived from the UV luminosity.  Applying a steeper attenuation curve will lead to a lower level of dust attenuation and thus a lower SFR for a given UV continuum.

Observationally, \citet{2015ApJ...800..108S} found that star-forming galaxies at $2<z<6.5$ have an average attenuation curve that is very similar to the \citet{2000ApJ...533..682C} curve in the overall shape and present a moderate UV bump feature.
On the other hand, there have been several claims that $z\sim2$ star-forming galaxies have an attenuation curve that is steeper than the \citet{2000ApJ...533..682C} curve \citep{2012A&A...545A.141B,2016ApJ...827...20S,2018ApJ...853...56R,2020ApJ...888..108B}.
For example, \citet{2012A&A...545A.141B} found an average $\left<B_k\right>=1.6$ and $\left<\delta\right>=-0.27$ for galaxies at $1\lesssim z\lesssim2$ applying a ``modified'' \citet{2000ApJ...533..682C} curve,
\begin{equation}
    k_\mathrm{mod}(\lambda)=\frac{R_V}{4.05} k_\mathrm{Cal}(\lambda)\left(\frac{\lambda}{5500\textrm{\AA}}\right)^\delta + k_\mathrm{bump}(\lambda),
    \label{eq:kmod}
\end{equation}
where $\delta$ modifies the slope.  Their $\left<B_k\right>=1.6$ is higher than our measurements at face value.  If we adopted the modified baseline curve using their average value of $\delta=-0.27$, however, we obtained $B_k\approx1.5$ from the stack of the entire sample, which is in agreement with the average value of \citet{2012A&A...545A.141B}.

If all galaxies in the sample can be assumed to follow a common baseline attenuation, whatever the slope, then the dependence of $B_k$ that we found would qualitatively still hold.  
On the other hand, there are some claims that lower sSFR galaxies tend to have a steeper attenuation curve \citep{2013ApJ...775L..16K,2020ApJ...888..108B}.  If this is true, then our analysis may have underestimated $B_k$ and overestimated SFR in lower sSFR galaxies.  Subsequently, we may have artificially narrowed both the range in sSFR (or the width of the MS, i.e., the range of  $\Delta\log~\mathrm{sSFR}$) and the range in $B_\mathrm{k}$; thus, the effects on the coefficients in Equations (\ref{eq:B_k=M,sSFR}) and (\ref{eq:B_k=M,dMS}) would be small.

Though the conversion to the amplitude in $k(\lambda)$ depends on the assumption, we stress that the presence of the UV bump in the attenuation curves has been robustly confirmed through the pure observables such as $B_A$ and $\gamma_{34}$ for our sample.  Attenuation curves without a UV bump cannot explain the shape of the stacked spectra that are bent at $\lambda\approx2175$~{\AA}.  The observed trend in $B_A$ (Figures \ref{fig:beta_b_vs_gamma34_2Dbin}, \ref{fig:M_vs_sSFR_2Dbin}, and \ref{fig:M_vs_sSFR_wLocal}), therefore, is essentially independent of the assumption of the shape of the baseline attenuation curve and is thus useful for predicting the absolute excess absorption for galaxies at similar redshifts.  

The shape of the attenuation curve potentially has significant impacts on the SEDs of galaxies and thus derived fundamental quantities such as SFR, as mentioned above.  
In particular, the potential variations in the FUV slope of the attenuation curves also imply that the attenuation of the Lyman continuum photons may even largely vary from galaxy to galaxy.  This, for instance, may have a significant impact in estimating SFR from the H$\alpha$ flux (or whatever the recombination line flux), since dust absorption in the Lyman continuum reduces the number of produced H$\alpha$ photons \citep{2016A&A...586A..83P}.  Variations in the Lyman continuum absorption may also affect the measurements of the ionizing photon escape fraction of galaxies.  Accurate determination of the overall shape of the attenuation curve is thus essential for better understanding the evolution of galaxies.

\section{Summary}
\label{sec:summary}

We have investigated the strength of the 2175~{\AA} UV bump feature in the attenuation curves of a sample of 505 star-forming galaxies at $1.3\le z\le 1.8$ in the zCOSMOS-deep survey.  Approximately 30\% of the galaxies exhibit a robust signature of the UV bump ($\gamma_{34}<-2$) in their individual spectra (Section \ref{sec:results_UVcontinua}).  
Significant intrinsic scatter in the observed $\gamma_{34}$ at a given UV slope clearly indicates the presence of a real diversity in the bump strength in the attenuation curves across the sample (see Section \ref{sec:results_UVcontinua}).

To increase the S/N, we have also measured the UV bump profiles in stacked spectra representing the whole sample, in subsamples of galaxies selected in ($\gamma_{34}$, $\beta_\mathrm{b}$) space, and in stellar mass and sSFR.  The attenuation profiles are all well described by Drude profiles with a center wavelength of 2175~{\AA} but varying amplitudes.  The derived bump amplitudes vary across the range $B_k\approx 0.2\textrm{--}0.8$ (Section \ref{sec:results_bumpProfiles}) with an inverse correlation with $\gamma_{34}$.   Using the stacks in $M_\ast$ bins, we found a tight positive $M_\ast$--$B_k$ correlation across $9.8\lesssim \log M_\ast/M_\odot \lesssim 10.5$ along the MS, though the correlation may not hold at the low- and high-mass ends (Figure \ref{fig:mass_vs_Ebump}).

Binning the sample in the $M_\ast$--sSFR plane, we found that there is a strong negative trend between $B_k$ and sSFR at fixed $M_\ast$ while $B_k$ increases moderately with $M_\ast$ at fixed sSFR (Section \ref{sec:results_bump_vs_gal}).  
This correlation with sSFR in the high-redshift sample is strikingly counter to the observed absence of the UV bump in the average attenuation curve of local starburst galaxies, since these local sources actually have lower sSFRs than our high-$z$ MS sample.
We found, however, that the two samples empirically come into much better agreement if we plot the bump strength against the $\Delta\log~\mathrm{sSFR}$ relative to the evolving MS at the appropriate epoch, rather than against the sSFR itself (see Figure \ref{fig:M_vs_dMS_wLocal}).   

We have interpreted these findings in terms of the recent SFH of the galaxies, especially considering the changes in the underlying grain properties.  The bump strength is determined by the balance between the destruction and production of the bump carriers.  The former may be accelerated by a higher frequency of SNe and/or more intense radiation fields in galaxies with higher instantaneous sSFRs, whereas the latter only reflects the SFH 1~Gyr before, but not the current SFR, for the time delay in the appearance of carbon-rich AGB stars (which are here assumed to be the main suppliers of precursor carbonaceous grains of the bump carriers) from the onset of star formation.  This suggests that the bump strength should be largely determined by the ratio of the current SFR measured on order $10^{7\textrm{--}8}~\mathrm{yr}$ timescales to that of order 1~Gyr ago, which we denote as SF$_{79}$ (Section \ref{sec:discussion:interpretation}). 

We therefore explored how the SF$_{79}$ would be expected to vary with the sSFR of a galaxy in two different regimes: (i) that of a quasi-constant sSFR and (ii) that of a short-lived rapid increase in the SFR.  This reveals an interesting effect.  High-redshift MS galaxies with quasi-constant sSFRs of order of $\mathrm{sSFR}\sim1~\mathrm{Gyr}^{-1}$ (appropriate for the high redshift sample at $z \sim 2$) and local starburst galaxies that are undergoing a short-lived sharp increase in SFR should both exhibit the same relations between SF$_{79}$ and $\Delta$sSFR but quite different relations between SF$_{79}$ and sSFR.  

The fact that the bump strength is observed to behave in this same way therefore adds weight to the idea that it is the variation in SFR over the last roughly 1~Gyr, as parameterized by SF$_{79}$, that is responsible for the observed variations in the bump strength through the creation and destruction of the carrier grains responsible for the bump.  

A prediction of this no doubt oversimplified picture is that MS galaxies at low redshift should show a very much weaker trend of bump strength with $\Delta \log~\mathrm{sSFR}$ than their high-redshift counterparts.  This prediction, however, appears to be counter to recent results of \citet{2018ApJ...859...11S}.  This apparent conflict may indicate that the radiative transfer effects in different dust-to-star geometries and dust column densities play roles in shaping the attenuation curves.

In the future, next-generation multi-object spectrographs, such as VLT/MOONS and Subaru/Prime Focus Spectrograph, will provide us with rest-frame UV spectra of order $10^{5\textrm{--}6}$ galaxies at high redshifts.  This will enable us to investigate the attenuation curves and the bump feature against various galaxy properties using both individual spectra and stacks and thus to understand much better the nature of dust through cosmic time.

\acknowledgments
{
We are grateful to Irene Shivaei for useful discussions and thank the anonymous referee for constructive comments and suggestions.
This research is based on observations undertaken at the European Southern Observatory (ESO) Very Large Telescope (VLT) under Large Program 175.A-0839 and has been supported by the Swiss National Science Foundation (SNF).
This work was supported in part by KAKENHI (JDS 26400221) through the Japan Society for the Promotion of Science (JSPS) and the World Premier International Research Center Initiative (WPI), MEXT, Japan.
This work uses data collected at the Subaru telescope, which is operated by the National Astronomical Observatory of Japan.
}


\bibliography{ads}
\bibliographystyle{aasjournal}



\appendix
\section{Reanalysis with the UV$+$IR-based SFRs}
\label{sec:Appendix:reanalysis}

In our main analysis, we used total SFRs that were estimated from the UV luminosity with appropriate correction for dust absorption (see Section \ref{sec:M_and_SFR}).
However, for a nonnegligible fraction of the galaxies, the dust correction becomes quite large (attenuation at rest frame 1600~{\AA}, $A_{1600} \gtrsim 2.5$); thus, the dust-corrected UV luminosity may be uncertain.  Another concern is that the rest-frame UV emission does not trace highly dust-obscured star formation \citep{2017ApJ...838L..18P}; thus, dusty starburst galaxies could be included in our sample as normal star-forming galaxies.  Therefore, here we reestimate the total SFRs of our sample galaxies by incorporating the rest-frame far-IR--to--millimeter photometry and present the results from reanalysis using the new SFR estimates.

\subsection{SFR from IR Luminosity}
\label{sec:Appendix:SFR_UVIR}

We utilized a ``super-deblended'' far-IR--to--millimeter photometric catalog presented by \citet{2018ApJ...864...56J}.  This catalog contains point spread function fitting photometry at fixed prior positions including 88,008 galaxies detected in VLA 1.4~GHz, 3~GHz, and/or MIPS 24~$\mu$m images.  To derive the total IR luminosity, we use the available photometry in the five Herschel PACS (100 and 160~$\mu$m) and SPIRE (250, 350, and 500~$\mu$m) complemented with JCMT/SCUBA2 850~$\mu$m, ASTE/AzTEC 1.1~mm, and IRAM/MAMBO 1.2~mm.  We fit these photometric fluxes with a coupled modified blackbody plus mid-IR truncated power-law component using the prescription given in \citet{2012MNRAS.425.3094C}.  The total IR luminosity, $L_\mathrm{IR}$, is then taken from the rest frame 8--1000~$\mu$m and converted to the IR-based SFR by employing a relation in \citet{2014ARA&A..52..415M} converted to a \citet{2003PASP..115..763C} IMF:
\begin{equation}
    \mathrm{SFR}_\mathrm{IR}~(M_\odot~\mathrm{yr}^{-1}) = 2.64 \times 10^{-44} L_\mathrm{IR}~(\mathrm{erg~s^{-1}}).
\end{equation}
The total SFR is then computed as $\mathrm{SFR}_\mathrm{UV+IR}=\mathrm{SFR}_\mathrm{UV}+\mathrm{SFR}_\mathrm{IR}$ where $\mathrm{SFR}_\mathrm{UV}$ is computed from the observed UV luminosity not corrected for dust absorption.

\subsection{The Bump Strength as a Function of $M_\ast$ and sSFR}
\label{sec:Appendix:bump_vs_gal}

\begin{figure}[tbp]
\begin{center}
\includegraphics[width=3.3in]{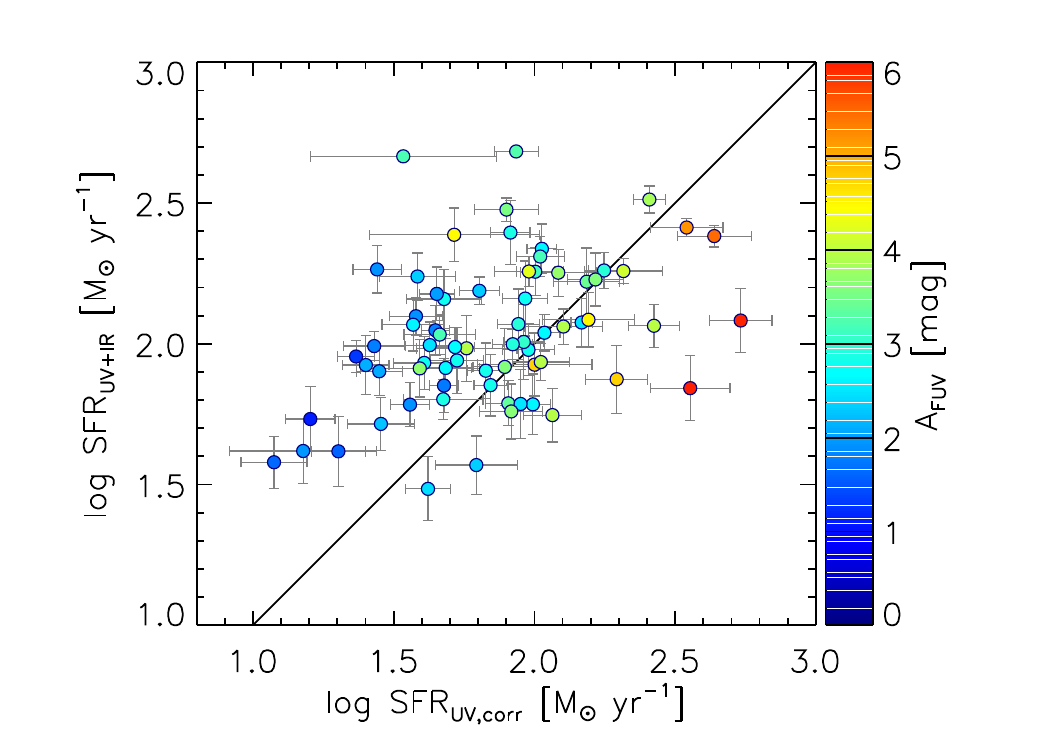}
\includegraphics[width=3.3in]{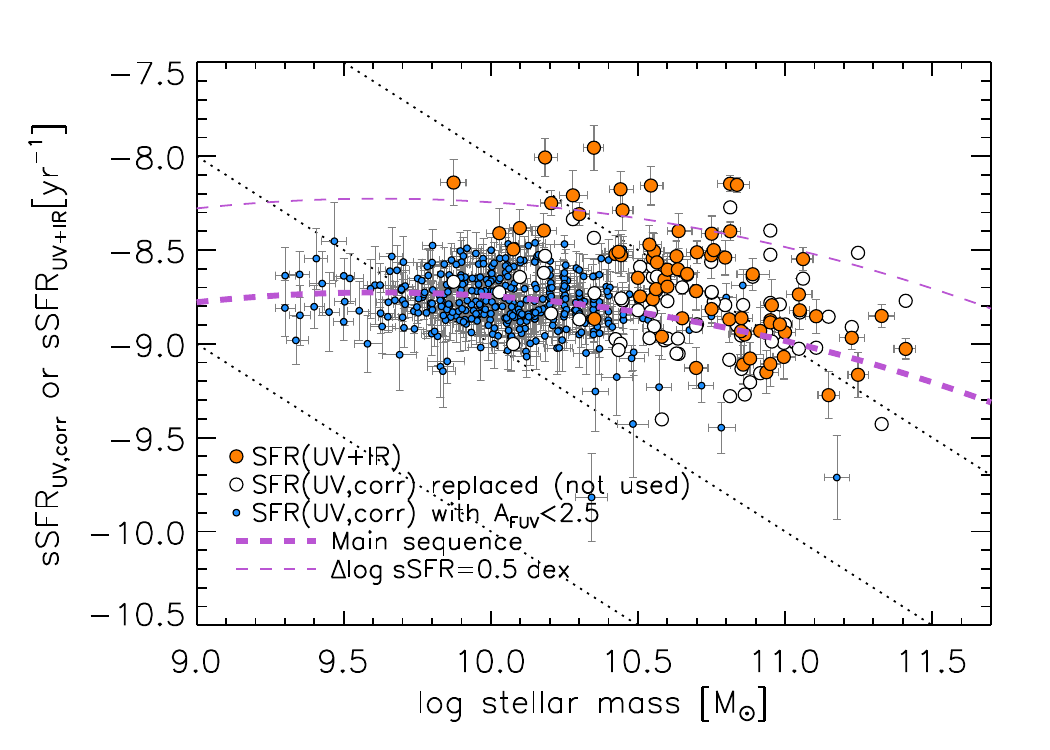}
\caption{
Upper panel: comparison between SFR$_\mathrm{UV+IR}$ and SFR$_\mathrm{UV,corr}$ for 70 galaxies in our sample for which $L_\mathrm{IR}$ is measured at $\mathrm{S/N}\ge3.0$.  Each symbol is color-coded with $A_{1600}$.  The diagonal line indicates a one-to-one relation.
Lower panel: sSFR as a function of $M_\ast$.  Small blue circles indicate galaxies with $A_{1600}\le 2.5~\mathrm{mag}$ for which SFR$_\mathrm{UV,corr}$ is adopted for reanalysis.  Orange circles indicate those for which $L_\mathrm{IR}$ is measured at $\mathrm{S/N}\ge3.0$ and thus SFR is replaced with SFR$_\mathrm{UV+IR}$, while white circles indicate the original SFR$_\mathrm{UV,corr}$, which is not used for reanalysis.
\label{fig:SFR_UVIR}}
\end{center}
\end{figure}

Cross-matching our sample with the far-IR catalog, we found a counterpart for 429 among our sample of 505 galaxies.  For 70 of these, we measured $L_\mathrm{IR}$ at $\mathrm{S/N}\ge3.0$, ranging across $\log L_\mathrm{IR}/L_\odot\sim11.5\textrm{--}12.5$. In Figure \ref{fig:SFR_UVIR}, we compare the new SFR$_\mathrm{UV+IR}$ with the fiducial SFR$_\mathrm{UV,corr}$ for these 70 galaxies.  The data points are color-coded by the dust attenuation at rest frame 1600~{\AA}, $A_{1600}$.  Although these two SFRs are in broad agreement with each other, there is a substantial scatter.  Particularly, there are some objects having relatively low $A_{1600}$ whose SFR$_\mathrm{UV+IR}$ exceeds SFR$_\mathrm{UV,corr}$ by $\gtrsim0.4$~dex, suggesting that these galaxies contain star-forming regions that contribute largely to the total SFR but are heavily dust-obscured, and thus their total SFR is not fully recovered in SFR$_\mathrm{UV,corr}$ even by applying dust absorption correction.  
In contrast, a couple of sources are located well below the one-to-one relation with relatively large $A_{1600}$, suggesting that their SFR$_\mathrm{UV,corr}$ may be overestimated due to overcorrection for dust.  This implies that the uncertainties in the total SFR$_\mathrm{UV,corr}$ could, in general, be larger than the nominal error bars, especially in those with larger $A_{1600}$.

\begin{figure}[tbp]
\begin{center}
\includegraphics[width=3.3in]{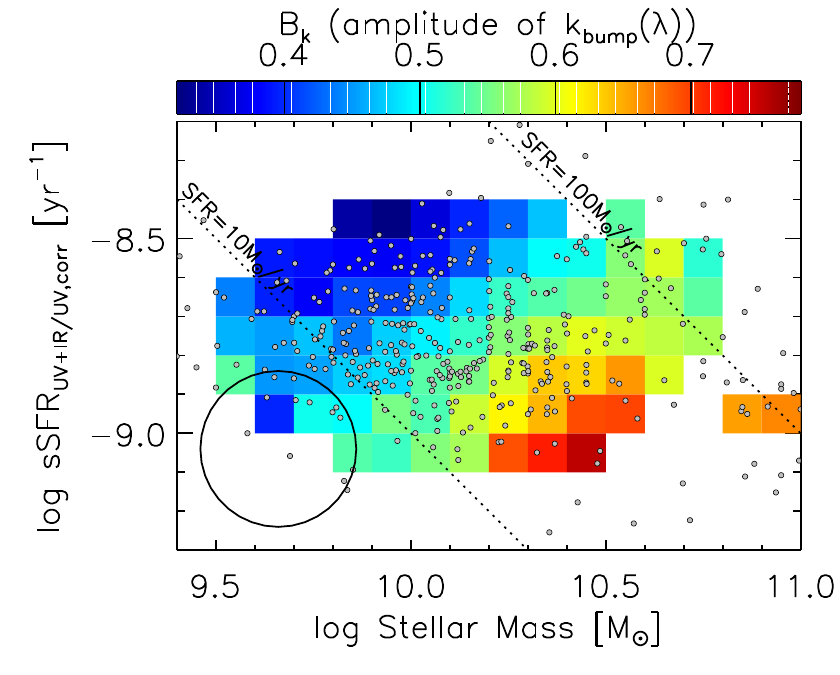}
\includegraphics[width=3.3in]{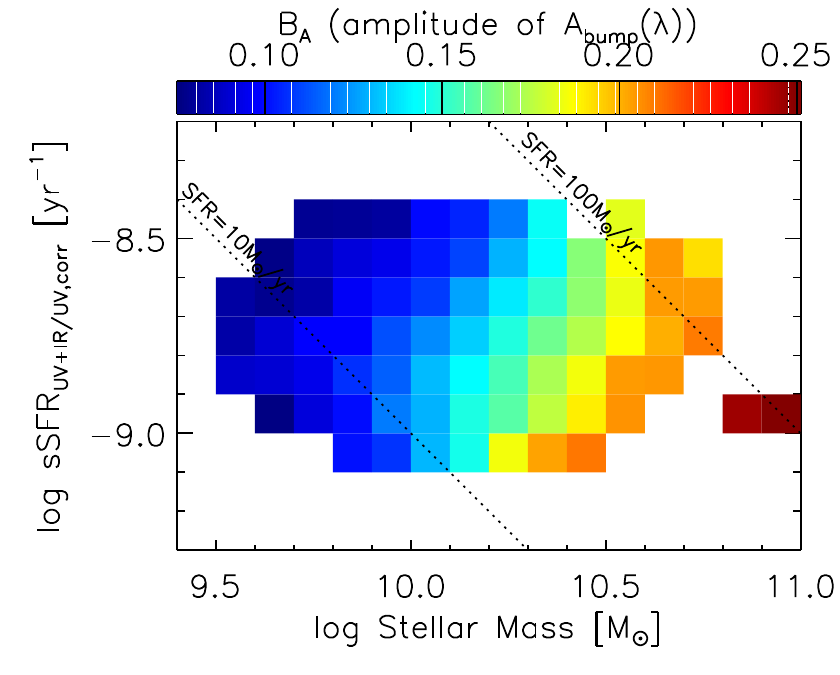}
\includegraphics[width=3.3in]{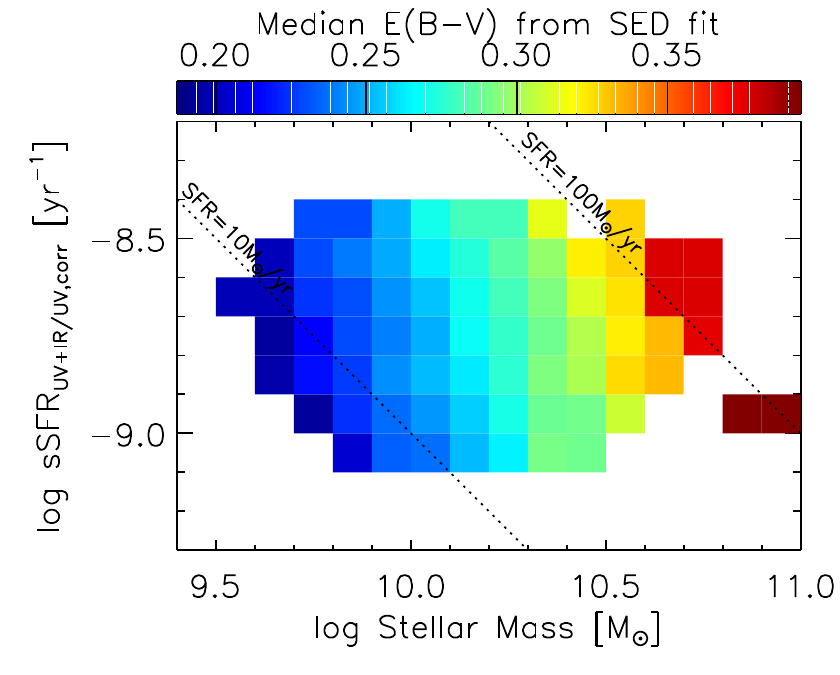}
\caption{Same as the right panels in Figure \ref{fig:M_vs_sSFR_2Dbin} but showing the results from the reanalysis using SFR$_\mathrm{UV+IR}$ as described in the text.  The ranges of the color bars are the same as in Figure \ref{fig:M_vs_sSFR_2Dbin}.
\label{fig:M_vs_sSFR_2Dbin_UVIR}}
\end{center}
\end{figure}

For the reanalysis, we replaced the fiducial SFR$_\mathrm{UV,corr}$ with the total $\mathrm{SFR}_\mathrm{UV+IR}$ for these 70 objects.  For the remainder, we adopted SFR$_\mathrm{UV,corr}$ if they have $A_{1600}<2.5~\mathrm{mag}$ (see Equation \ref{eq:AFUV}) but excluded the other 74 sources with $A_{1600}\ge2.5~\mathrm{mag}$ from the sample.  The final sample here contains 431 galaxies.  In the lower panel of Figure \ref{fig:SFR_UVIR}, we show the sample in the $M_\ast$ versus sSFR plane.  After replacing SFR$_\mathrm{UV,corr}$ with SFR$_\mathrm{UV+IR}$, the sample remains largely consistent with the same MS, with only a handful of objects located well above the MS ($\Delta \log~\mathrm{sSFR}\gtrsim0.5~\mathrm{dex}$).
These objects could be heavily obscured starburst galaxies.  In this paper, we do not specifically treat this type of galaxy because of their minor contribution to the whole sample and thus to the conclusions.

Using this sample, we carried out the same analysis described in Section \ref{sec:results_bump_vs_gal}.  An exception is that we stacked galaxies at each grid point in the $M_\ast$ versus sSFR plane within a radius of 0.2~dex in both axes, instead of 0.1~dex in the sSFR axis, because the number density of the data points in the $M_\ast$ versus sSFR plane is reduced (but this does not change the results anyway).  Figure \ref{fig:M_vs_sSFR_2Dbin_UVIR} shows the results.  The linear fits are given as 
\begin{eqnarray}
    \log B_k = && -0.242+0.167\log (M_\ast/10^{10}M_\odot) \nonumber \\ 
    && -0.299~\log (\mathrm{sSFR}/\mathrm{Gyr^{-1}}),\\
    \log B_A = && -0.888+0.399\log (M_\ast/10^{10}M_\odot) \nonumber \\ 
    && -0.211~\log (\mathrm{sSFR}/\mathrm{Gyr^{-1}}),\\
    \log E(B-V) = && -0.628+0.231\log (M_\ast/10^{10}M_\odot) \nonumber \\ 
    && +0.067~\log (\mathrm{sSFR}/\mathrm{Gyr^{-1}}).
    \end{eqnarray} 
The rms residuals in $B_k$, $B_A$, and $E(B-V)$ are, respectively, 0.040, 0.010, and 0.017.
The dependence of the bump amplitude $B_k$ appears to be very similar to what is seen in Figure \ref{fig:M_vs_sSFR_2Dbin}: $B_k$ is in a tight negative correlation with sSFR while moderately increasing with $M_\ast$.  This is also the case for $B_A$ and the median $E(B-V)$, as compared with the corresponding panels in Figure \ref{fig:M_vs_sSFR_2Dbin}.  The consistent result from the reanalysis makes our statements in this paper further robust.

\end{document}